\pdfoutput=1
\documentclass[12pt,letterpaper]{article}

\usepackage{amsmath}
\usepackage{amsfonts}
\usepackage{amssymb}
\usepackage{amsthm}
\usepackage{graphicx}
\usepackage{mathrsfs}
\usepackage{natbib}
\usepackage{url}
\usepackage{todonotes}
\usepackage{verbatim}
\usepackage{titletoc}

\usepackage[letterpaper]{geometry}
\usepackage{setspace}
\newtheoremstyle{break}
  {\topsep}{\topsep}%
  {\itshape}{}%
  {\bfseries}{}%
  {\newline}{}%
\theoremstyle{break}
\newtheorem{theorem}{Theorem}

\newtheorem{Ass}{Assumption}
\newtheorem{lem}{Lemma}
\newtheorem{cor}{Corollary}
\newtheorem{defi}{Definition}

\usepackage[font=small, labelfont=bf]{caption}

\DeclareMathOperator*{\Var}{Var}
\DeclareMathOperator*{\sign}{sign}

\author{Isaiah Andrews\footnote{Department of Economics, MIT, iandrews@mit.edu} \and Maximilian Kasy\footnote{Department of Economics, Harvard University, maximiliankasy@fas.harvard.edu}}

\title{Identification of and correction for publication bias\thanks{We thank Josh Angrist, Ellora Derenoncourt, Gary Chamberlain, Xavier D'Haultfoeuille, Gary King, Jesse Shapiro, Jann Spiess, and seminar participants at Brown, Carlos III, CEMFI, Columbia, CREST, Harvard/MIT, the Harvard applied statistics seminar, the Harvard development retreat, Microsoft Research, the Montreal econometrics seminar, Penn, Penn State, Princeton, and TSE for many helpful comments and suggestions.  We also thank Paul Wolfson and Dale Belman as well as Michael Kremer for sharing their data. This research was funded in part by the Silverman (1968) Family Career Development Chair at MIT, and by the National Science Foundation under grant number 1654234.}}

\begin{document}

\maketitle

\begin{abstract}
Some empirical results are more likely to be published than others.
Such selective publication leads to biased estimates and distorted inference.
This paper proposes two approaches for identifying the conditional probability of publication as a function of a study's results, the first based on systematic replication studies and the second based on meta-studies. 
For known conditional publication probabilities, we  propose median-unbiased estimators and associated confidence sets that correct for selective publication.
We apply our methods to recent large-scale replication studies in experimental economics and psychology, and to meta-studies of the effects of minimum wages and de-worming programs.\\

\noindent \textsc{Keywords: Publication bias, replication, meta-studies,\\ 
\hspace*{2.3cm}identification\\
JEL Codes: C18, C12, C13}

\clearpage

\end{abstract}

\section{Introduction}

Despite following the same protocols, replications of published experiments frequently find effects  of smaller magnitude or opposite sign than those in the initial studies \citep[cf.][]{open2015estimating, camerer2016evaluating}.
One leading explanation for replication failure is publication bias \citep[cf.][]{Ioannidis2005, ioannidis2008most, mccrary2016conservative, ChristensenMiguel2016}.  Journal editors and referees may be more likely to publish results that are  statistically significant, that confirm some prior belief or, conversely, that are surprising. 
Researchers in turn face strong incentives to select which findings to write up and submit to journals based on the likelihood of ultimate publication.
Together, these forms of selectivity lead to severe bias in published estimates and confidence sets.

This paper provides, to the best of our knowledge, the first nonparametric identification results for 
 the conditional publication probability as a function of the empirical results of a study.  Once the conditional publication probability is known, we derive bias-corrected estimators and confidence sets.  Finally, we apply the proposed methods to several empirical literatures.

\paragraph{Identification of publication bias}
Section \ref{sec:identification} considers two approaches to identification. 
The first uses data from systematic replications of a collection of original studies, each of which applies the same experimental protocol to a new sample from the same population as the corresponding original study. 
Absent selectivity, the joint distribution of initial and replication estimates is symmetric. 
Asymmetries in this joint distribution nonparametrically identify conditional publication probabilities, assuming the latter depend only on the initial estimate.

The second approach uses data from meta-studies.
Meta-studies statistically combine the estimates from multiple (published) studies to derive pooled estimates. Meta-studies are based on estimates and standard errors from these studies.
Absent
 selectivity the distribution of estimates for high variance studies is a noisier version of the distribution for low variance studies, under an independence assumption common in the meta-studies literature.
Deviations from this prediction again identify conditional publication probabilities.

Both approaches identify conditional publication probabilities up to scale. Multiplying publication probabilities by a constant factor does not change the distribution of published estimates, and likewise does not affect publication bias and size distortions.  
%Hence, identification up to scale does not pose any problems for our proposed bias corrections.

\paragraph{Correcting for publication bias}
Section \ref{sec:inference} discusses the consequences of selective publication for statistical inference.
For selectivity known (up to scale), we propose median unbiased estimators and valid confidence sets for scalar parameters. These results allow valid inference on the parameters of each study, rather than merely on average effects across a given literature.  For settings where we must estimate the degree of selectivity, we further propose Bonferroni-corrected confidence intervals which account for estimation error in the selection model.
The supplement derives optimal quantile-unbiased estimators for scalar parameters of interest in the presence of nuisance parameters, as well as results on Bayesian inference.

\paragraph{Applications}
Section \ref{sec:applications} applies the theory developed in this paper to four empirical literatures.  We first use data from the experimental economics and psychology replication studies of \cite{camerer2016evaluating} and \cite{open2015estimating}, respectively. Estimates based on our replication approach suggest that results significant at the 5\% level are over 30 times more likely to be published than are insignificant results, providing strong evidence of selectivity.
Estimation based on our meta-study approach, which uses only the originally published results, yields similar conclusions.

We then consider two settings where no replication estimates are available.
The first is the literature on the impact of minimum wages on employment.  Estimates based on data from the meta-study by \cite{wolfson201515} suggest 
that results corresponding a negative significant effect of the minimum wage on employment are about 3 times more likely to be published than are insignificant results. Positive and significant effects might also be less likely to be published than negative and significant effects, but the corresponding coefficient estimates are rather noisy.
Second, we consider the literature on the impact of mass deworming on child body weight.
Estimates based on data from the meta-study by \cite{deworming2016} find that results appear more likely to be included in this meta-study when they do not find a significant impact of deworming, though the standard errors are large and we cannot reject the null hypothesis of no selectivity.

\paragraph{Literature}
There is a large literature on publication bias; good reviews are provided by \cite{rothstein2006publication} and \cite{ChristensenMiguel2016}. 
We will discuss some of the approaches from this literature in the context of our framework below.
One popular method, used in e.g. \cite{CardKrueger1995} and \cite{egger1997bias}, regresses z-statistics on the inverse of the standard error and takes a non-zero intercept as evidence of publication bias. Our approach using meta-studies builds on related intuitions.
Another approach in the literature considers the distribution of p-values or z-statistics across studies, and takes bunching, discontinuities, or non-monotonicity in this distribution as indication of selectivity or estimate inflation \citep[cf.][]{DelongLang1992, brodeur2016star}.
Other approaches include the ``trim and fill'' method \citep{duval2000nonparametric} and parametric selection models \citep{iyengar1988selection, hedges1992modeling}.
Some precedent for our proposed corrections to inference can be found in \cite{mccrary2016conservative}, while the parametric models in our applications are  related to those of \cite{hedges1992modeling}.  

Further recent work on publication bias includes 
\cite{stanley2014meta}, who propose to use power as a weighting criterion for meta-analyses to increase robustness to selective publication. \cite{schuemie2014interpreting} suggest empirical calibration of p-values in medical research.
\cite{bruns2016p} and \cite{bruns2017meta} discuss meta analysis in observational settings with possibly biased estimates.
\cite{stanley2017finding} consider non-linear meta-regressions.
\cite{carter2017correcting} compare different meta-analytic methods for psychological research. 
Recent empirical studies exploring publication bias in economics and finance include  \cite{ioannidis2017power}, \cite{ChenZimmermann2017}, \cite{havranek2015measuring} and \cite{hou2017replicating}. Finally, 
\cite{Furukawa2017} proposes an economic model of publication bias.

\paragraph{Road map}
Section \ref{sec: setup} introduces the setting we consider, as well as a running example.
Section \ref{sec:identification} presents our main identification results, and discusses approaches from the literature.
Section \ref{sec:inference} discusses bias-corrected estimators and confidence sets, assuming conditional publication probabilities are known.
Section \ref{sec:applications} presents results for our empirical applications.
All proofs are given in the supplement, which also contains details of our applications, additional empirical and theoretical results, and a stylized model of optimal publication decisions.

\paragraph{Notation}
Throughout the paper, upper case letters denote random variables and lower case letters denote realizations.
The latent parameter governing the distribution of observables for a given study is $\Theta$. We condition on $\Theta$ whenever frequentist objects are considered, while unconditional expectations, probabilities, and densities integrate over the population distribution of $\Theta$ across studies.
Estimates are denoted by $X$, while estimates normalized by their standard deviation are denoted by $Z$.
Latent studies (published or unpublished) are indexed by $i$ and marked by a superscript $*$, while published studies are indexed by $j$. Subscripts $i$ and $j$ will sometimes be omitted when clear from context.

\section{Setting}
\label{sec: setup}

Throughout this paper we consider variants of the following data generating process.  Within an empirical literature of interest, there is a population of latent studies $i$. The true effect $\Theta_i^*$ in study $i$ is drawn from distribution $\mu$. Thus, different latent studies may estimate different true parameters. The case where all latent studies estimate the same parameter is nested by taking the distribution $\mu$ to be degenerate.

Conditional on the true effect, the result $X^*_i$ in latent study $i$ is drawn from a
known continuous distribution with density $f_{X^*|\Theta^*}$.
We take both $X_i^*$ and $\Theta_i^*$ to be scalar unless otherwise noted.
Studies are published if $D_i=1$, which occurs with probability $p(X^*_i)$, and we observe the truncated sample of published studies (that is, we observe $X_i^*$ if and only if $D_i=1$).  Publication decisions reflect both researcher and journal decisions; we do not attempt to disentangle the two. Let $I_j$ denote the index $i$ corresponding to the $j$th published study.
We obtain the following model:

\begin{defi}[Truncated sampling process]
\label{defi:DGP}
Consider the following data generating process for latent (unobserved) variables.\\
$(\Theta^*_i,X_i^*,D_i)$ are jointly i.i.d. across $i$, with
\begin{align*}
\Theta^*_i&\sim \mu\\
X^*_i | \Theta^*_i &\sim f_{X^*|\Theta^*}(x|\Theta^*_i)\\
D_i | X^*_i, \Theta^*_i &\sim Ber(p(X^*_i))
\end{align*}
Let $I_0=0$, $I_j = \min \{i:\; D_i=1,\; i>I_{j-1}\}$ and $\Theta_j = \Theta_{I_j}^*$.
We observe i.i.d. draws
$$
X_j=X^*_{I_j}.
$$
\end{defi}

Section \ref{sec:identification}  considers extensions of this model that allow us to identify and estimate  $p(\cdot)$.
Section \ref{sec:inference} discusses how to use knowledge of $p(\cdot)$ to  perform inference on $\Theta_j$ when $X_j$ is observed.
Of central importance throughout is the likelihood of observing $X_j$ given $\Theta_j$:
\begin{lem}[Truncated likelihood]
\label{lem:likelihood}
The truncated sampling process of Definition \ref{defi:DGP} implies the following likelihood:
\begin{equation}
f_{X|\Theta}\left(x|\theta\right) =   f_{X^*|\Theta^*, D}(x|\theta,1)= \frac{p\left(x\right)}{E\left[p\left(X^*_i\right)|\Theta^*_i=\theta\right]}f_{X^*|\Theta^*}\left(x|\theta\right). \label{eq: truncated density}
\end{equation}
\end{lem}

For fixed $\theta,$ selective publication reweights the distribution of published results by $p(\cdot).$  As we consider different values of $\theta$ for fixed $x,$ by contrast, the likelihood is scaled by the publication probability for a latent study with true effect $\theta$, $E\left[p\left(X^*_i\right)|\Theta^*_i=\theta\right].$

\paragraph{Study-level covariates}

The model of Definition \ref{defi:DGP}, and in particular independence between publication decisions and $\Theta^*$ given $X^*,$ may only hold conditional on some set of observable study characteristics $W^*$.
For example, journals may treat studies on particular topics, or using particular research designs, differently. 
Likewise, the distribution of true effects may differ across these categories.  
In this case Equation \eqref{eq: truncated density} would have to be modified to 
$$f_{X|\Theta, W}\left(x|\theta, w\right) =   f_{X^*|\Theta^*, W^*, D}(x|\theta,w,1)= \frac{p\left(x, w\right)}{E\left[p\left(X^*_i, W_i^*\right)|\Theta^*_i=\theta\right]}f_{X^*|\Theta^*}\left(x|\theta,w\right).$$
In our applications, for example, we consider conditioning on journal of publication and year of initial circulation of a study.
For simplicity of notation, however, we suppress such additional conditioning throughout our theoretical discussion.

\subsection{An illustrative example}

\label{ssec:illustrativeexample}

\begin{figure}
\includegraphics{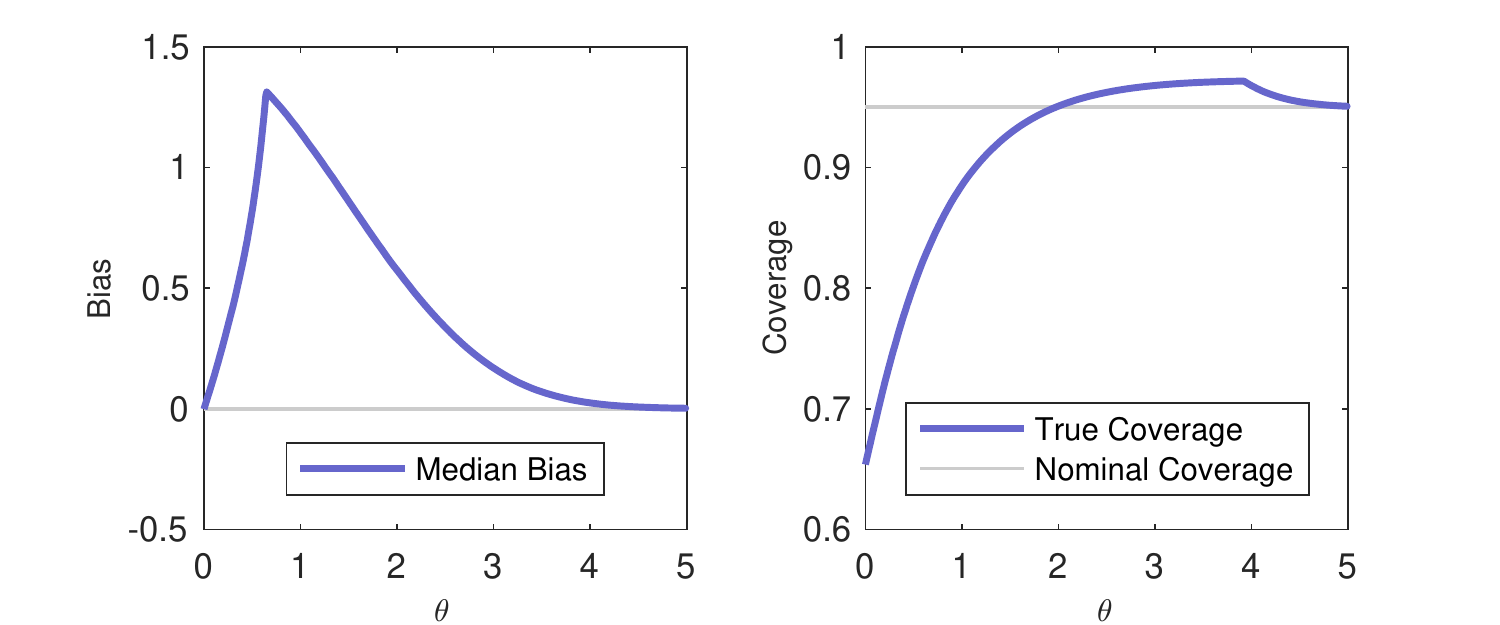}
\caption{The left panel plots the median bias of the conventional estimator $\hat{\Theta}_{j}=Z_{j}$, while the right panel plots the true coverage of the conventional 95\% confidence interval, both for $p(z)=.1 + .9 \cdot \mathbf{1}(|Z|>1.96)$.
 \label{fig:Truncated bias}}
\end{figure}

To illustrate our setting we consider a simple example to which we will return throughout the paper.
A journal receives a stream of studies $i=1,2,\ldots$ reporting experimental estimates $Z_i^*\sim N(\Theta_i^*, 1)$ of treatment effects $\Theta_i^*$, where each experiment examines a different treatment.  We denote the estimates by $Z^*$ rather than $X^*$ here to emphasize that they can be interpreted as z-statistics.  
Denote the distribution of treatment effects across latent studies by $\mu$.  Normality is in many cases a plausible asymptotic approximation; $\Var(Z^*|\Theta^*)=1$ is a scale normalization.  The journal publishes studies with $Z_i^*$ in the interval $\left[-1.96, 1.96 \right]$ with probability $p(Z_i^*)=.1$, while results outside this interval are published with probability $p(Z_i^*)=1$.  
This publication policy reflects a preference for 
``significant results,''  where a two-sided z-test rejects the null hypothesis $\Theta^*=0$ at the 5\% level.  This journal is ten times more likely to publish significant results than insignificant ones.
This selectivity results in publication bias: published results, whose distribution is given by Lemma \ref{lem:likelihood} above,   tend to over-estimate the magnitude of the treatment effect. Published confidence intervals under-cover the true parameter value for small values of $\Theta$ and over-cover for somewhat larger values.
This is demonstrated by Figure \ref{fig:Truncated bias}, which plots the median bias, $med(\hat\Theta_j|\Theta_j=\theta)-\theta,$ of the usual estimator $\hat\Theta_j=Z_j$, as well as the coverage of the conventional 95\% confidence interval $[Z_j-1.96, Z_j+1.96]$.  

\subsection{Alternative data generating processes}

To clarify the implications of our model, we contrast it with two alternative data generating processes.

\paragraph{Observability}
The setup of Definition \ref{defi:DGP} assumes that we only observe the draws $X^*$ for which $D = 1$.
Alternative assumptions about observability might be appropriate, however, if additional information is available.
First, we might know of the existence of unpublished studies, for example from experimental preregistrations, without observing their results $X^*$. 
In this case, called censoring, we observe i.i.d. draws of $(Y,D)$, where
$Y= D \cdot X^*$.
The corresponding censored likelihood is
\[
f_{Y,D|\Theta^*}(x, d|\theta^*) = d\cdot p(x) \cdot f_{X^*|\Theta^*}\left(x|\theta\right) + (1-d) \cdot (1-E[D_i| \Theta_i^*=\theta^*]).
\]
Second, we might additionally observe the results $X^*$ from unpublished working papers as in \cite{Francoetal2014}.
The likelihood in this case is 
\[f_{X^*,D| \Theta^*}(x,d | \theta)=  p(x)^d (1-p(x))^{1-d} \cdot f_{X^*|\Theta^*}(x|\theta).\]
Even under these alternative observability assumptions, the truncated likelihood (\ref{eq: truncated density}) arises as a limited information (conditional) likelihood, so identification and inference results based on this likelihood remain valid.  Specifically, this likelihood conditions on publication decisions in the model with censoring, and on both publication decisions and unpublished results in the model with $X^*$ observed. Thus, while additional information about the existence or content of unpublished studies might be used to gain additional insight, the results developed below continue to apply.

\paragraph{Manipulation of results}

Our analysis assumes that the distribution of the results $X^*$ in latent studies given the true effects $\Theta^*$, $f_{X^*|\Theta^*},$ is known.  This implicitly restricts the scope for researchers to  inflate the results of latent studies, cf. \cite{brodeur2016star}.
%, since if researchers directly manipulate their results this can imply an unknown form for $f_{X^*|\Theta^*},$ which is inconsistent with our framework.  
There are, however, many forms of manipulation or ``p-hacking'' \citep{Simonsohnetal2014} which are accommodated by our model.  
In particular, if researchers conduct many independent analyses (where the results of each analysis follow known $f_{X^*|\Theta^*}$) but write up and submit only significant analyses, this is a special case of our model.  More broadly, essentially any form of manipulation can be represented in a more general model where $p$ depends on both $X^*$ and $\Theta^*.$  This extension is discussed in Section \ref{sssec: selection on theta} below.

\section{Identifying selection}
\label{sec:identification}

This section proposes two approaches for identifying $p(\cdot).$  The first uses systematic replication studies.  By a ``replication'' we mean what \cite{Clemens2015} terms a ``reproduction,'' obtained by applying the same experimental protocol or analysis to a new sample from the same population as the original study. 
For each published $X$ in a given set of studies, such replications provide an independent estimate $X^r$ governed by the same parameter $\Theta$ as the original study.
Under the assumption that selectivity operates only on $X$ and not on $X^r$, we prove nonparametric identification of $p(\cdot)$ up to scale.  Under the additional assumption of normally distributed estimates we also establish identification of the latent distribution $\mu$ of true effects $\Theta^*.$
The distribution $\mu$ of $\Theta^*$, and more specifically the average $E[\Theta^*]$, is the key object of interest in most meta-studies; cf. \cite{rothstein2006publication}.
When the studies under consideration are on the same topic, for example the effect of minimum wage increases on employment, then the average provides a natural summary of the findings of this literature.

The second approach considers  meta-studies where there is variation across published studies in the standard deviation $\sigma$ of normally distributed estimates $X$ of $\Theta$, where normality can again be understood as arising from the usual asymptotic approximations.
Under the assumption that the standard deviation $\sigma^*$ is independent of $\Theta^*$ in the population of latent studies, and that publication probabilities are a function of the z-statistic $Z^*=X^* / \sigma$ alone, we again show nonparametric identification of $p(\cdot)$ up to scale, as well as of $\mu$.

%The results in this section prove non-parametric identification, without restrictions on functional form.
%To actually estimate $p(\cdot)$, as well as the distribution of $\Theta^*$, we then propose some parametric models which can be estimated using maximum likelihood.\footnote{Matlab and Python code which implements these estimators is available from the authors' webpages.}
%Such parametric specifications are indicated in practice given the limited number of observations in the available datasets -- the unit of observation are published studies. 

Identification based on systematic replication studies is considered in Section \ref{ssec:identificationreplication}.
Identification based on meta-studies is considered in Section \ref{ssec:identificationmeta}.  In both sections, we return to our treatment effect example to illustrate results and develop intuition.
Approaches in the literature, including meta-regressions and bunching of p-values, are discussed in the context of our assumptions in Section \ref{ssec:identificationliterature}. 

%\clearpage
\subsection{Systematic replication studies}

\label{ssec:identificationreplication}

We first consider the case of systematic replication studies, where both $X^*$ and $X^{*r}$ are drawn independently from the same known distribution $f_{X^*|\Theta^*}$, conditional on $\Theta^*$.
In this setting the joint density $f_{X^*,X^{*r}}$, integrating out $\Theta^*$, is symmetric in its arguments. Deviations from symmetry of $f_{X,X^r}$ identify $p(\cdot)$ up to scale.
We then extend this result in several ways, allowing different sample sizes for the original and replication studies as well as selection on $\Theta.$

%In practice this setting has to be modified in several ways before we can take it to the data provided by recent systematic replication studies such as \cite{open2015estimating} and \cite{camerer2016evaluating}.
%First, replication sample sizes deviate from the original sample sizes in these studies, and are chosen based on power calculations informed by the initial estimates $X$.
%Second, the sign of the initial estimates is normalized to be positive in these replication studies.
%We modify our results accordingly, and show that nonparametric identification is preserved.
%We propose estimators based on these modifications; these are the estimators we take to the data in section \ref{sec:applications}.

%A more fundamental violation of our assumptions arises from selectivity based on unobserved results, such as robustness checks and placebo tests. Such additional selectivity can break the conditional independence $D^* \perp \Theta^* | X^*$, so that the probability of publication given $X^*$ and $\Theta^*$ becomes $p(X^*, \Theta^*).$  In this setting, we prove that we can still identify the truncated distribution $f_{X|\Theta}$ using replication studies, which is enough to let us apply our selection-corrected inference procedures.

\subsubsection{The symmetric baseline case}
\label{sssec: symmetric baseline}

We extend the model in Definition \ref{defi:DGP} above to incorporate a conditionally independent replication draw $X^{*r}$ which is observed whenever $X^*$ is.
The key implications of our model are symmetry of the joint distribution of $(X^*, X^{*r})$, and that selectivity of publication operates only on $X^*$ and not on $X^{*r}$.
The latter assumption is plausible for systematic replication studies such as \cite{open2015estimating} and \cite{camerer2016evaluating}, but may fail in non-systematic replication settings, for instance if replication studies are published only when they ``debunk'' prior published results.

%A distinct but  related concern is that studies for which replication is attempted may not be representative of the published literature as a whole, for example if replication is attempted only for controversial or surprising results.  In this case our assumptions continue to hold but $p(\cdot)$ should be interpreted as the probability that a latent study is both published and chosen for replication.  %We discuss this issue in the context of our application below.

\begin{defi}[Replication data generating process]
\label{defi:replicationDGP}
Consider the following data generating process for latent (unobserved) variables.\\
$(\Theta^*_i, X_i^*, D_i, X_i^{*r}, )$ are jointly i.i.d. across $i$, with
\begin{align*}
\Theta^*_i&\sim \mu\\
X^*_i | \Theta^*_i &\sim f_{X^*|\Theta^*}(x|\Theta^*_i)\\
D_i | X^*_i, \Theta^*_i &\sim Ber(p(X^*_i))\\
X^{*r}_i | D_i, X^*_i, \Theta^*_i &\sim f_{X^*|\Theta^*}(x|\Theta^*_i).
\end{align*}
Let  $I_0=0$, $I_j = \min \{i:\; D_i=1,\; i>I_{j-1}\}$ and $\Theta_j = \Theta_{I_j}$.
We observe i.i.d. draws of
$$
(X_j, X^r_j) = (X^*_{I_j}, X^{*r}_{I_j}).
$$
\end{defi}

The next result extends Lemma \ref{lem:likelihood} to derive the joint density of $(X, X^r)$.

\begin{lem}[Replication Density]
\label{lem:replicationlikelihood}
Consider the setup of Definition \ref{defi:replicationDGP}. In this setup, the conditional density of $(X,X^r)$ given $\Theta$ is
\begin{align*}
f_{X,X^r|\Theta}(x, x^r|\theta)&= f_{X^*,X^{*r}|\Theta^*,D}(x, x^r|\theta,1)\\
&= \frac{p(x)}{E[p(X^*_i)|\Theta^*_i=\theta]}f_{X^*|\Theta^*}\left(x|\theta\right) f_{X^*|\Theta^*}\left(x^r|\theta\right).
\end{align*} 
The marginal density of $(X,X^r)$ is
$$
f_{X,X^r}(x, x^r) = \frac{p(x)}{E[p(X^*_i)]}  \int f_{X^*|\Theta^*}\left(x|\theta_i^*\right) f_{X^*|\Theta^*}\left(x^r|\theta_i^*\right) d\mu(\theta^*_i).
$$
\end{lem}

This lemma immediately implies that any asymmetries in the joint distribution of $X,X^r$ must arise from the publication probability $p(\cdot).$  In particular,
\[  \frac{f_{X, X^r}(b,a)}{f_{X, X^r}(a,b) } =\frac{p(b)}{p(a)},\]
whenever the denominators on either side are non-zero.
Using this fact, we prove that $p(\cdot)$ is nonparametrically identified up to scale.

\begin{theorem}[Nonparametric identification using replication experiments]
\label{theo:identificationreplication}
Consider the setup for replication experiments of Definition \ref{defi:replicationDGP}, and assume that the support of $f_{X^*, X^{*r}}$ is of the form $A \times A$ for some measurable set $A$.
In this setup $p(\cdot)$ is nonparametrically identified on $A$ up to scale.
\end{theorem}

\paragraph{Testable restrictions}
The density derived in Lemma \ref{lem:replicationlikelihood} shows that the model of Definition \ref{defi:replicationDGP} implies testable restrictions. Specifically, define $h(a,b) = \log(f_{X, X^r}(b,a)) - \log(f_{X, X^r}(a,b))$.
By Lemma \ref{lem:replicationlikelihood}, $h(a,b) = \log(p(b)) - \log(p(a))$, and therefore
\[h(a,b) + h(b,c) + h(c,a) = 0\]
for any three values $a,b,c$.  One could construct a nonparametric test of the model based on these restrictions and an estimate of $f_{X,X^r}$.
In the applications below we opt for an alternative approach, and test restrictions on an identified model which nests the setup of Definition \ref{defi:replicationDGP}, detailed in Section \ref{sssec: selection on theta}  below.

\paragraph{Illustrative example (continued)}
To illustrate our identification approach using replication studies, we return to the illustrative example introduced in Section \ref{ssec:illustrativeexample}.
In this setting, suppose that the true effect $\Theta^*$ is distributed $N(1, 1)$ across latent studies. As before, assume that $Z^*$ is $N(\Theta^*, 1)$ distributed conditional on $\Theta^*,$ that $p(Z^*)=1$ when $|Z^*| > 1.96$,  and that $p(Z^*)=.1$ otherwise.  Hence, results that are significantly different from zero at the 5\% level based on a two-sided z-test are ten times more likely to be published than are insignificant results. 
 
\begin{figure}[t!]
\includegraphics{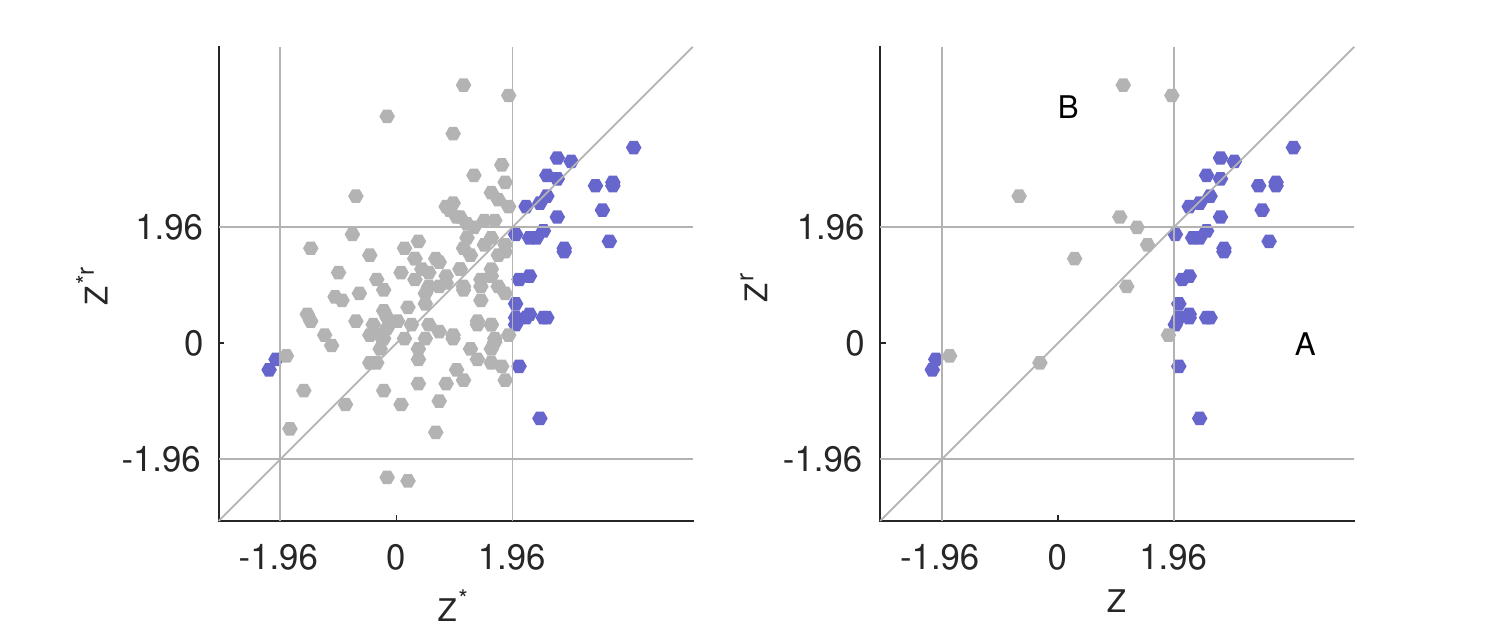}
\protect\caption{This figure illustrates the effect of selective publication in the replication experiments setting using simulated data, where selection is on statistical significance, as described in the text. The left panel shows the joint distribution of a random sample of latent estimates and replications; the right panel shows the subset which are published.  Results where the original estimates are significantly different from zero at the 5\% level are plotted in blue, while insignificant results are plotted in grey.
 \label{fig:ReplicationIllustration}}
\end{figure}

This setting is illustrated in Figure \ref{fig:ReplicationIllustration}. The left panel of this figure shows 100 random draws $(Z^*, Z^{*r})$; draws where $|Z^*|\leq 1.96$ are marked in grey, while draws where $|Z^*|> 1.96$ are marked in blue.
The right panel shows the subset of draws $(Z,Z^r)$ which are published. These are the same draws as $(Z^*, Z^{*r})$, except that 90\% of the draws for which $Z^*$ is statistically insignificant are deleted.

Our identification argument in this case proceeds by considering deviations from symmetry around the diagonal $Z=Z^r.$  Let us compare what happens in the regions marked $A$ and $B$.  In $A$, $Z$ is statistically significant but $Z^r$ is not; in $B$ it is the other way around. By symmetry of the data generating process, the latent $(Z^*, Z^{*r})$ fall in either area with equal probability. The fact that the observed $(Z,Z^r)$ lie in region $A$ substantially more often than in region $B$ thus provides evidence of selective publication, and the exact deviation of the distribution of $(Z,Z^r)$ from symmetry identifies $p(\cdot)$ up to scale.

\subsubsection{Generalizations and practical complications}

In practice we need to modify the assumptions above to fit our applications, where the sample size for the replication often differs from that in the initial study, and the sign of the initial estimate $X$ is normalized to be positive.
We thus extend our identification results to accommodate these issues.

\paragraph{Differing variances}

To account for the impact of differing sample sizes on the distribution of $X^{*r}$ relative to $X^*$, we need to be more specific about the form of these distributions. We assume that both $X^*$ and $X^{*r}$ are normally distributed unbiased estimates of the same latent parameter $\Theta^*$, and that their variances are known.  The assumption of approximate normality with known variance is already implicit in the inference procedures used in most applications.  Since we require normality of only the final estimate from each study, rather than the underlying data, this assumption can be justified using standard asymptotic results even in settings with non-normal data, heteroskedasticity, clustering, or other features commonly encountered in practice.
Normalizing the variance of the initial estimate to one yields the following setup, where we again denote the estimate by $Z$ rather than $X$ to emphasize the normalization of the variance.
\begin{align}
\Theta^*_i&\sim \mu \nonumber\\
Z^*_i | \Theta^*_i &\sim N(\Theta_i^*, 1) \nonumber \\
D_i | Z^*_i, \Theta^*_i &\sim Ber(p(Z^*_i)) \nonumber \\
\sigma_i^* |  Z^*_i, D_i, \Theta^*_i &\sim f_{\sigma|  Z^*} \nonumber \\
Z^{*r}_i | \sigma_i^*, Z^*_i, D_i, \Theta^*_i &\sim N(\Theta_i^*, \sigma_i^{*2})
\label{eq:replicationexperimentsSigma}
\end{align}
We use $\sigma$ to denote both the standard deviation as a random variable and the realized standard deviation.
We again assume that results are published whenever $D_i=1$, so that
\[f_{Z,Z^r,\sigma}(z, z^r, \sigma) = f_{Z^*,Z^{*r},\sigma^* | D}(z, z^r, \sigma|1).\]
Allowing the replication variance $\sigma_i^*$ to differ from one takes us out of the symmetric framework of Definition \ref{defi:replicationDGP}.
Display \ref{eq:replicationexperimentsSigma} also allows the possibility that the distribution of $\sigma_i^*$ might depend on $Z_i^*$. Dependence of $\sigma^*_i$ on $Z^*_i$ is present, for example, if power calculations are used to determine replication sample sizes, as in both \cite{open2015estimating} and \cite{camerer2016evaluating}. In that case, $\sigma_i^*$ is positively related to the magnitude of $Z_i^*$, but conditionally unrelated to $\Theta_i^*$.

The following corollary states that identification carries over to this setting.
The proof relies on the fact that we can recover the symmetric setting by (de)convolution of $Z^r$ with normal noise, given $Z$ and $\sigma$, which then allows us to apply Theorem \ref{theo:identificationreplication}.
The assumption of normality further allows recovery of $\mu$, the distribution of $\Theta^*$.
\begin{cor}
\label{corr:replicationexperimentsSigma}
Consider the setup for replication experiments in display \eqref{eq:replicationexperimentsSigma}. Suppose we observe i.i.d. draws of $(Z, Z^r)$.
In this setup $p(\cdot)$ is nonparametrically identified on $\mathbb{R}$ up to scale,
and $\mu$ is identified as well.
\end{cor}

\paragraph{Normalized sign}
A further complication is that the sign of the estimates $Z$ in our replication datasets is normalized to be positive, with the sign of $Z^r$ adjusted accordingly: see Section \ref{sec: econ experiments} below for further discussion.
The following corollary shows that under this sign normalization identification of $p(\cdot)$ still holds, so long as $p(\cdot)$ is symmetric.

\begin{cor}
\label{corr:replicationexperimentsNormalizedSign}
Consider the setup for replication experiments of display \eqref{eq:replicationexperimentsSigma}.
Assume additionally that $p(\cdot)$ is symmetric, $p(z)=p(-z)$, and that $f_{\sigma|Z^*}(\sigma|z)=f_{\sigma|Z^*}(\sigma|-z)$ for all $z$.
Suppose that we observe i.i.d. draws of
$$(W,W^r) = \sign(Z) \cdot (Z, Z^r). \label{eq: W def}$$
In this setup $p(\cdot)$ is non-parametrically identified on $\mathbb{R}$ up to scale,
and the distribution of $|\Theta^*|$ is identified as well.
\end{cor}

\subsubsection{Selection depending on $\Theta^*$ given $X^*$}
\label{sssec: selection on theta}

Selection of an empirical result $X$ for publication might depend not only on $X$ but also on other empirical findings reported in the same manuscript, or on unreported results obtained by the researcher. If that is the case, our assumption that publication decisions are independent of true effects conditional on reported results, $D\perp \Theta^* | X^*$, may fail.
Allowing for a more general selection probability $p(X^*,\Theta^*)$, we can still identify $f_{X|\Theta}$, which is the key object for bias-corrected inference as discussed in Section \ref{sec:inference}.
Consider the following setup.
\begin{align}
\Theta^*_i&\sim f_{\Theta^*} \nonumber\\
Z^*_i | \Theta^*_i &\sim N(\Theta_i^*, 1) \nonumber \\
D_i | Z^*_i, \Theta^*_i &\sim Ber(p(Z^*_i, \Theta_i^*)) \nonumber\\
\sigma^*_i |  D_i, Z^*_i,  \Theta^*_i &\sim f_{\sigma|  Z^*} \nonumber \\
Z^{*r}_i | \sigma^*_i, D_i,  Z^*_i, \Theta^*_i &\sim N(\Theta_i^*, \sigma_i^2) \label{eq:replicationexperimentsDependence}
\end{align}
Assume again that results are published whenever $D_i=1$.  The assumption $D_i | Z^*_i, \Theta^*_i \sim Ber(p(Z^*_i, \Theta_i^*))$ is the key generalization relative to the setup considered before. This allows publication decisions to depend on both the reported estimate and the true effect, and allows a wide range of models for the publication process.  In particular, this accommodates models where publication decisions depend on a variety of additional variables, including alternative specifications and robustness checks not reported in the replication dataset.  Publication probabilities conditional on $Z^*$ and $\Theta^*$ then implicitly average over these variables, resulting in additional dependence on $\Theta^*.$  For a worked-out example of this form, see Section \ref{suppsec: latent selection model} of the supplement.

\begin{theorem}
\label{theo:identificationDependence}
Consider the setup for replication experiments of display \eqref{eq:replicationexperimentsDependence}.
In this setup $f_{Z|\Theta}$ is nonparametrically identified.
\end{theorem}

The proof of Theorem \ref{theo:identificationDependence} implies that the joint density $f_{Z,Z^r,\sigma,\Theta}$ is identified.
Under the assumptions of display \eqref{eq:replicationexperimentsDependence} the joint density of $(Z,Z^r,\sigma,\Theta)$ is
\[f_{Z,Z^r,\sigma,\Theta}(z,z^r,\sigma,\theta) = \frac{p(z,\theta)}{E[p(Z^*,\Theta^*)]} \varphi(z-\theta) \tfrac{1}{\sigma}\varphi\left(\tfrac{z^r-\theta}{\sigma}\right)f_{\sigma|  Z^*}(\sigma|z) \tfrac{d\mu}{d\nu}(\theta),\]
where we use $\nu$ to denote a dominating measure on the support of $\Theta$.
%\footnote{Here we consider the density of $(X,\Theta)$ with respect to the product of Lebesgue measure on the support of $X$ with the dominating measure $\nu$ on the support of $\Theta$.}
Without further restrictions $p(z,\theta)$ is not identified; we can always divide $p(z,\theta)$ by some function $g(\theta)$ and multiply $\frac{d\mu}{d\nu}(\theta)$ by the same function to get an observationally equivalent model.
Theorem \ref{theo:identificationDependence} implies, however, that $p(z,\theta)$ is identified up to a normalization given $\theta,$ since
$$\frac{f_{Z|\Theta}(z,\theta)}{f_{Z^*|\Theta^*}(z,\theta)}=\frac{p(z,\theta)}{E[p(Z^*,\Theta^*)|\Theta^*=\theta]}.$$
We can for instance impose $\sup_z p(z,\theta) =1$ for all $\theta$ to get an identified model.
In our applications we consider a parametric version of this model and test $p(z,\theta)=p(z)$ as a specification check on our baseline model.

\subsection{Meta-studies}

\label{ssec:identificationmeta}

We next consider identification using meta-studies.
Suppose that studies report normally distributed estimates $X^*$ with mean $\Theta^*$ and standard deviation $\sigma^{*}$, and that selectivity of publication is based on the z-statistic $Z^*=X^*/\sigma^*$.
The key identifying assumption is that $\Theta^*$ is statistically independent of $\sigma^{*}$ across studies, so studies with larger sample sizes do not have systematically different estimands.
Under this assumption, the distribution of $X^*$ conditional on a larger value $\sigma^{*}=\sigma_1$ is equal to the 
convolution of normal noise of variance $\sigma_1^2-\sigma_2^2$ with the
distribution of $X^*$ conditional on a smaller value $\sigma^{*}=\sigma_2$.
Deviations from this equality for the observed distribution $f_{X|\sigma}$ identify $p(\cdot)$ up to scale.

\begin{defi}[Meta-study data generating process]
\label{defi:independentsigmaDGP}
Consider the following data generating process for latent (unobserved) variables.\\
 $(\sigma_i^*, \Theta^*_i,X_i^*,D_i)$ are jointly i.i.d. across $i$, such that
\begin{align*}
\sigma_i^* &\sim \mu_\sigma\\
\Theta_i^* | \sigma_i^* & \sim \mu_\Theta\\
X_i^* | \Theta_i^*, \sigma_i^* &\sim N(\Theta_i^*, \sigma_i^{*2})\\
D_i | X^*_i, \Theta_i^*, \sigma_i^* &\sim Ber(p(X^*_i/\sigma_i^*))
\end{align*}
Let  $I_0=0$, $I_j = \min \{i:\; D_i=1,\; i>I_{j-1}\}$ and $\Theta_j = \Theta_{I_j}$.
We observe i.i.d. draws of
$$
(X_j, \sigma_j) = (X^*_{I_j}, \sigma^*_{I_j}).
$$
Define $Z_i^*=\frac{X_i^*}{\sigma_i^*}$ and $Z_j=\frac{X_j}{\sigma_j}$.
\end{defi}

A key object for identification of $p(\cdot)$ in this setting is the conditional density $f_{Z|\sigma}$.

\begin{lem}[Meta-study density]
\label{lem:metalikelihood}
Consider the setup of definition \ref{defi:independentsigmaDGP}. The conditional density of $Z$ given $\sigma$ is
\[f_{Z|\sigma}(z|\sigma) = \frac{p(z)}{E[p(Z^*)|\sigma]} \int \varphi(z-\theta / \sigma) d\mu(\theta). \]
\end{lem}

We build on Lemma \ref{lem:metalikelihood} to prove our main identification result for the meta-studies setting.
Lemma \ref{lem:metalikelihood} implies that, for $\sigma_2>\sigma_1$,
\[
\frac{f_{Z|\sigma}(z|\sigma_2)}{f_{Z|\sigma}(z|\sigma_1)} =\frac{E[p(Z^*)|\sigma=\sigma_1]}{E[p(Z^*)|\sigma=\sigma_2]} \cdot \frac{\int \varphi(z-\theta / \sigma_2) d\mu(\theta)}{\int \varphi(z-\theta / \sigma_1) d\mu(\theta)},
\]
where the  first term on the right hand side does not depend on $z$.
Since $f_{Z|\sigma}(z|\sigma_2) / f_{Z|\sigma}(z|\sigma_1)$ is identified, this suggests we might be able to invert this equality to recover $\mu$, which would then immediately allow us to identify $p(\cdot)$.
The proof of Theorem \ref{theo:identificationsigmanonpar} builds on this idea, considering $\partial_\sigma \log(f_{Z|\sigma}(z|\sigma))$.

\begin{theorem}[Nonparametric identification using meta-studies]
\label{theo:identificationsigmanonpar}
Consider the setup for experiments with independent variation in $\sigma$, described by Definition \ref{defi:independentsigmaDGP}.
Suppose that the support of $\sigma$ contains an open interval.
Then $p(\cdot)$ is identified up to scale,
and $\mu$ is identified as well.
\end{theorem}

\paragraph{Illustrative example (continued)}

\begin{figure}
\includegraphics{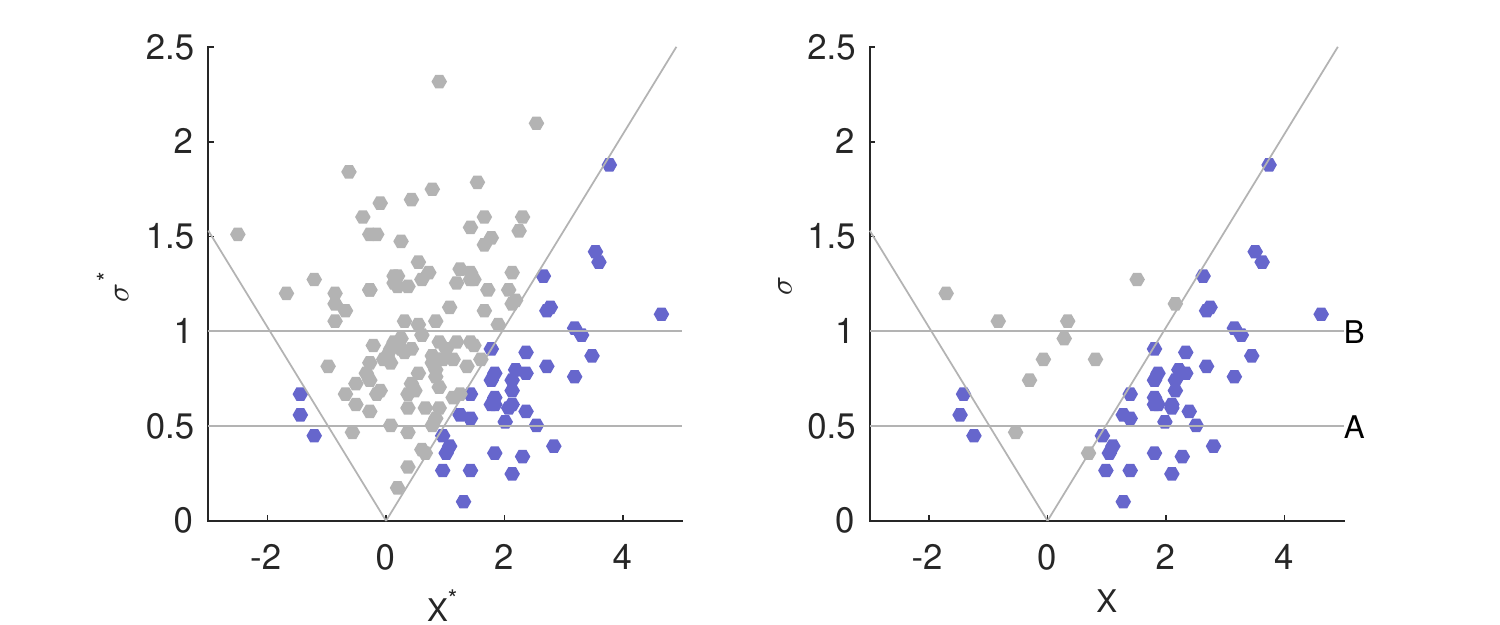}
\protect\caption{This figure illustrates the effect of selective publication in the meta-studies setting using simulated data, where selection is on statistical significance, as described in the text. The left panel shows a random sample of latent estimates; the right panel shows the subset of estimates which are published.  Results which are significantly different from zero at the 5\% level are plotted in blue, while insignificant results are plotted in grey.
 \label{fig:VariationSigmaIllustration}}
\end{figure}

As before, assume 
that $\Theta^*$ is $N(1, 1)$ distributed.
Suppose further that $\sigma^*$ is independent of $\Theta^*$ across latent studies, and that 
$X^*$ is $N(\Theta^*, \sigma^*)$ distributed conditional on $\Theta^*,$ $\sigma^*$.
Let  $p(X^*/\sigma^*)=1$ when $|X^*/\sigma^*| > 1.96$, $p(X^*/\sigma^*)=.1$ otherwise.  Thus, results which differ significantly from zero at the 5\% level are again ten times as likely to be published as insignificant results.
This setting is illustrated in Figure \ref{fig:VariationSigmaIllustration}.
The left panel of this figure shows 100 random draws $(X^*, \sigma^*)$; draws where $|X^*/\sigma^*|\leq 1.96$ are marked in grey, while draws where $|X^*/\sigma^*|> 1.96$ are marked in blue.
The right panel shows the subset of draws $(X,\sigma)$ which are published, where 90\% of statistically insignificant draws are deleted.

Compare what happens for two different values of the standard deviation $\sigma$, marked by $A$ and $B$ in Figure  \ref{fig:VariationSigmaIllustration}.
By the independence of $\sigma^*$ and $\Theta^*$, the distribution of $X^*$ for larger values of $\sigma^*$ is a noised up version of the distribution for smaller values of $\sigma^*$. To the extent that the same does not hold for the distribution of published $X$ given $\sigma$, this must be due to selectivity in the publication process.  In this example, statistically insignificant observations are ``missing'' for larger values $\sigma.$ Since publication is more likely when $|X^*/\sigma^*| > 1.96$, the estimated values $X$ tend to be larger on average for larger values of $\sigma$, and the details of how the conditional distribution of $X$ given $\sigma$ varies with $\sigma$ will again allow us to identify $p(\cdot)$ up to scale.

\paragraph{Normalized sign}
In some of our applications the sign of the reported estimates $X$ is again normalized to be positive.
The following corollary shows that $p(\cdot)$ remains identified under this sign normalization provided it is symmetric in its argument.

\begin{cor}
\label{corr:metastudyNormalizedSign}
Consider the setup of Definition \ref{defi:independentsigmaDGP}.
Assume additionally that $p(\cdot)$ is symmetric, i.e., $p(x/\sigma)=p(-x/\sigma)$.
Suppose that we observe i.i.d. draws of $(|X|, \sigma)$.
In this setup $p(\cdot)$ is nonparametrically identified on $\mathbb{R}$ up to scale,
and the distribution of $|\Theta^*|$ is identified as well.
\end{cor}

\paragraph{Dependence on $\sigma^*$}
Publication decisions might depend not only on the z-statistic $Z^*$, but also on the standard deviation $\sigma^*$.
Consider the setup of Definition \ref{defi:independentsigmaDGP} modified such that
$$D_i | X^*_i, \Theta_i^*, \sigma_i^* \sim Ber(p(X^*_i/\sigma_i^*) \cdot q(\sigma_i^*)).$$
Theorem \ref{theo:identificationsigmanonpar} immediately implies identification of the function $p(\cdot)$ for this generalized setup.
The generalized setup is observationally equivalent to the model of Definition \ref{defi:independentsigmaDGP} with the distribution of $\sigma^*$ reweighted by $q(\cdot)/E[q(\sigma^*)]$.

\subsection{Relation to approaches in the literature}

\label{ssec:identificationliterature}

Various approaches to detect selectivity and publication bias have been proposed in the literature.
We briefly analyze some of the these approaches in our framework.
First, we discuss to what extent we should expect the results of significance tests to 
``replicate'' in a sense considered in the literature, and show that the probability of such replication may be low even in the absence of publication bias.  
Second, we discuss meta-regressions, and show that while they provide a valid test of the null of no selectivity under our meta-study assumptions, they are difficult to interpret under the alternative.
Third, we consider approaches based on the distribution of p-values or z-statistics, and analyze the extent to which bunching or discontinuities of this distribution provide evidence for selectivity or inflation of estimates.

\paragraph{Should results ``replicate?''}
The findings of recent systematic replication studies such as \cite{open2015estimating} and \cite{camerer2016evaluating} are sometimes interpreted as indicating an inability to ``replicate the  results'' of published research.
In this setting, a ``result'' is understood to ``replicate'' if both the original study and its replication find a statistically significant effect in the same direction. The share of results which replicate in this sense is prominently discussed in  \cite{camerer2016evaluating}.
Our framework suggests, however, that the probability of replication in this sense might be low even without selective publication or other sources of bias.

Consider the setup for replication experiments in display \eqref{eq:replicationexperimentsSigma} with constant publication probability $p(\cdot)$, so that publication is not selective and $f_{Z,Z^r} = f_{Z^*, Z^{r*}}$. 
For illustration, assume further that $\sigma^{*}\equiv 1$.
The probability that a result replicates in the sense described above is
\begin{multline*}
P(Z^{*r}\cdot sign\{Z^*\} > 1.96 | |Z^*| > 1.96)\\
= \frac{P(Z^{*r} <- 1.96,\; Z^* <-  1.96) + P(Z^{*r}  > 1.96,\; Z^* > 1.96)}{P(Z^* <-  1.96) + P(Z^* > 1.96) }\\
= \frac{\int \left [\Phi(-1.96-\theta)^2 +  \Phi(-1.96+\theta)^2\right ] d\mu(\theta)}{\int \left [\Phi(-1.96-\theta) +  \Phi(-1.96+\theta)\right ] d\mu(\theta)}.
\end{multline*}
If the true effect is zero in all studies then this probability is $0.025.$
If the true effect in all studies is instead large, so that $|\Theta^*| > M$ with probability one for some large $M$, then the probability of replication is approximately one.
Thus, 
the probability that results replicate in this sense gives little indication of whether selective publication or some other source of bias for published research is present unless we either restrict the distribution of true effects or observe replication frequencies less than 0.025.  Strengths and weaknesses of alternative measures of replication are discussed in \cite{Simonsohn2015}, \cite{Gilbertetal2016}, and \cite{Patiletal2016}.

\paragraph{Meta-regressions}

A popular test for publication bias in meta-studies \citep[cf.][]{CardKrueger1995, egger1997bias} uses regressions of either of the following forms:
\begin{align*}
E^*[X | 1, \sigma] &= \gamma_0 + \gamma_1 \cdot \sigma,~~~~
 E^*\left [Z | 1, \tfrac{1}{\sigma} \right ] = \beta_0 + \beta_1 \cdot \tfrac{1}{\sigma},
\end{align*}
where we use $E^*$ to denote best linear predictors.
The following lemma is immediate.
\begin{lem}
Under the assumptions of Definition \ref{defi:independentsigmaDGP}, if $p(\cdot)$ is constant then 
\begin{align*}
E^*[X | 1, \sigma] = E[\Theta^*], ~~~E^*\left [Z | 1, \tfrac{1}{\sigma}  \right ] = E[\Theta^*]\cdot \tfrac{1}{\sigma}
\end{align*}
\end{lem}
As this lemma confirms, meta-regressions can be used to construct tests for the null of no publication bias. In particular, absent publication bias $\beta_0 = 0$ and $\gamma_1=0$, so tests for these null hypotheses allow us to test the hypothesis of no publication bias, though there are some forms of selectivity against which such tests have no power.
As also noted in the previous literature, absent publication bias the coefficients $\beta_1$ and $\gamma_0$ recover the average of $\Theta^*$ in the population of latent studies.  While these coefficients are sometimes interpreted as selection-corrected estimates of the mean effect across studies  \citep[cf. ][]{DoucouliagosStanley2009, ChristensenMiguel2016}, this interpretation is potentially misleading in the presence of publication bias.  In particular, the conditional expectation $E[X | 1, \sigma]$ is nonlinear in both $\sigma$ and $1/\sigma$, which implies that $\beta_0$, $\gamma_1$ are generally biased as estimates of $E[\Theta^*]$.\footnote{\cite{Stanley2008} and \cite{DoucouliagosStanley2009} note this bias but suggest that one can still use $H_0:\gamma_1=0$ to test the hypothesis of zero true effect if there is no heterogeneity in the true effect $\Theta^*$ across latent studies.} 
 To illustrate the resulting complications, we discuss a simple example with one-sided significance testing in Section \ref{suppsec: meta-regression coefficients} of the supplement.\footnote{A further complication is that meta-regression coefficients are not interpretable in settings with sign-normalized estimates, as in two of our applications.  See Section \ref{suppsubsec: Meta-studies} of the supplement for further discussion.}

\paragraph{The distribution of p-values and z-statistics}
Another approach in the literature considers the distribution of p-values, or the corresponding z-statistics, across published studies.  For example, \cite{Simonsohnetal2014} analyze whether the distribution of p-values in a given literature is right- or left-skewed.
\cite{brodeur2016star} compile 50,000 test results from all papers published in the American Economic Review, the Quarterly Journal of Economics, and the Journal of Political Economy between 2005 and 2011, and analyze their distribution to draw conclusions about distortions in the research process.

Under our model, absent selectivity of the publication process the distribution $f_Z$ is equal to $f_{Z^*}$.
If  we additionally assume that $Z^*|\Theta^* \sim N(\Theta^*,1)$ and $\Theta^* \sim \mu$, this implies that
$$
f_Z(z) = f_{Z^*}(z) = (\pi \ast \varphi)(z) = \int \varphi(z-\theta) d\mu(\theta).
$$
This model has testable implications, and requires that the deconvolution of $f_Z$ with a standard normal density $\varphi$ yield a probability measure $\mu$.
This implies that the density $f_{Z^*}$ is infinitely differentiable.
If selectivity is present, by contrast, then
\[f_Z(z) = \frac{p(z)}{E[p(Z^*)]} \cdot f_{Z^*}(z), \]
and any discontinuity of $f_Z(z)$ (for instance at critical values such as $z=1.96$) identifies a corresponding discontinuity of $p(z)$ and indicates the presence of selectivity:
$$\frac{\lim_{z \downarrow z_0} f_Z(z)}{\lim_{z \uparrow z_0} f_Z(z)}=
\frac{\lim_{z \downarrow z_0} p(z)}{\lim_{z \uparrow z_0} p(z)}.$$
If we impose that $p(\cdot)$ is a step function, for example, then this argument allows us to identify $p(\cdot)$ up to scale.

The density $f_{Z^*}$ also precludes excessive bunching, since for all $k\geq 0$ and all $z,$
$\partial_z^k f_{Z^*}(z) \leq \sup_z \partial_z^k \varphi(z)$ and $\partial_z^k f_{Z^*}(z) \geq \inf_z \partial_z^k \varphi(z)$
so that in particular $f_{Z^*}(z) \leq \varphi(0)$ and $f''_{Z^*}(z) \geq \varphi''(0)=-\varphi(0)$ for all $z$.
Spikes in the distribution of $Z$ thus likewise indicate the presence of selectivity or inflation.

Unlike our model, which focuses on selection, \cite{brodeur2016star} are interested in potential inflation of test results by researchers, and in particular in non-monotonicities of $f_Z$ which cannot be explained by monotone publication probabilities $p(z)$ alone. They construct tests for such non-monotonicities based on parametrically estimated distributions $f_{Z^*}$.

\section{Corrected inference}
\label{sec:inference}

This section derives median unbiased estimators and valid confidence sets for scalar parameters $\theta.$ For most of the section we assume $p(\cdot)$ is known; corrections accounting for estimation error in $p(\cdot)$ are discussed at the end of the section.  As in our identification results, $f_{X^*|\Theta^*}$ is assumed known throughout. The supplement extends these results to derive optimal estimators for scalar components of vector-valued $\theta,$ and analyzes Bayesian inference under selective publication. While our identification results in the last section relied on an empirical Bayes perspective, which assumed that $\Theta_i^*$ was drawn from some distribution $\mu,$ this section considers standard frequentist results which hold conditional on $\Theta$.
This reflects the different question at hand: Estimability of $p(\cdot)$, as considered in Section \ref{sec:identification}, requires multiple observations of studies $j$ with potentially heterogeneous estimands $\Theta_j$. In this section, by contrast, we are interested in valid inference on $\Theta$ for a given study $j$, and so condition on $\Theta_j$. Conditioning on $\Theta_j$ corresponds to standard notions of bias and size control.

Selective publication reweights the distribution of $X$ by $p(\cdot).$  To obtain valid estimators and confidence sets, we need to correct for this reweighting.  To define these corrections denote the cdf for published results $X$ given true effect $\Theta$ by $F_{X|\Theta}.$ For $f_{X|\Theta}$ the density of published results derived in Lemma \ref{lem:likelihood},
$$
F_{X|\Theta}(x|\theta)= \int_{-\infty}^x f_{X|\Theta}(\tilde x|\theta) d\tilde x=\frac{1}{E[p(X^*) | \Theta^*=\theta]} \int_{-\infty}^x p(\tilde x)f_{X^*|\Theta^*}(\tilde x|\theta)d\tilde x. 
$$

For many distributions $f_{X^*|\Theta^*},$ and in particular in the leading normal case (see Lemma \ref{lem: quantile unbiased sufficient condition} below) this cdf is strictly decreasing in $\theta$.  Using this fact we can adapt an approach previously applied by, among others, D. \cite{Andrews1993} and  \cite{StockWatson1998} and invert the cdf as a function of $\theta$ to construct a quantile-unbiased estimator. 
In particular, if we define
$\hat{\theta}_{\alpha}\left(x\right)$ as the solution to 
\begin{equation}
F_{X|\Theta}\left (x|\hat{\theta}_{\alpha}\left(x\right)\right)=\alpha\in(0,1),   \label{eq: alphahat def}
\end{equation}
so $x$ lies at the $\alpha$-quantile of the distribution implied by $\hat{\theta}_{\alpha}\left(x\right)$, then
 $\hat{\theta}_{\alpha}\left(X\right)$ is an $\alpha$-quantile unbiased estimator for $\theta$.  
%For simplicity, we assume the parameter space for $\theta$ is $\mathbb{R}.$

\begin{theorem} \label{thm: quantile unbiased}
If for all $x$, $F_{X|\Theta}(x|\theta)$ is continuous and strictly decreasing in $\theta,$ tends to one as $\theta\to-\infty$, and tends to zero as $\theta\to\infty,$ then $\hat{\theta}_{\alpha}(x)$ as defined in  (\ref{eq: alphahat def}) exists, is unique, and is continuous and strictly increasing for all $x.$ If, further, $F_{X|\Theta}(x|\theta)$ is continuous in $x$ for all $\theta$ then $\hat{\theta}_{\alpha}(X)$ is $\alpha$-quantile unbiased for $\theta$ under the truncated sampling setup of Definition \ref{defi:DGP},
\[
P \left( \hat{\theta}_{\alpha}\left(X\right)\le\theta | \Theta=\theta \right) =\alpha~\text{for all}~\theta.
\]
\end{theorem}

If $f_{X^{*}|\Theta^{*}}\left(x|\theta\right)$ is normal, as in our applications, then the assumptions of Theorem \ref{thm: quantile unbiased} hold whenever $p(x)$ is strictly positive for all $x$ and almost everywhere continuous.%\footnote{Note that these requirements can be considerably weakened, but are imposed for simplicity.}

\begin{lem} \label{lem: quantile unbiased sufficient condition}
If the distribution of latent draws $X^*$ conditional on $(\Theta^*,\sigma^*)$ is $N(\Theta^*,\sigma^{*2})$, 
$p(x)>0$ for all $x$, and $p(\cdot)$ is almost everywhere continuous, then the assumptions of Theorem \ref{thm: quantile unbiased} are satisfied.
\end{lem}

These results allow straightforward frequentist inference that corrects for selective publication.
In particular, using Theorem \ref{thm: quantile unbiased} we can consider the median-unbiased estimator $\hat{\theta}_{\frac{1}{2}}\left(X\right)$
for $\theta$, as well as the equal-tailed level $1-\alpha$ confidence
interval 
\begin{equation}
\left[\hat{\theta}_{\frac{\alpha}{2}}\left(X\right), \hat{\theta}_{1-\frac{\alpha}{2}}\left(X\right)\right].   \nonumber
\end{equation}
This estimator and confidence set fully correct the bias and coverage distortions induced by selective publication.  Other selection-corrected confidence intervals are also possible in this setting.  For example, provided the density $f_{X^*|\Theta^*}(x|\theta)$ belongs to an exponential family one can form confidence intervals by inverting uniformly most powerful unbiased tests as in \cite{fithian2014optimal}.  Likewise, one can consider alternative estimators, such as the weighted average risk-minimizing unbiased estimators considered in \cite{muellerwang2015}, or the MLE based on the truncated likelihood $f_{X|\Theta}.$

\paragraph{Illustrative example (continued)}

\begin{figure}[t!]
\begin{center}
\includegraphics{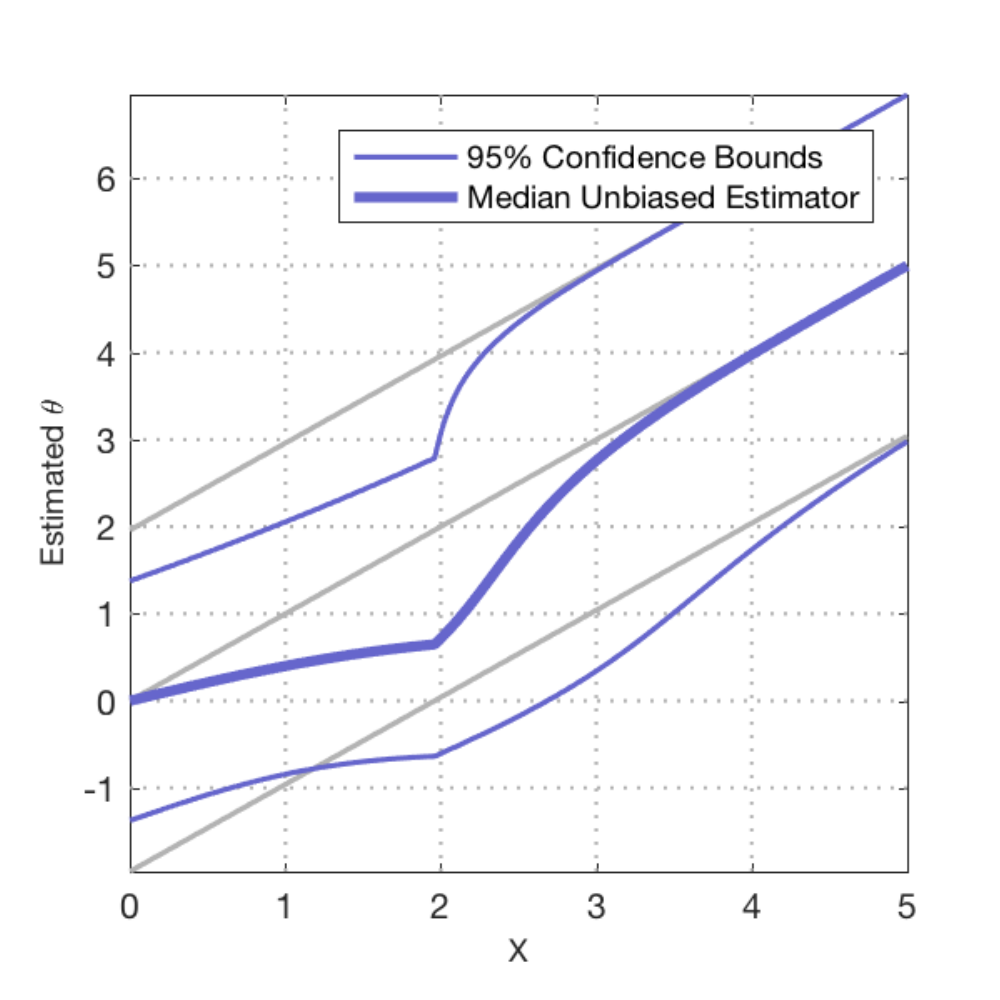}
\end{center}
\caption{This figure plots frequentist 95\% confidence bounds and the median unbiased estimator for the normal model where results that are significant at the 5\% level are published with probability one, while insignificant results are published with probability 10\%. The usual (uncorrected) estimator and confidence bounds are plotted in grey for comparison.\label{fig: Confidence bounds}}
\end{figure}

To illustrate these results, we return to the treatment effect example discussed above.   Figure \ref{fig: Confidence bounds}
plots the median unbiased estimator, as well as upper and lower 95\% confidence
bounds as a function of $X$ for the same publication probability $p(\cdot)$ considered
above. We see that the median unbiased estimator lies below the usual
estimator $\hat{\theta}=X$ for small positive $X$ but that the difference is eventually decreasing in $X$.
The truncation-corrected
confidence interval shown in Figure \ref{fig: Confidence bounds} has exactly correct coverage, is smaller than the usual interval for small $X$,
wider for moderate values $X$, and essentially the same for $X\ge5$. 

%Figure \ref{fig: Confidence bounds} provides useful guidance for readers of published papers interested in the magnitude of true effects. 
%Suppose that the illustrative example is a reasonable approximation of how selection works in practice, as our empirical findings below suggest is the case for experimental economics.
%Then the following ``rule of thumb'' adjustments correspond roughly to median-unbiased estimates. (i) If reported effects are close to zero, or very far from zero (z-statistics bigger than 4), then these estimates can be taken at face value. (ii) In intermediate ranges, magnitudes should be adjusted downwards. 
%A reported z-statistic of 1 should be taken to indicate an effect (relative to the standard error) of about 0.4. A reported z-statistic of 2 should be taken to indicate an effect of about 0.7, and a reported z-statistic of 3 should be taken to indicate an effect of about 2.75.  Likewise, two-sided tests reject zero when z-statistics are larger than about 2.7 in absolute value.

We do not recommend adjusting publication standards to reflect these corrections. If publication probabilities in this example were based on more stringent critical values, for instance, then the corrections discussed above would need to be adjusted.  Instead, the purpose of these corrections is to allow readers of published research to draw valid inferences, taking the publication rule as given.  The publication rule itself can then be chosen on other grounds, for example to maximize social welfare or provide incentives to researchers.  We briefly discuss the question of optimal publication rules in the conclusion, as well as in Section \ref{suppsec:Optimal selection in a simple model} supplement.

In this example, our approach is closely related to the correction for selective publication proposed by \cite{mccrary2016conservative}.  There, the authors propose conservative tests derived under an extreme form of publication bias in which insignificant results are never published.  If we consider testing the null hypothesis that $\theta$ is equal to zero, and calculate our equal-tailed confidence interval under the publication probability $p(\cdot)$ implied by the model of \cite{mccrary2016conservative}, then our confidence interval contains zero if and only if the test of \cite{mccrary2016conservative} fails to reject.

\paragraph{Estimation Error in $p(\cdot)$}
Thus far our corrections have assumed the conditional publication
probability is known. If $p\left(\cdot\right)$ is instead estimated with error,
median unbiased estimation is challenging but constructing valid
confidence sets for $\theta$ is straightforward. 

Suppose we
parameterize the conditional publication probability by a finite dimensional parameter $\beta$,
and let $\hat{\theta}_{\alpha}\left(X_{i};\beta\right)$ be the $\alpha$-quantile
unbiased estimator under $\beta$.  For many specifications of $p\left(\cdot\right)$, and in particular
for those used in our applications below, $\hat{\theta}_{\alpha}\left(x;\beta\right)$
is continuously differentiable in $\beta$ for all $x$. If we have
a consistent and asymptotically normal estimator $\hat{\beta}$ for
$\beta$, for $0<\delta<\alpha$, consider the interval
\[
\left[\hat{\theta}_{\frac{\alpha-\delta}{2}}\left(X;\hat{\beta}\right)-c_{1-\frac{\delta}{2}}\hat{\sigma}_{L}\left(X\right),\hat{\theta}_{1-\frac{\alpha-\delta}{2}}\left(X;\hat{\beta}\right)+c_{1-\frac{\delta}{2}}\hat{\sigma}_{U}\left(X\right)\right]
\]
where $c_{1-\frac{\delta}{2}}$ is the level $1-\frac{\delta}{2}$
quantile of the standard normal distribution while $\hat{\sigma}_{L}\left(x\right)$
and $\hat{\sigma}_{U}\left(x\right)$ are delta-method standard
errors for $\hat{\theta}_{\frac{\alpha-\delta}{2}}\left(x;\hat{\beta}\right)$
and $\hat{\theta}_{1-\frac{\alpha-\delta}{2}}\left(x;\hat{\beta}\right)$,
respectively.  If our model for $p(\cdot)$ is correctly specified, Bonferroni's inequality implies that this interval
covers $\theta$ with probability at least
$1-\alpha$ in large samples.\footnote{Even in cases where we do not have an asymptotically normal estimator for $\beta,$ for example because we consider a fully nonparameteric model for $p(\cdot)$, given an initial level $1-\delta$ confidence set $CS_\beta$ for $\beta$ we can form a Bonferroni confidence set for $\theta$ as $\left[\inf_{\beta\in CS_\beta}\hat{\theta}_{\frac{\alpha-\delta}{2}}\left(X;\hat{\beta}\right),\sup_{\beta\in CS_\beta}\hat{\theta}_{1-\frac{\alpha-\delta}{2}}\left(X;\hat{\beta}\right)\right].$}

\section{Applications}
\label{sec:applications}

This section applies the results developed above to estimate the degree of selectivity in several empirical literatures.  

\paragraph{Key identifying assumptions}
The results of Section \ref{sec:identification} imply nonparametric identification of both $p(\cdot)$ and $\mu$.
Using systematic replication studies, we identify $p(\cdot)$ based on asymmetries in the joint distribution of original and replication estimates.
This approach is based on the assumption that selection for publication depends only on the original estimates and not on the replication estimates. This assumption is highly plausible by design in the two replication settings we consider.

Identification using meta-studies identifies $p(\cdot)$ by comparing the distribution of estimates with different associated standard errors across studies.
This approach is based on the assumption that studies with different sample sizes on a given topic do not have systematically different estimands.
While we cannot guarantee validity of this assumption by design, plausibility of this assumption is enhanced by our finding that it yields estimates very similar to 
 the approach based on replication studies.
It should also be noted that variants of this assumption are imposed in the vast majority of existing meta-studies.

\paragraph{Maximum likelihood estimation}
The sample sizes in our applications are limited.
For our main analysis, we specify parsimonious parametric models for both the conditional publication probability $p(\cdot)$ and the  distribution $\mu$ of true effects across latent studies, which we then fit by maximum likelihood.
Parametric specifications of the nonparametrically identified model lead to intuitive and tractable estimators.  In the supplement we consider alternative, moment-based estimators which build on our identification arguments in Section \ref{sec:identification}.  These estimators are nonparametric in $\mu$ and yield similar results  to the parametric specifications reported here.

We consider step function models for $p(\cdot),$ with jumps at conventional critical values, and possibly at zero.  
Since $p(\cdot)$ is only identified up to scale, we impose the normalization $p(z)=1$ for $z>1.96$ throughout. This is without loss of generality, since $p(\cdot)$ is allowed to be larger than $1$ for other cells.
 
We assume different parametric models for the distribution of latent effects $\Theta^*$, discussed case-by-case below.  In our first two applications the sign of the original estimates is normalized to be positive.\footnote{The studies in these datasets consider different outcomes, so the relative signs of effects across studies are arbitrary.  Setting the sign of the initial estimate in each study to be positive has the desirable effect of ensuring invariance to the sign normalization chosen by the authors of each study.}  We denote these normalized estimates by $W=|Z|,$ and in these settings we impose that $p(\cdot)$  is symmetric.

\paragraph{Details and extensions}
Details and  further motivation for our specifications, as well as a specification for the model of Section \ref{sssec: selection on theta} which we use to develop specification tests,  are discussed in Section \ref{suppsec: Likelihood and parametric specifications} of the supplement.

In the present section we assume that our identifying assumptions hold unconditionally for the samples considered.
In Section \ref{suppsec: Additional empirical results} of the Supplement, we explore robustness of our results to additional conditioning on covariates $W$, including year of first circulation and journal of publication.  However, in no case do we reject our baseline (unconditional) specification at conventional significance levels.
To check the robustness of our findings, we report additional empirical results based on further additional specifications in Section \ref{suppsec: Additional empirical results} of the supplement. To check robustness to our parametric assumptions, we report estimates based on an alternative GMM estimation approach that does not rely on parametric specifications of $\mu$ in Section \ref{suppsec:momentresults} of the supplement. This alternative approach yields broadly similar estimates of $p(\cdot)$.

%We then discuss our results based on the experimental economics replications of \cite{camerer2016evaluating}, the experimental psychology replications of \cite{open2015estimating}, the minimum-wage meta-study of \cite{wolfson201515}, and the deworming meta-study  of \cite{deworming2016}.

\subsection{Economics laboratory experiments}\label{sec: econ experiments}

Our first application uses data from a recent large-scale replication of experimental economics papers by \cite{camerer2016evaluating}.  The authors replicated all 18 between-subject laboratory experiment papers published in the American Economic Review and Quarterly Journal of Economics between 2011 and 2014.\footnote{In their supplementary materials, \cite{camerer2016evaluating} state that ``To be part of the study a published paper needed to report at least one significant between subject treatment effect that was referred to as statistically significant in the paper.''  However, we have reviewed the issues of the American Economic Review and Quarterly Journal of Economics from the relevant period, and confirmed that no studies were excluded due to this restriction.}  Further details on the selection and replication of results can be found in \cite{camerer2016evaluating}, while details on our handling of the data are discussed in the supplement.

A strength of this dataset for our purposes, beyond the availability of replication estimates, is the fact that it replicates results from all papers in a particular subfield published in two leading economics journals over a fixed period of time.  This mitigates concerns about the selection of which studies to replicate.  Moreover, since the authors replicate 18 such studies, it seems reasonable to think that they would have published their results regardless of what they found, consistent with our assumption that selection operates only on the initial studies and not on the replications.   

A caveat to the interpretation of our results is that \cite{camerer2016evaluating} select the most important statistically significant finding from each paper, as emphasized by the original authors, for replication. 
This selection changes the interpretation of $p(\cdot)$, which has to be interpreted as the probability that a result was published \emph{and} selected for replication.  In this setting, our corrected estimates and confidence intervals provide guidance for interpreting the headline results of published studies.

\begin{figure}[t!]
\begin{center}

\includegraphics{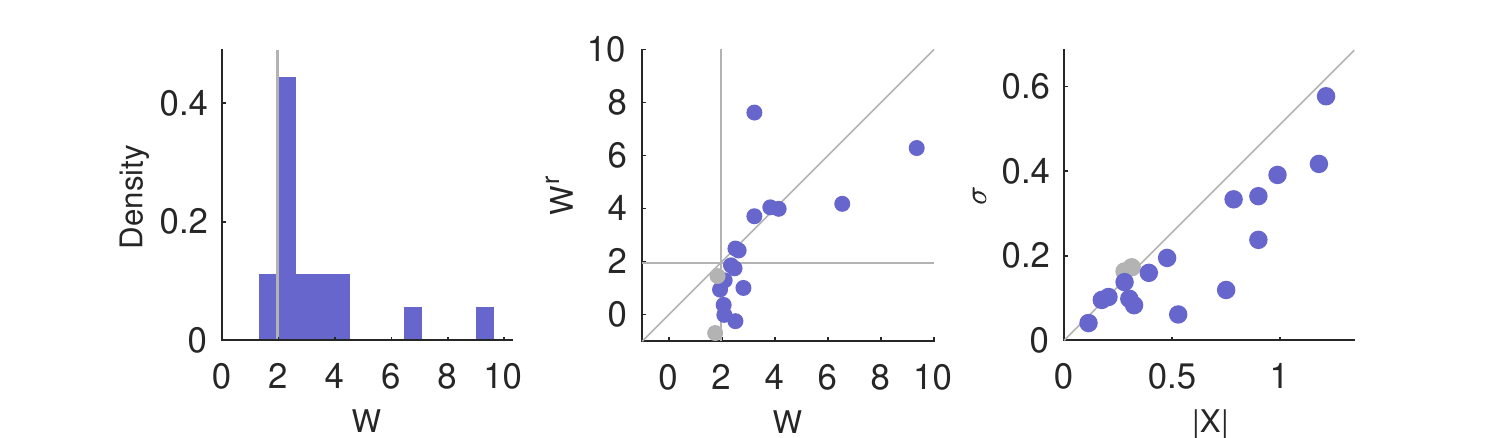}
\end{center}
\caption{The left panel shows a binned density plot for the normalized z-statistics $W=|X|/\sigma$ using data from \cite{camerer2016evaluating}.  The grey line marks $W=1.96.$ The middle panel plots the  z-statistics $W$ from the initial study against the estimate $W^r$ from the replication study.  The grey lines mark $W$ and $W^r=1.96,$ as well as $W=W^r.$  The right panel plots the initial estimate $|X|=W\cdot\sigma$ against its standard error $\sigma.$  The grey line marks $|X|/\sigma=1.96.$ \label{fig:EconScatter}}
\end{figure}

\paragraph{Histogram}
Before we discuss our formal estimation results, consider the distribution of originally published estimates $W=|Z|$, shown by the histogram in the left panel of Figure \ref{fig:EconScatter}.
This histogram suggests of a large jump in the density $f_W(\cdot)$ at the cutoff $1.96$, and thus of a corresponding jump of the publication probability $p(\cdot)$ at the same cutoff; cf. the discussion in Section \ref{ssec:identificationliterature}.
Such a jump is confirmed by both our replication and meta-study approaches.

\paragraph{Results from replication specifications}

The middle panel of Figure  \ref{fig:EconScatter} plots the joint distribution of $W,$  $W^r$ in the replication data of \cite{camerer2016evaluating}, using the same conventions as in Figure \ref{fig:ReplicationIllustration}.  To estimate the degree of selection in these data we consider the model
\begin{align*}
|\Theta^*| \sim \Gamma(\kappa,\lambda), ~~~ p(Z) \propto \begin{cases}
\beta_p & |Z|< 1.96\\
1 & |Z|\geq 1.96.
\end{cases}
\end{align*}
This assumes that the absolute value of the true effect $\Theta^*$ follows a gamma distribution with shape parameter $\kappa$ and scale parameter $\lambda.$  This nests a wide range of cases, including $\chi^2$ and exponential distributions, while keeping the number of parameters low.  Our model for $p(\cdot)$ allows a discontinuity in the publication probability at $|Z|=1.96,$ the critical value for a 5\% two-sided z-test.
Fitting this model by maximum likelihood yields the estimates reported in the left panel of Table \ref{tab:EconEstim}.  Recall that $\beta_p$ in this model can be interpreted as the publication probability for a result that is insignificant at the 5\% level based on a two-sided z-test, relative to a result that is significant at the 5\% level. These estimates therefore imply that significant results are more than thirty times more likely to be published than insignificant results.  
Moreover, we strongly reject the hypothesis of no selectivity, $H_0:\beta_p=1$. 
 
To test the validity of our baseline assumption of selection on $z$ and not on $\theta$, $p(z,\theta)=p(z),$ we calculate a score test using a model discussed in Section \ref{sssec: replication specifications} of the supplement. This yields a p-value of 0.53, so we find no evidence that the assumption $P\left(D=1|Z^*,\Theta^*\right) = p(Z^*)$ imposed in our baseline model is violated.

\begin{table}[!t]
    \begin{minipage}{.5\linewidth}
      \centering
      \textsc{Replication}\\[5pt]
\begin{tabular}{cc |c}
$ \kappa$ & $\lambda$ & $\beta_p$ \\ 
 \hline 
0.373 & 2.153 & 0.029 \\  
(0.266) & (1.024) & (0.027) \\  
\end{tabular}

    \end{minipage}%
    \begin{minipage}{.5\linewidth}
      \centering
      \textsc{Meta-study}\\[5pt]
\begin{tabular}{cc |c}
$\tilde \kappa$ & $\tilde \lambda$ & $\beta_p$ \\ 
 \hline 
1.343 & 0.157 & 0.038 \\  
(1.310) & (0.076) & (0.051) \\  
\end{tabular}

    \end{minipage}
\caption{Selection estimates from lab experiments in economics, with robust standard errors in parentheses. The left panel reports estimates from replication specifications, while the right panel reports results from meta-study specifications.  Publication probability $\beta_p$ is measured relative to the omitted category of studies significant at 5\% level, so an estimate of 0.029 implies that results which are insignificant at the 5\% level are 2.9\% as likely to be published as significant results. The parameters $(\kappa,\lambda)$ and $(\tilde\kappa,\tilde\lambda)$ are not comparable.\label{tab:EconEstim}} 
\end{table}

\paragraph{Results from meta-study specifications}

While the \cite{camerer2016evaluating} data include replication estimates, we can also apply our meta-study approach using just the initial estimates and standard errors.  Since this approach relies on additional independence assumptions, comparing these results to those based on replication studies provides a useful check of the reliability of our meta-analysis estimates.

We begin by plotting the data used by our meta-analysis estimates in the right panel of Figure  \ref{fig:EconScatter}.  We consider the model
\begin{align*}
|\Theta^*| \sim \Gamma(\tilde\kappa,\tilde\lambda), ~~~ p(X/\sigma) \propto \begin{cases}
\beta_p & |X/\sigma|< 1.96\\
1 & |X/\sigma|\geq 1.96,
\end{cases}
\end{align*}
noting that $\Theta^*$ is now the mean of $X^*,$ rather than $Z^*,$ and thus that the interpretation of  $(\tilde\kappa,\tilde\lambda)$ differs from that of $(\kappa,\lambda)$ in our replication specifications. 
Fitting this model by maximum likelihood yields the estimates reported in the right panel of Table \ref{tab:EconEstim}.  Comparing these estimates to those in the left panel, we see that the estimates from the two approaches are similar, though the metastudy estimates suggest a somewhat smaller degree of selection.  Hence, we find that in the \cite{camerer2016evaluating} data we obtain similar results from our replication and meta-study specifications.

\paragraph{Bias correction}

\begin{figure}[p!]
\begin{center}
\includegraphics{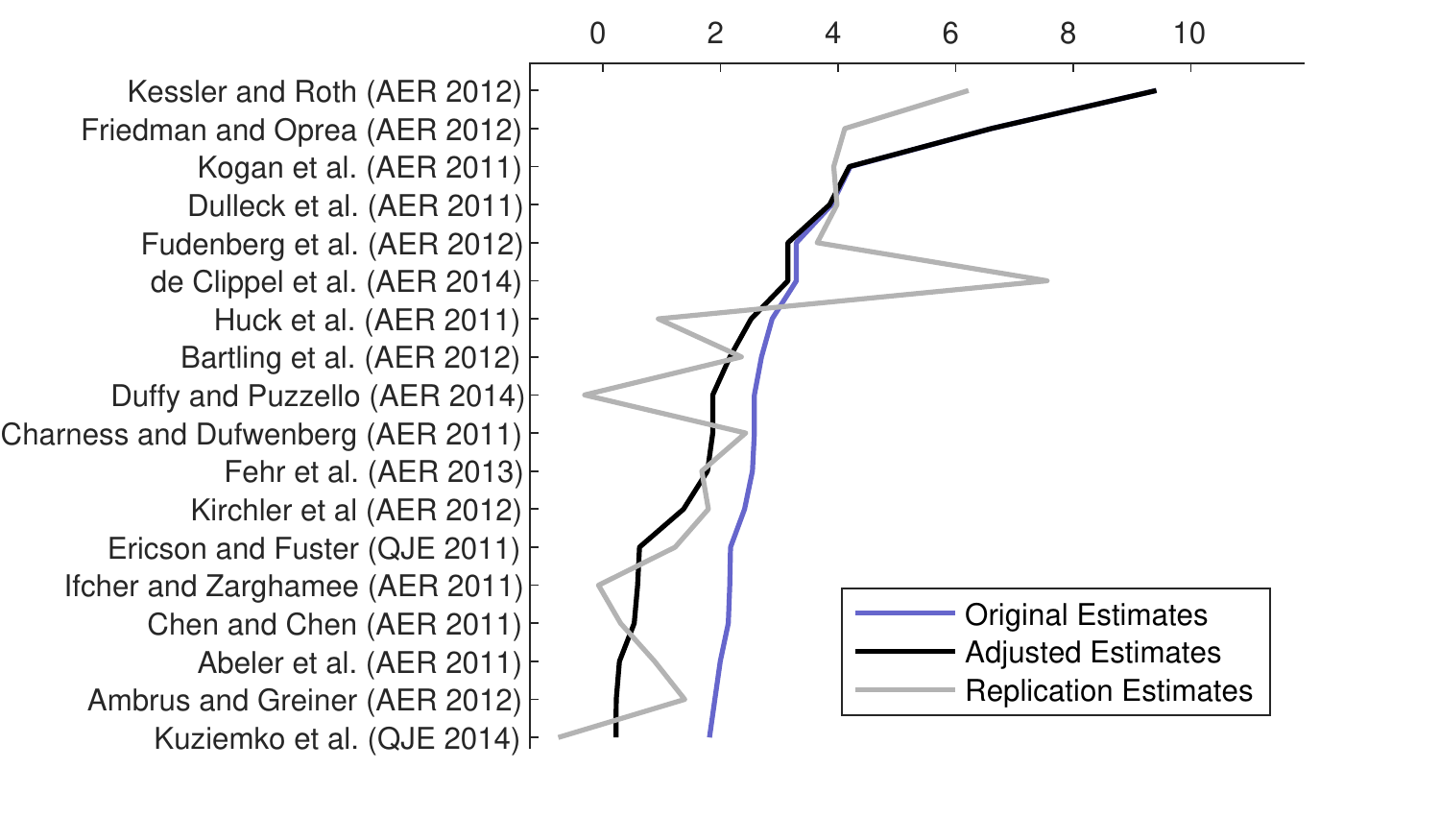}
\includegraphics{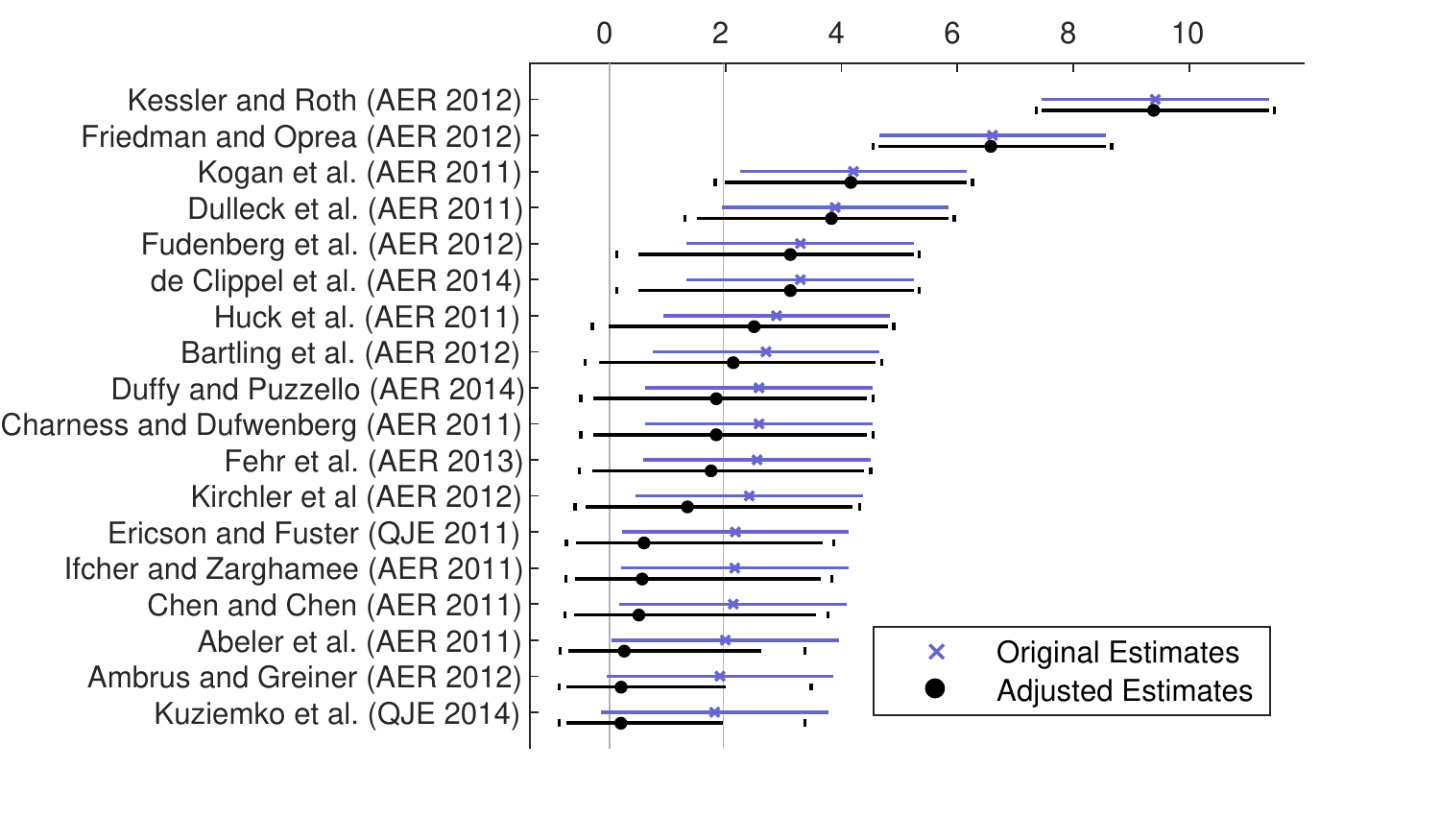}
\end{center}
\caption{The top panel plots the estimates $W$ and $W^r$ from the original and replication studies in \cite{camerer2016evaluating}, along with the median unbiased estimate $\hat\theta_{\frac{1}{2}}$ based on the estimated selection model and the original estimate.  The bottom panel plots the original estimate and 95\% confidence interval, as well as the median unbiased estimate and adjusted 95\% confidence interval $\left[\hat{\theta}_{0.025}\left(W\right), \hat{\theta}_{0.975}\left(W\right)\right]$ based on the estimated selection model.   Adjusted intervals not accounting for estimation error in the selection model are plotted with solid lines, while endpoints of Bonferroni-corrected intervals are marked with ``$\mathbf{\shortmid}$''. \label{fig:ReplicationEconOriginalAndAdjusted}}
\end{figure}

To interpret our estimates, we calculate our median-unbiased estimator and confidence sets based on our replication estimate $\beta_p=.029.$  Figure \ref{fig:ReplicationEconOriginalAndAdjusted} plots the median unbiased estimator, as well as the original and adjusted confidence sets (with and without bonferroni corrections), for the 18 studies included in \cite{camerer2016evaluating}.  Considering the first panel, which plots the median unbiased estimator along with the original and replication estimates, we see that the adjusted estimates track the replication estimates fairly well but are smaller than the original estimates in many cases.  The second panel plots the original estimate and conventional 95\% confidence set in blue, and the adjusted estimate and 95\% confidence set in black.  As we see from this figure, even without Bonferroni corrections twelve of the adjusted confidence sets include zero, compared to just two of the original confidence sets.  Hence, adjusting for the estimated degree of selection substantially changes the number of significant results in this setting.

\subsection{Psychology laboratory experiments}
\label{ssec: psych application}

Our second application is to data from \cite{open2015estimating}, who conducted a large-scale replication of experiments in psychology.  The authors considered studies published in three leading psychology journals, Psychological Science, Journal of Personality and Social Psychology, and Journal of Experimental Psychology: Learning, Memory, and Cognition, in 2008.  They assigned papers to replication teams on a rolling basis, with the set of available papers determined by publication date.
Ultimately, 158 articles were made available for replication, 111 were assigned, and 100 of those replications were completed in time for inclusion in \cite{open2015estimating}.  
Replication teams were instructed to replicate the final result in each article as a default, though deviations from this default were made based on feasibility and the recommendation of the authors of the original study.  Ultimately, 84 of the 100 completed replications consider the final result of the original paper.

As with the economics replications above, the systematic selection of results for replication in \cite{open2015estimating} is an advantage from our perspective.  
A complication in this setting, however, is that not all of the test statistics used in the original and replication studies are well-approximated by z-statistics (for example, some of the studies use $\chi^2$ test statistics with two or more degrees of freedom).  To address this, we limit attention to the subset of studies which use z-statistics or close analogs thereof, leaving us with a sample of 73 studies. Specifically, we limit attention to studies using z- and t-statistics, or $\chi^2$ and F-statistics with one degree of freedom (for the numerator, in the case of F-statistics), which can be viewed as the squares of z- and t-statistics, respectively. To explore sensitivity of our results to denominator degrees of freedom for t- and F-statistics, in the supplement we limit attention to the 52 observations with denominator degrees of freedom of at least 30 in the original study and find quite similar results.  

\begin{figure}[t!]
\begin{center}
\includegraphics{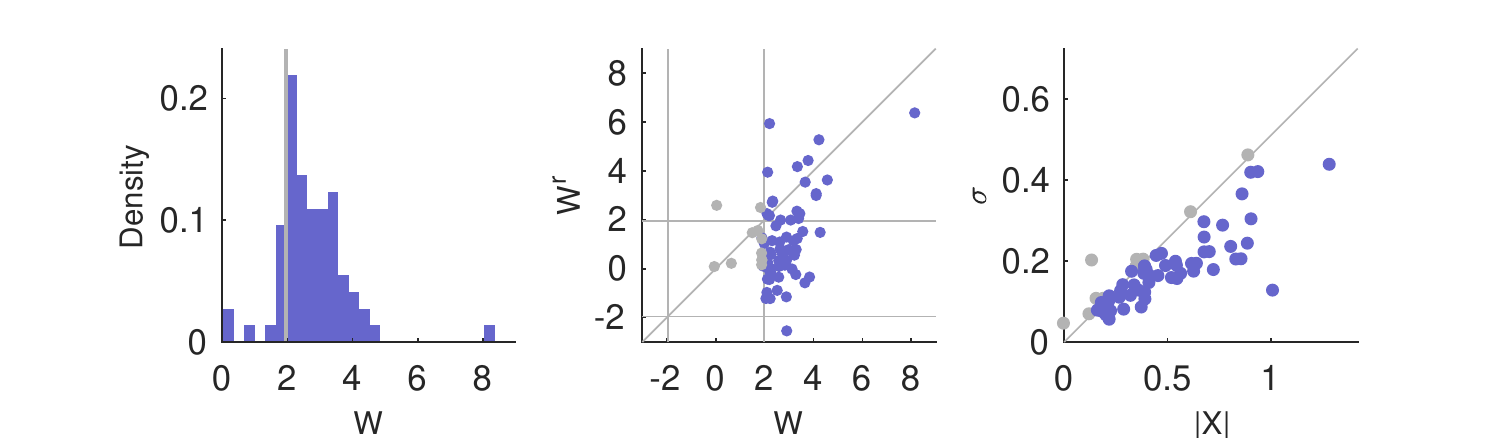}
\end{center}
\caption{The left panel shows a binned density plot for the normalized z-statistics $W=|X|/\sigma$ using data from \cite{open2015estimating}.  The grey line marks $W=1.96.$ The middle panel plots the  z-statistics $W$ from the initial study against the estimate $W^r$ from the replication study.  The grey lines mark $|W|$ and $|W^r|=1.96,$ as well as $W=W^r.$  The right panel plots the initial estimate $|X|=W\cdot\sigma$ against its standard error $\sigma.$  The grey line marks $|X|/\sigma=1.96.$ \label{fig:PsychScatter}}
\end{figure}

\paragraph{Histogram}
The distribution of originally published estimates $W$ is shown by the histogram in the left panel of Figure \ref{fig:PsychScatter}.
This histogram is suggestive of a large jump in the density $f_W(\cdot)$ at the cutoff $1.96$, as well as possibly a jump at the cutoff $1.64$, and thus of  corresponding jumps of the publication probability $p(\cdot)$ at the same cutoffs.
Such jumps are again confirmed by the estimates from both our replication and meta-study approaches.

\paragraph{Results from replication specifications}

The middle panel of Figure  \ref{fig:PsychScatter} plots the joint distribution of $W,$  $W^r$ in the replication data of \cite{open2015estimating}.  We fit the model
\begin{align*}
|\Theta^*| \sim \Gamma(\kappa,\lambda), ~~~ p(Z) \propto \begin{cases}
\beta_{p,1} & |Z|< 1.64\\
\beta_{p,2} & 1.64 \le|Z|< 1.96\\
1 & |Z|\geq 1.96.
\end{cases}
\end{align*}
This model again assumes that the absolute value of the true effect $|\Theta^*|$ follows a gamma distribution across latent studies.  Given the larger sample size, we consider a slightly more flexible model than before and allow discontinuities in the publication probability at the critical values for both 5\%  and 10\%  two-sided z-tests.

\begin{table}[!t]
    \begin{minipage}{.5\linewidth}
      \centering
      \textsc{Replication}\\[5pt]
\begin{tabular}{cc |cc}
$ \kappa$ & $\lambda$ & $\beta_{p,1}$ & $\beta_{p,2}$ \\ 
 \hline 
0.315 & 1.308 & 0.009 & 0.205 \\  
(0.143) & (0.334) & (0.005) & (0.088) \\  
\end{tabular}

    \end{minipage}%
    \begin{minipage}{.5\linewidth}
      \centering
      \textsc{Meta-study}\\[5pt]
\begin{tabular}{cc |cc}
$\tilde \kappa$ & $\tilde \lambda$ & $\beta_{p,1}$ & $\beta_{p,2}$ \\ 
 \hline 
0.974 & 0.153 & 0.017 & 0.306 \\  
(0.549) & (0.053) & (0.009) & (0.135) \\  
\end{tabular}

    \end{minipage}
\caption{Selection estimates from lab experiments in psychology, with robust standard errors in parentheses. The left panel reports estimates from replication specifications, while the right panel reports results from meta-study specifications.  Publication probabilities $\beta_p$ are measured relative to the omitted category of studies significant at 5\% level.   The parameters $(\kappa,\lambda)$ and $(\tilde \kappa,\tilde \lambda)$ are not comparable.\label{tab:PsychEstim}} 
\end{table}

Fitting this model by maximum likelihood yields the estimates reported in the left panel of Table \ref{tab:PsychEstim}. These estimates  imply that results that are significantly different from zero at the 5\% level are over a hundred times more likely to be published than results that are insignificant at the 10\% level, and nearly five times more likely to be published than results that are significant at the 10\% level but insignificant at the 5\% level.  We strongly reject the hypothesis of no selectivity.

A score test of the null hypothesis $p(z,\theta)=p(z)$ yields a p-value of 0.42. Thus, we again find no evidence that the assumption $P\left(D=1|Z^*,\Theta^*\right) = p(Z^*)$ imposed in our baseline model is violated.  

\paragraph{Results from meta-study specifications}
As before, we re-estimate our model using our meta-study specifications, and plot the joint distribution of estimates and standard errors in the right panel of Figure \ref{fig:PsychScatter}.  Fitting the model  yields the estimates reported in the right panel of Table \ref{tab:PsychEstim}.   As in the last section, we find that the meta-study and replication estimates are broadly similar, though the meta-study estimates again suggest a somewhat more limited degree of selection.

\paragraph{Bias corrections}

\begin{figure}[t!]
\begin{center}
\includegraphics{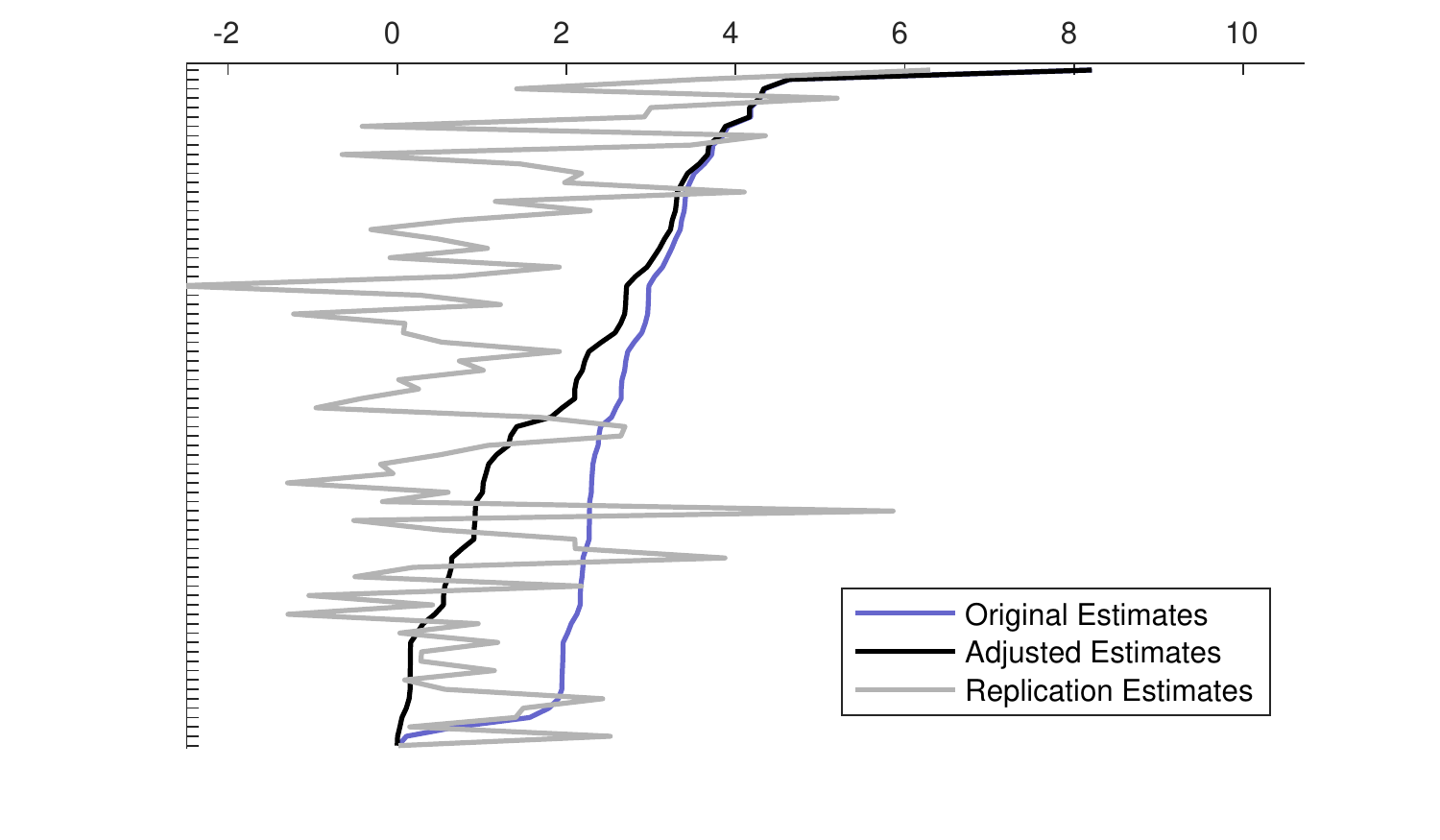}
\end{center}
\caption{This figure plots the estimates $W$ and $W^r$ from the original and replication studies in \cite{open2015estimating}, along with the median unbiased estimate $\hat\theta_{\frac{1}{2}}$ based on the estimated selection model and the original estimate.  
%The bottom panel plots the original estimate and 95\% confidence interval, as well as the median unbiased estimate and adjusted 95\% confidence interval $\left[\hat{\theta}_{0.025}\left(X\right), \hat{\theta}_{0.975}\left(X\right)\right]$ based on the estimated selection model.
\label{fig:ReplicationPsychOriginalAndAdjusted}}
\end{figure}

To interpret our results, we plot our median-unbiased estimates based on the \cite{open2015estimating} data in Figure \ref{fig:ReplicationPsychOriginalAndAdjusted}.  We see that our adjusted estimates track the replication estimates fairly well for studies with small original z-statistics, though the fit is worse for studies with larger original z-statistics.  
Our adjustments again dramatically change the number of significant results, with 62 of the 73 original 95\% confidence sets excluding zero, and only 28 of the adjusted confidence sets (not displayed) doing the same.
 
\paragraph{Approved replications}
\cite{Gilbertetal2016} argue that the protocols in some of the \cite{open2015estimating}  replications differed substantially from the initial studies.  To explore robustness with respect to this critique, in the supplement we report results from further restricting the sample to the subset of replications which used protocols approved by the original authors prior to the replication.
Doing so we find roughly similar estimates, though the estimated degree of selection is smaller.

\subsection{Effect of minimum wage on employment}

Our third application uses data from \cite{wolfson201515}, who conduct a meta-analysis of studies on the elasticity of employment with respect to the minimum wage. In particular,  \cite{wolfson201515} collect analyses of the effect of minimum wages on employment that use US data and were published or circulated as working papers after the year 2000.  They collect estimates from all studies fitting their criteria that report both estimated elasticities of employment with respect to the minimum wage and standard errors, resulting in a sample of a thousand estimates drawn from 37 studies, and we use these estimates as the basis of our analysis.  For further discussion of these data, see \cite{wolfson201515}.

Since the \cite{wolfson201515} sample includes both published and unpublished papers, we evaluate our estimators based on both the full sample and the sub-sample of published estimates.  We find qualitatively similar answers for the two samples, so we report results based on the full sample here and discuss results based on the subsample of published estimates in the supplement.
We define $X$ so that $X>0$ indicates a negative effect of the minimum wage on employment.  
%we estimate our model based on both the full set of studies and the subset of published studies, which contains 705 estimates drawn from 31 studies.  We find very similar estimates for the two samples, which might be interpreted as evidence of of limited selection operating at the journal level.  We hesitate to over-interpret these results, however, since many of the working papers in the sample may be published in the future, as suggested by the fact that working papers are more common towards the end of the sample.  For consistency with our other applications, here we discuss results limiting attention to the subset of published papers, but results based on the full set of papers are presented in the supplement.

%We believe that assuming that $D_i$ is independent of $\Theta_i^*$ and $\sigma_i^*$ conditional on $X_i^*$ is a reasonable assumption, or at least a good approximation.
%To clarify what this assumption entails, let us however sketch a setting where it would be violated. Suppose that each study contains two conditionally statistically independent estimates $i=1,2$ of the same latent parameter $\Theta_1^*=\Theta_2^*$, with corresponding z-statistics $Z_1^*,Z_2^*$.
%Suppose the study is published if and only if both of these exceed some critical value $c$.
%Then $D_i$ is positively correlated with $\Theta_i^*$ conditional on $Z_i^*$.

\begin{figure}[t!]
\begin{center}
\includegraphics{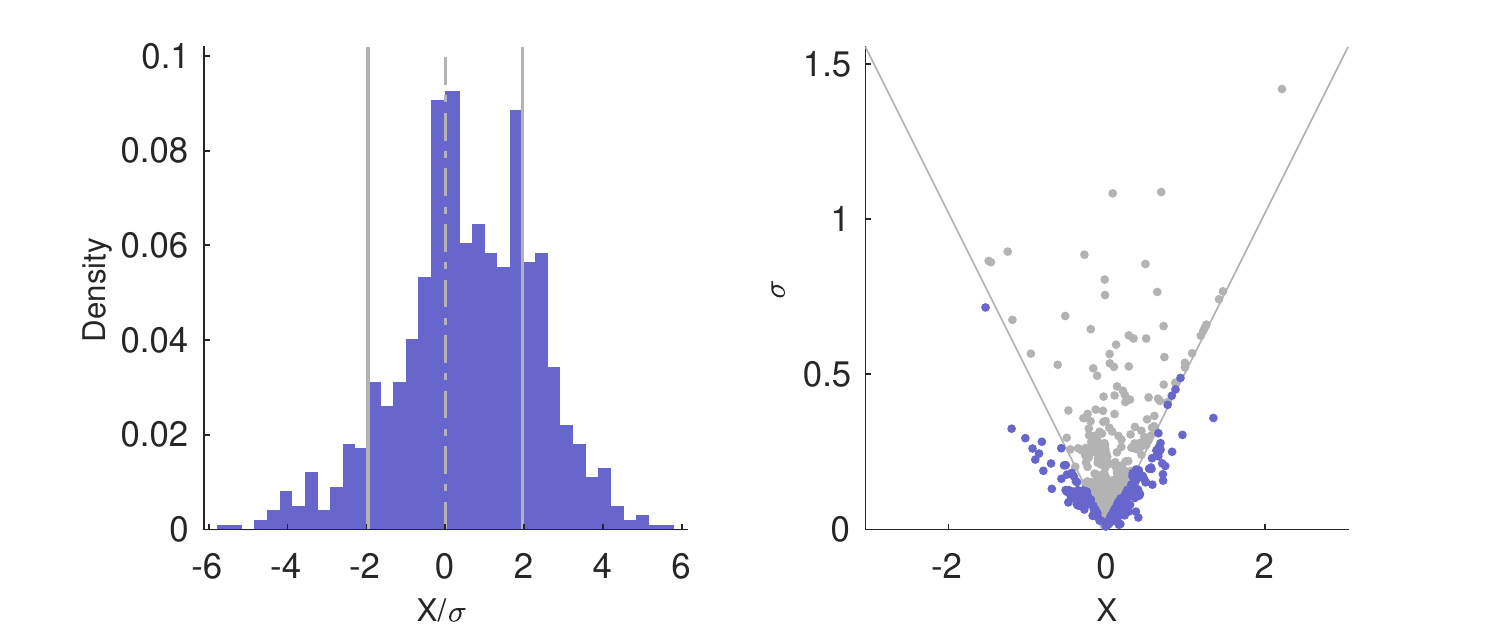}
\end{center}
\caption{The left panel shows a binned density plot for the z-statistics $X/\sigma$ in the \cite{wolfson201515} data.  The solid grey lines mark $|X|/\sigma=1.96,$ while the dash-dotted grey line marks $X/\sigma=0.$
The right panel  plots the estimate $X$ against its standard error $\sigma$.  The grey lines mark  $|X|/\sigma=1.96.$   \label{fig:MinimumWageScatter}}
\end{figure}

\paragraph{Histogram}
Consider first the distribution of the normalized estimates $Z$, shown by the histogram in the left panel of Figure \ref{fig:MinimumWageScatter}.
This histogram is somewhat suggestive of jumps in the density $f_Z(\cdot)$ around the cutoffs $-1.96$, $0$, and $1.96$, and thus of  corresponding jumps in the publication probability $p(\cdot)$ at the same cutoffs; these jumps seem less pronounced than in our previous applications, however.

\paragraph{Results from meta-study specifications}

For this application we do not have any replication estimates, and so move directly to our meta-study specifications.  The right panel of Figure  \ref{fig:MinimumWageScatter} plots the joint distribution of $X$, the estimated elasticity of employment with respect to decreases in the minimum wage, and the standard error $\sigma$ in the \cite{wolfson201515} data.

As a first check, we run meta-regressions as discussed in section \ref{ssec:identificationliterature}, clustering standard errors by study.
A regression of $X$ on $\sigma$ yields a slope of $0.408$ with a standard error of $0.372$.
A regression of $Z$ on $1/\sigma$  yields an intercept of $0.343$ with a standard error of $0.283$.
Both of these estimates suggest selection favoring results finding a negative effect of minimum wages on employment, but neither allows us to reject the null of no selection at conventional significance levels.

We next consider the model 
\begin{align*}
\Theta^* \sim \bar\theta+t(\nu)\cdot\tilde\tau, ~~~ p(X/\sigma) \propto \begin{cases}
\beta_{p,1} & X/\sigma< -1.96\\
\beta_{p,2} & -1.96 \le X/\sigma < 0\\
\beta_{p,3} & 0 \le X/\sigma < 1.96\\
1 & X/\sigma \geq 1.96.
\end{cases}
\end{align*}
Since the data are not sign-normalized, we model $\Theta^*$ using a t distribution with degrees of freedom $\tilde\nu$ and location and scale parameters $\bar\theta$ and $\tilde\tau,$ respectively.
Unlike in our previous applications, we allow the probability of publication to depend on the sign of the z-statistic $X/\sigma$ rather than just on its absolute value.  This is important, since it seems plausible that the publication prospects for a study could differ depending on whether it found a positive or negative effect of the minimum wage on employment.  
Recall that $X>0$ indicates a negative effect of the minimum wage on employment.
Our estimates based on these data are reported in Table \ref{tab:MinimumWageMeta}, where we find that results which are insignificant at the 5\% level are about 30\% as likely to be published as are significant estimates finding a negative effect of the minimum wage on employment.  
 Our point estimates also suggest that studies finding a  positive and significant effect of the minimum wage on employment may be less likely to be published, but this estimate is quite noisy and we cannot reject the hypothesis that selection depends only on signficance and not on sign.  

\begin{table}[h!]
\begin{center}
\begin{tabular}{ccc |ccc}
$\bar \theta$ & $\tilde \tau$ & $\tilde \nu$ & $\beta_{p,1}$ & $\beta_{p,2}$ & $\beta_{p,3}$ \\ 
 \hline 
0.018 & 0.019 & 1.303 & 0.697 & 0.270 & 0.323 \\  
(0.009) & (0.011) & (0.279) & (0.350) & (0.111) & (0.094) \\  
\end{tabular}

\caption{Meta-study estimates from minimum wage data, with standard errors clustered by study in parentheses. Publication probabilities $\beta_p$ measured relative to omitted category of estimates positive and significant at 5\% level.\label{tab:MinimumWageMeta}}
\end{center}
\end{table}

These results are consistent with the meta-analysis estimates of \cite{wolfson201515}, who found evidence of some publication bias towards a negative employment effect, as well as the results of \cite{CardKrueger1995}, who focused on an earlier, non-overlapping set of studies.  

%Our results suggest that there is a systematic preference for publishing results indicating a negative impact of minimum wages; this preference may be driven by a prior belief in competitive (as opposed to monopsonistic) models of the labor market, which would predict such a negative impact; cf. section VIII in \cite{card1994}

Since the studies in this application estimate related parameters, it is also interesting to consider the estimate $\bar\theta$ for the mean effect in the population of latent estimates.  
The point estimate suggests that the average latent study finds a small but statistically significant negative effect of the minimum wage on employment.  This effect is about half as large as the ``naive'' average effect $\bar\theta$ we would estimate by ignoring selectivity,  $.041$ with a standard error of $0.011$.

%
%%\paragraph{Bias corrections}
%
%Given the large number of estimates, plotting the original and corrected estimates and confidence intervals as in Figure \ref{fig:ReplicationEconOriginalAndAdjusted} is impractical.  Our corrections also make a noticeable, though not enormous, difference here.  In particular, the percentage of results finding a negative and significant effect of the minimum wage (at the 5\% level) on employment drops from 23\% to 19\%, while the fraction finding a positive and significant effect rises from 8\% to 10.5\%.  Over 64\% of the initial estimates indicate a negative effect of the minimum wage on employment, while 57.5\% of our median-unbiased estimates do.  Thus, correcting for publication bias in this setting leads to a modest decrease in the fraction of specifications finding negative results.

\paragraph{Multiple estimates}
A complication arises in this application, relative to those considered so far, due to the presence of multiple estimates per study.   Since it is difficult to argue that a given estimate in each of these studies constitutes the ``main'' estimate, restricting attention to a single estimate per study would be arbitrary.  This somewhat complicates inference and identification.

For inference, it is implausible that estimate standard-error pairs $X_j, \sigma_j$ are independent within study.  To address this, we cluster our standard errors by study.

For identification, the problem is somewhat more subtle.
Our model assumes that the latent parameters $\Theta_i^*$ and $\sigma_i^*$ are statistically independent across estimates $i$, and that $D_i$ is independent of $(\Theta_i^*, \sigma_i^*)$ conditional on $X_i^*/\sigma_i^*$.  It is straightforward to relax the assumption of independence across $i,$ provided the marginal distribution of $(\Theta_i^*, \sigma_i^*,X_i^*,D_i)$ is such that $D_i$ remains independent of $(\Theta_i^*, \sigma_i^*)$ conditional on $X_i^*/\sigma_i^*$.
This conditional independence assumption is justified if we believe that both researchers and referees consider the merits of each estimate on a case-by-case basis, and so decide whether or not to publish each estimate separately.
Alternatively, it can also be justified if the estimands $\Theta_i^*$ within each study are statistically independent (relative to the population of estimands in the literature under consideration).
As discussed in Section \ref{sssec: selection on theta}, however, if these assumptions fail our model is misspecified.

\subsection{Deworming meta-study}

Our final application uses data from the recent meta-study  \cite{deworming2016} on the effect of mass drug administration for deworming on child body weight.  They collect results from randomized controlled trials which report child body weight as an outcome, and focus on intent-to-treat estimates from the longest follow-up reported in each study.  They include all studies identified by the previous review of \cite{Cochrane2015}, as well as additional trials identified by \cite{Campbell2017}.  They then extract estimates as described in \cite{deworming2016} and obtain a final sample of 22 estimates drawn from 20 studies, which we take as the basis for our analysis.  For further discussion of sample construction, see \cite{Cochrane2015}, \cite{deworming2016}, and \cite{Campbell2017}.
To account for the presence of multiple estimates in some studies, we again cluster by study.

\begin{figure}[h!]
\begin{center}
\includegraphics{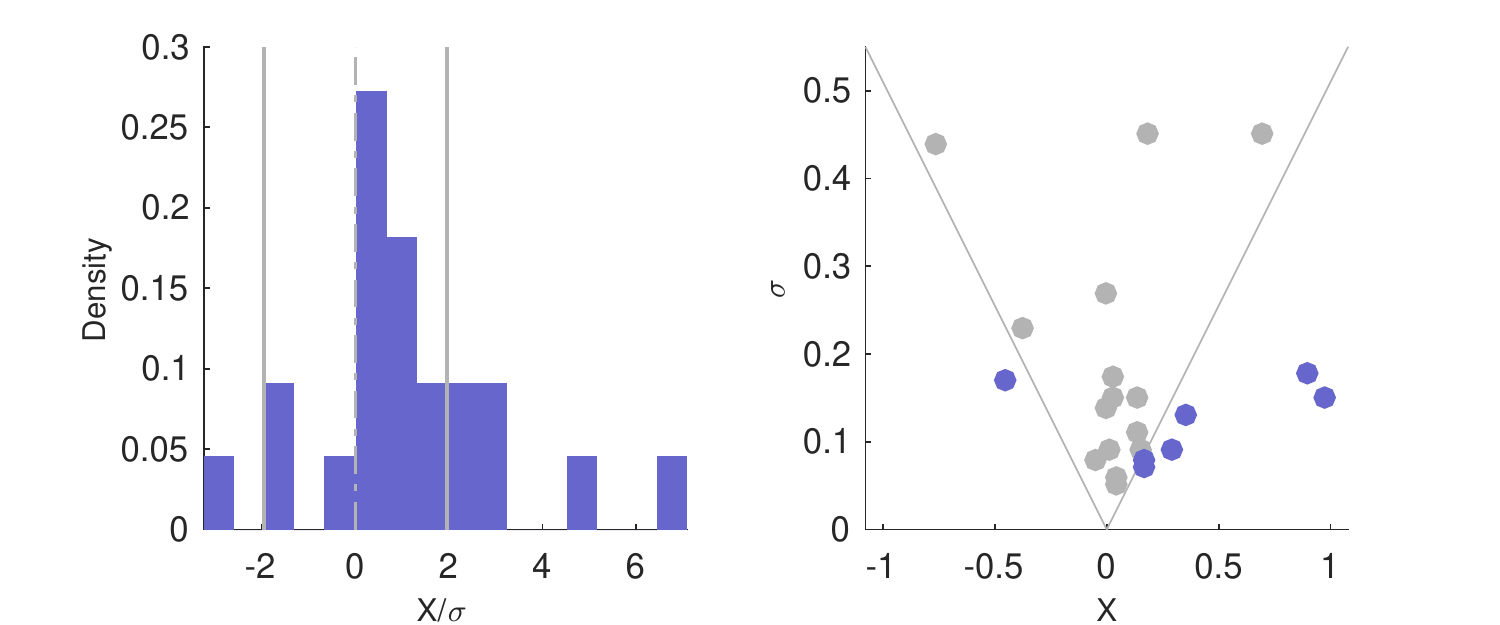}
\end{center}
\caption{The left panel shows a binned density plot for the z-statistics $X/\sigma$ in the \cite{deworming2016} data.  The solid grey lines mark $|X|/\sigma=1.96,$ while the dash-dotted grey line marks $X/\sigma=0.$
The right panel  plots the estimate $X$ against its standard error $\sigma$.  The grey lines mark  $|X|/\sigma=1.96.$\label{fig:DewormingScatter}}
\end{figure}

\paragraph{Histogram}
Consider first the distribution of the normalized estimates $Z$, shown by the histogram in the left panel of Figure \ref{fig:DewormingScatter}.
Given the small sample size of 22 estimates, this histogram should not be interpreted too strongly. That said, the density of $Z$ appears to jump up at $0$, which suggests selection toward positive estimates.

\paragraph{Results from meta-study specifications}

The right panel of Figure  \ref{fig:DewormingScatter} plots the joint distribution of $X$, the estimated intent to treat effect of mass deworming on child weight, along with the standard error $\sigma$ in the \cite{deworming2016} data.  

As a first check, we again run meta-regressions as discussed in Section \ref{ssec:identificationliterature}, clustering standard errors by study.
A regression of $X$ on $\sigma$ yields a slope of $-0.296$ with a standard error of $0.917$.
A regression of $Z$ on $1/\sigma$  yields an intercept of $0.481$ with a standard error of $0.889$.
Neither of these estimates allows rejection of the null of no selection at conventional significance levels.

We next consider the model  
\begin{align*}
\Theta^* \sim N(\bar\theta, \tilde{\tau}^2), ~~~ p(X/\sigma) \propto \begin{cases}
\beta_{p} & |X/\sigma|< -1.96\\
1 & |X/\sigma| \geq 1.96,
\end{cases}
\end{align*}
where we constrain the the distribution of $\Theta^*$ to be normal and the function $p(\cdot)$ to be symmetric to limit the number of free parameters, which is important since we have only 22 observations.  Fitting this model yields the estimates reported in Table \ref{tab:DewormingMeta}.
The point estimates here suggest that statistically significant results are less likely to be included in the meta-study of \cite{deworming2016} than are insignificant results. \\

\begin{table}[h!]
\begin{center}
\begin{tabular}{cc |c}
$\bar \theta$ & $\tilde \tau$ & $\beta_p$ \\ 
 \hline 
0.190 & 0.343 & 2.514 \\  
(0.120) & (0.128) & (1.869) \\  
\end{tabular}

\caption{Meta-study estimates from deworming data, with robust standard errors in parentheses. Publication probabilities $\beta_p$ measured relative to omitted category of studies significant at 5\% level.\label{tab:DewormingMeta}}
\end{center}
\end{table}

However, the standard errors are quite large, and the difference in publication (inclusion) probabilities between significant and insignificant results is itself not significant at conventional levels, so there is no basis for drawing a firm conclusion here.   Likewise, the estimated  $\bar\theta$ suggests a positive average effect in the population, but is not significantly different from zero at conventional levels.

In the supplement we report results based on alternative specifications which allow the function $p(\cdot)$ to be asymmetric.  These specifications suggest selection against negative estimates.

Our findings here are potentially relevant in the context of the controversial debate surrounding mass deworming; see for instance \cite{wormwars2015}.
The point estimates for our baseline specification suggest that insignificant results have a higher likelihood of being included in \cite{deworming2016} relative to significant ones.
In light of the large standard errors and limited robustness to changing the specification of $p(\cdot)$, however, these findings should not be interpreted too strongly.

\section{Conclusion}
\label{sec:conclusion}

%We conclude by briefly summarizing our contributions, and then discuss some implications for empirical researchers and the organization of the publication process.

%\paragraph{Summary of main contributions}
This paper contributes to the literature in three ways.
First, we provide nonparametric identification results for 
 selectivity (in particular, the conditional publication probability) as a function of the empirical findings of a study. 
Second, we provide methods to calculate bias-corrected estimators and confidence sets when the form of selectivity is known. 
Third, we apply the proposed methods to several literatures, documenting the varying scale and kind of selectivity.

\paragraph{Implications for empirical research}
What can researchers and readers of empirical research take away from this paper?
First, when conducting a meta-analysis of the findings of some literature, researchers may wish to apply our methods to assess the degree of selectivity in this literature, and to apply appropriate corrections to individual estimates, tests, and confidence sets.
We provide code on our webpages which implements the proposed methods for a flexible family of selection models.

Second, when reading empirical research, readers may wish to adjust the published point estimates and confidence sets along the lines discussed in Section \ref{sec:inference}.
Suppose for instance that for a given field publication probabilities increase considerably when estimates exceed the 5\% significance threshold, but publication does not otherwise depend on findings.
In that case, if reported effects are close to zero, or very far from zero (z-statistic bigger than 4, say), then these estimates can be taken at face value. In intermediate ranges, in particular for z-statistics around 2, magnitudes should be adjusted downwards.

It should be emphasized that we do not advocate adjusting publication standards to reflect our corrected critical values.
If these cutoffs were to be systematically used in the publication process, this would simply entail an ``arms race'' of selectivity, rendering the more stringent critical values invalid again.

\paragraph{Optimal publication rules}
One might take the findings in this paper, and the debate surrounding publication bias more generally, to indicate that the publication process should be non-selective with respect to findings. This might for instance be achieved by instituting some form of result-blind review.
The hope would be that non-selectivity of the publication process might restore the validity (unbiasedness, size control) of standard inferential methods.

Note, however, that optimal publication rules may depend on results. Consider for instance a setting where policy decisions are made based on published findings, policy makers have a limited capacity to read publications, and journal editors maximize the same social welfare function as policy makers.  In a stylized model of such a setting, detailed in Section \ref{suppsec:Optimal selection in a simple model} of the supplement, we show that expected social welfare is maximized by publishing the results which allow policy makers to update the most relative to their prior beliefs. The corresponding publication rule favors the publication of surprising findings, thus violating non-selectivity.  A more general theory of optimal publication is of considerable interest for future research.

%extension of identification results to multi-result and spec checks case

%\clearpage
\bibliographystyle{apalike}
%\bibliography{../../../library}
\bibliography{PublicationBias}

\begin{thebibliography}{}

\bibitem[Andrews, 1993]{Andrews1993}
Andrews, D.~W. (1993).
\newblock Exactly median-unbiased estimation of first order autoregressive/unit
  root models.
\newblock {\em Econometrica}, 61(1):139--165.

\bibitem[Baricz, 2008]{baricz_mills_2008}
Baricz, {\'A}. (2008).
\newblock Mills' ratio: Monotonicity patterns and functional inequalities.
\newblock {\em Journal of Mathematical Analysis and Applications},
  340(2):1362--1370.

\bibitem[Brodeur et~al., 2016]{brodeur2016star}
Brodeur, A., L{\'e}, M., Sangnier, M., and Zylberberg, Y. (2016).
\newblock Star wars: The empirics strike back.
\newblock {\em American Economic Journal: Applied Economics}, 8(1):1--32.

\bibitem[Bruns, 2017]{bruns2017meta}
Bruns, S.~B. (2017).
\newblock Meta-regression models and observational research.
\newblock {\em Oxford Bulletin of Economics and Statistics}.

\bibitem[Bruns and Ioannidis, 2016]{bruns2016p}
Bruns, S.~B. and Ioannidis, J.~P. (2016).
\newblock P-curve and p-hacking in observational research.
\newblock {\em PLoS One}, 11(2):e0149144.

\bibitem[Camerer et~al., 2016]{camerer2016evaluating}
Camerer, C.~F., Dreber, A., Forsell, E., Ho, T.-H., Huber, J., Johannesson, M.,
  Kirchler, M., Almenberg, J., Altmejd, A., and Chan, T. (2016).
\newblock Evaluating replicability of laboratory experiments in economics.
\newblock {\em Science}, 351(6280):1433--1436.

\bibitem[Card and Krueger, 1995]{CardKrueger1995}
Card, D. and Krueger, A.~B. (1995).
\newblock Time-series minimum-wage studies: A meta-analysis.
\newblock {\em American Economic Review}, 85(2):238--243.

\bibitem[Carter et~al., 2017]{carter2017correcting}
Carter, E., Sch{\"o}nbrodt, F., Gervais, W.~M., and Hilgard, J. (2017).
\newblock Correcting for bias in psychology: A comparison of meta-analytic
  methods.

\bibitem[Chen and Zimmermann, 2017]{ChenZimmermann2017}
Chen, A.~Y. and Zimmermann, T. (2017).
\newblock Selection bias and the cross-section of expected returns.
\newblock Unpublished Manuscript.

\bibitem[Christensen and Miguel, 2016]{ChristensenMiguel2016}
Christensen, G.~S. and Miguel, E. (2016).
\newblock Transparency, reproducibility, and the credibility of economics
  research.
\newblock NBER Working Paper No. 22989.

\bibitem[Clemens and Sandefur, 2015]{wormwars2015}
Clemens, M. and Sandefur, J. (2015).
\newblock Mapping the worm wars: What the public should take away from the
  scientific debate about mass deworming.

\bibitem[Clemens, 2015]{Clemens2015}
Clemens, M.~A. (2015).
\newblock The meaning of failed replications: a review and proposal.
\newblock {\em Journal of Economic Surveys}, Forthcoming.

\bibitem[Croke et~al., 2016]{deworming2016}
Croke, K., Hicks, J.~H., Hsu, E., Kremer, M., and Miguel, E. (2016).
\newblock Does mass deworming affect child nutrition? {Meta-analysis,}
  cost-effectiveness, and statistical power.
\newblock Technical Report 22382, National Bureau of Economic Research.

\bibitem[{De Long} and Lang, 1992]{DelongLang1992}
{De Long}, J.~B. and Lang, K. (1992).
\newblock Are all economic hypotheses false?
\newblock {\em Journal of Political Economy}, 100(6):1257--1272.

\bibitem[Doucouliagos and Stanley, 2009]{DoucouliagosStanley2009}
Doucouliagos, H. and Stanley, T. (2009).
\newblock Publication selection bias in minimum-wage research? {A}
  meta-regression analysis.
\newblock {\em British Journal of Industrial Relations}, 47(2):406--428.

\bibitem[Duval and Tweedie, 2000]{duval2000nonparametric}
Duval, S. and Tweedie, R. (2000).
\newblock A nonparametric ``trim and fill'' method of accounting for
  publication bias in meta-analysis.
\newblock {\em Journal of the American Statistical Association},
  95(449):89--98.

\bibitem[Egger et~al., 1997]{egger1997bias}
Egger, M., Smith, G.~D., Schneider, M., and Minder, C. (1997).
\newblock Bias in meta-analysis detected by a simple, graphical test.
\newblock {\em Bmj}, 315(7109):629--634.

\bibitem[Fithian et~al., 2014]{fithian2014optimal}
Fithian, W., Sun, D., and Taylor, J. (2014).
\newblock Optimal inference after model selection.
\newblock {\em arXiv preprint arXiv:1410.2597}.

\bibitem[Franco et~al., 2014]{Francoetal2014}
Franco, A., Malhotra, N., and Simonovits, G. (2014).
\newblock Publication bias in the social sciences: Unlocking the file drawer.
\newblock {\em Science}, 345(6203):1502--1505.

\bibitem[Furukawa, 2017]{Furukawa2017}
Furukawa, C. (2017).
\newblock Unbiased publication bias: Theory and evidence.
\newblock Unpublished Manuscript.

\bibitem[Gilbert et~al., 2016]{Gilbertetal2016}
Gilbert, D.~T., King, G., Pettigrew, S., and Wilson, T.~D. (2016).
\newblock {Comment on ``Estimating the reproducibility of psychological
  science''}.
\newblock {\em Science}, 351(6277):1037.

\bibitem[Havr{\'a}nek, 2015]{havranek2015measuring}
Havr{\'a}nek, T. (2015).
\newblock Measuring intertemporal substitution: The importance of method
  choices and selective reporting.
\newblock {\em Journal of the European Economic Association}, 13(6):1180--1204.

\bibitem[Hedges, 1992]{hedges1992modeling}
Hedges, L.~V. (1992).
\newblock Modeling publication selection effects in meta-analysis.
\newblock {\em Statistical Science}, pages 246--255.

\bibitem[Honore and Powell, 1994]{HonorePowell1994}
Honore, B. and Powell, J. (1994).
\newblock Pairwise difference estimators for censored and truncated regression
  models.
\newblock {\em Journal of Econometrics}, 64:241--278.

\bibitem[Hou et~al., 2017]{hou2017replicating}
Hou, K., Xue, C., and Zhang, L. (2017).
\newblock Replicating anomalies.
\newblock Technical report, National Bureau of Economic Research.

\bibitem[Ioannidis et~al., 2017]{ioannidis2017power}
Ioannidis, J., Stanley, T.~D., and Doucouliagos, H. (2017).
\newblock The power of bias in economics research.
\newblock {\em The Economic Journal}, 127(605).

\bibitem[Ioannidis, 2008]{ioannidis2008most}
Ioannidis, J.~P. (2008).
\newblock Why most discovered true associations are inflated.
\newblock {\em Epidemiology}, 19(5):640--648.

\bibitem[Ioannidis, 2005]{Ioannidis2005}
Ioannidis, J. P.~A. (2005).
\newblock Why most published research findings are false.
\newblock {\em PLoS Med}, 2(8).

\bibitem[Iyengar and Greenhouse, 1988]{iyengar1988selection}
Iyengar, S. and Greenhouse, J.~B. (1988).
\newblock Selection models and the file drawer problem.
\newblock {\em Statistical Science}, pages 109--117.

\bibitem[Lee et~al., 2016]{lee2016exact}
Lee, J.~D., Sun, D.~L., Sun, Y., Taylor, J.~E., et~al. (2016).
\newblock Exact post-selection inference, with application to the lasso.
\newblock {\em The Annals of Statistics}, 44(3):907--927.

\bibitem[McCrary et~al., 2016]{mccrary2016conservative}
McCrary, J., Christensen, G., and Fanelli, D. (2016).
\newblock Conservative tests under satisficing models of publication bias.
\newblock {\em PloS one}, 11(2):e0149590.

\bibitem[Meister, 2009]{meister2009}
Meister, A. (2009).
\newblock {\em Deconvolution Problems in Nonparametric Statistics}.
\newblock Springer.

\bibitem[Mueller and Wang, 2015]{muellerwang2015}
Mueller, U. and Wang, Y. (2015).
\newblock Nearly weighted risk minimal unbiased estimation.
\newblock Unpublished Manuscript.

\bibitem[Murphy, 2011]{murphy2011ordinary}
Murphy, G.~M. (2011).
\newblock {\em Ordinary differential equations and their solutions}.
\newblock Courier Corporation.

\bibitem[{Open Science Collaboration}, 2015]{open2015estimating}
{Open Science Collaboration} (2015).
\newblock Estimating the reproducibility of psychological science.
\newblock {\em Science}, 349(6251):aac4716.

\bibitem[Patil and Peng, 2016]{Patiletal2016}
Patil, P. and Peng, R.~D. (2016).
\newblock What should researchers expect when they replicate studies? {A}
  statistical view of replicability in psychological science.
\newblock {\em Perspectives on Psychological Science}, 11(4):539--44.

\bibitem[Pfanzagl, 1994]{Pfanzagl1994}
Pfanzagl, J. (1994).
\newblock {\em Parametric Statistical Theory}.
\newblock De Gruyter.

\bibitem[Rothstein et~al., 2006]{rothstein2006publication}
Rothstein, H.~R., Sutton, A.~J., and Borenstein, M. (2006).
\newblock {\em Publication bias in meta-analysis: Prevention, assessment and
  adjustments}.
\newblock John Wiley \& Sons.

\bibitem[Schuemie et~al., 2014]{schuemie2014interpreting}
Schuemie, M.~J., Ryan, P.~B., DuMouchel, W., Suchard, M.~A., and Madigan, D.
  (2014).
\newblock Interpreting observational studies: why empirical calibration is
  needed to correct p-values.
\newblock {\em Statistics in medicine}, 33(2):209--218.

\bibitem[Simonsohn, 2015]{Simonsohn2015}
Simonsohn, U. (2015).
\newblock Small telescopes detectability and the evaluation of replication
  results.
\newblock {\em Psychological Science}, 26(5):559--569.

\bibitem[Simonsohn et~al., 2014]{Simonsohnetal2014}
Simonsohn, U., Nelson, L.~D., and Simmons, J.~P. (2014).
\newblock P-curve: A key to the file-drawer.
\newblock {\em Journal of Experimental Psychology: General}, 143(2):534--547.

\bibitem[Stanley et~al., 2017]{stanley2017finding}
Stanley, T., Doucouliagos, H., and Ioannidis, J. (2017).
\newblock Finding the power to reduce publication bias.
\newblock {\em Statistics in medicine}, 36(10):1580--1598.

\bibitem[Stanley, 2008]{Stanley2008}
Stanley, T.~D. (2008).
\newblock Meta-regression methods for detecting and estimating empirical
  effects in the presence of publication selection.
\newblock {\em Oxford Bulletin of Economics and Statistics}, 70(103-127).

\bibitem[Stanley and Doucouliagos, 2014]{stanley2014meta}
Stanley, T.~D. and Doucouliagos, H. (2014).
\newblock Meta-regression approximations to reduce publication selection bias.
\newblock {\em Research Synthesis Methods}, 5(1):60--78.

\bibitem[Stock and Watson, 1998]{StockWatson1998}
Stock, J. and Watson, M. (1998).
\newblock Median unbiased estimation of coefficient variance in a time-varying
  parameter model.
\newblock {\em Journal of the American Statistical Association},
  93(441):349--358.

\bibitem[Stock and Wright, 2000]{StockWright2000}
Stock, J.~H. and Wright, J.~H. (2000).
\newblock Gmm with weak identification.
\newblock {\em Econometrica}, 68(5):1055--1096.

\bibitem[Taylor-Robinson et~al., 2015]{Cochrane2015}
Taylor-Robinson, D.~C., Maayan, N., Soares-Weiser, K., Donegan, S., and Garner,
  P. (2015).
\newblock Cochrane database of systematic reviews.

\bibitem[Wasserman, 2006]{wasserman2006all}
Wasserman, L. (2006).
\newblock {\em All of nonparametric statistics}.
\newblock Springer Science \& Business Media.

\bibitem[Welch et~al., 2017]{Campbell2017}
Welch, V.~A., Ghogomu, E., Hossain, A., Awasthi, S., Bhutta, Z.~A.,
  Cumberbatch, C., Fletcher, R., McGowan, J., Krishnaratne, S., Kristjansson,
  E., Sohani, S., Suresh, S., Tugwell, P., White, H., and Wells, G.~A. (2017).
\newblock Mass deworming to improve developmental health and wellbeing of
  children in low-income and middle-income countries: a systematic review and
  network meta-analysis.
\newblock {\em The Lancet Global Health}, 5(1):e40--e50.

\bibitem[Wolfson and Belman, 2015]{wolfson201515}
Wolfson, P.~J. and Belman, D. (2015).
\newblock 15 years of research on us employment and the minimum wage.
\newblock {\em Available at SSRN 2705499}.

\bibitem[Yekutieli, 2012]{Yekutieli2012}
Yekutieli, D. (2012).
\newblock Adjusted bayesian inference for selected parameters.
\newblock {\em Journal of the Royal Statistical Society Series B},
  74(3):515--541.

\end{thebibliography}

\clearpage
\appendix
%\begin{appendices}

\noindent \begin{center}
{\large{}Supplement to the paper}
\par\end{center}{\large \par}

\noindent \begin{center}
{\LARGE{}Identification of and correction for publication bias}
\par\end{center}{\LARGE \par}

\medskip{}

\begin{center}
{\large{} ~~~~~~~Isaiah Andrews  ~~~~~~~~~~~~~~~  Maximilian Kasy}
\par\end{center}{\large \par}

\noindent \begin{center}
\today
\par\end{center}

\vspace*{10pt}

This appendix contains proofs and supplementary results for the paper ``Identification of and correction for publication bias.''  Section \ref{sec:proofs} collects proofs for the results stated in the main text.  Section \ref{suppsec: meta-regression coefficients} considers the behavior of meta-regression coefficients, discussed in Section \ref{ssec:identificationliterature} of the main text, in a simple example.  Section \ref{suppsec: Likelihood and parametric specifications} discusses the likelihoods used in our empirical applications. Section \ref{suppsec: latent selection model} states a simple model for selection on both $Z^*$ and a latent variable $V^*$.  Section \ref{suppsec: Application details} provides details on the empirical applications discussed in the main text, while Section \ref{suppsec: Additional empirical results} reports additional results.  Section \ref{suppsec:momentresults} discusses results based on moment estimators which are nonparametric in $\mu.$
Section \ref{suppsec: Inference corrections based on estimates} provides corrected inference plots, analogous to Figure \ref{fig: Confidence bounds} of the main text, based on our psychology, minimum wage, and deworming applications.  Section \ref{suppsec: Inference in multivariate normal models} generalizes the inference results discussed in the main text to multivariate normal settings, while Section \ref{suppsec: Bayesian inference} discusses the effect of selective publication on Bayesian inference.  Finally, Section \ref{suppsec:Optimal selection in a simple model} discusses optimal selection in a stylized model.

\subsubsection*{Appendix Contents}
\startcontents
\printcontents{atoc}{0}{\setcounter{tocdepth}{1}}
 
\section{Proofs}
\label{sec:proofs}

\paragraph{Proof of Lemma \ref{lem:likelihood}: }
By construction, and Bayes rule
\begin{align*}
f_{X|\Theta}\left(x|\theta\right) &= f_{X^*|\Theta,D}(x|\theta,1)\\
&= \frac{P(D_i=1| X^*_i=x, \Theta^*_i=\theta)}{ P( D_i=1| \Theta^*_i=\theta)} \cdot  f_{X^*|\Theta^*}(x| \theta) \\
&=  \frac{p\left(x\right)}{E\left[p\left(X^*_i\right)|\Theta^*_i=\theta\right]} \cdot f_{X^*|\Theta^*}\left(x|\theta\right).
\end{align*}
$\Box$

\paragraph{Proof of Lemma \ref{lem:replicationlikelihood}: }
The conditional density follows by the same argument used to derive the truncated likelihood in Lemma \ref{lem:likelihood}. As for the marginal density, by construction,
\begin{align*}
f_{X,X^r}\left(x, x^{r} \right) &= f_{X^*,X^{*r}|D_i}(x, x^r |  d=1)\\
&= \frac{P(D_i=1| X^*_i=x, X_i^{*r}=x^r)}{P( D_i=1)} \cdot  f_{X^*,X^{*r}}(x, x^r)\\
&=  \frac{p\left(x\right)}{E\left[p\left(X^*_i\right)\right]}f_{X^*,X^{*r}}(x, x^{r}),
\end{align*}
and, since $X^*_i \perp X^{*r} | \Theta_i^*$,
\begin{align*}
 f_{X^*,X^{*r}}(x, x^{r}) &=\int f_{X^*|\Theta^*}\left(x|\theta_i^*\right) f_{X^*|\Theta^*}\left(x^r|\theta_i^*\right) d\mu(\Theta^*_i).
\end{align*}
$\Box$

\paragraph{Proof of Theorem \ref{theo:identificationreplication}: }
The marginal likelihood $f_{X, X^r}$ derived in Lemma \ref{lem:replicationlikelihood} satisfies
\[f_{X, X^r}(a,b) \cdot p(b)   =  f_{X, X^r}(b,a) \cdot p(a)\]
for all $a,b$.

Let $(a, b) \in A \times A$ be any point such that $f_{X, X^r}(a,b) >0$, so that in particular $p(a) >0$.
By the assumptions on the support of $f_{X^*, X^{*r}}$ and the data generating process this implies that $f_{X, X^r}(a,c) >0$ for all $c \in A$.

This in turn implies that
\[p(c) = p(a) \cdot \frac{f_{X, X^r}(c,a)}{f_{X, X^r}(a,c) } \]
for all $c \in A$,
where $p(a)$ is the only unknown on the right hand side. We thus find that $p(x)$ is identified up to scale. $\Box$

%This in turn identifies the untruncated density from
%\[ f^*_{X, X^r}(a,b) = \frac{E[p(X)]}{p(a)} f_{X, X^r}(a,b)\]
%for all $a$ such that $p(a) >0$, and (exploiting symmetry)
%\[ f^*_{X, X^r}(a,b) = \frac{E[p(X)]}{p(b)} f_{X, X^r}(b,a)\]
%for all $b$ such that $p(b) >0$.

\paragraph{Proof of Corollary \ref{corr:replicationexperimentsSigma}: }
In the case where $\sigma\equiv 1$, this is a special case of Theorem \ref{theo:identificationreplication}, and the claim immediately follows. (Note that $(Z^*, Z^{*r})$ has full support $\mathbb{R}^2$.)
We will show that we can reduce the case where $\sigma\not\equiv 1$  to this special case.
Let $\tilde{Z}$ be such that 
\[\tilde{Z}_i | Z^*_i, D_i, \Theta^*_i \sim N(\Theta_i^*, 1).\]
If $f_{\tilde{Z} | Z}$ is identified, we are done.
Note that 
\[f_{\tilde{Z} | Z} = f_{\Theta | Z} \ast \varphi,\]
and
\[f_{Z^r | Z, \sigma} = f_{\Theta | Z, \sigma} \ast \varphi_\sigma.\]
Based on the last equation, $f_{\Theta | Z, \sigma}$ is identified using deconvolution (this is a standard result; see for instance \citealt{wasserman2006all}, Chapter 10.1, equation 10.18. An extensive discussion of deconvolution can be found in \citealt{meister2009}).
We then get
\[f_{\Theta | Z} = \int f_{\Theta | Z, \sigma} f_{\sigma|Z} d\sigma,\]
and identification of $p(\cdot)$ follows.

To show identification of $\mu$, note that knowledge of $p(\cdot)$ up to scale allows us to recover the density $f_{Z^*}$ via
\[f_{Z^*}(z) = \frac{E[p(Z^*)]}{p(z)} f_Z(z).\]
Deconvolution then identifies $\mu$, since $f_{Z^*} = \mu \ast \varphi$.
$\Box$

\paragraph{Proof of Corollary \ref{corr:replicationexperimentsNormalizedSign}: }
Let $S_i^*=\pm 1$ with probability $0.5$, independently of  $(Z^*_i, Z^{*r}_i,\sigma^*_i, \Theta^*_i)$, and $S_j= S^*_{I_j}$.
Define 
\[(V,V^r)= S\cdot(W,W^r).\]
We show that $(V,V^r)$ satisfies the assumptions of Corollary \ref{corr:replicationexperimentsSigma}, from which the claim then follows.

Define $\tilde{S}^* =  S^* \cdot \sign(Z^*)$, so that $(V,V^r)= \tilde{S}\cdot (Z,Z^r)$, and define $\tilde{\Theta}^* =\tilde{S}^* \cdot \Theta^*$.
Since $\tilde{S}$ is independent of $(Z,Z^r,\sigma,\Theta)$, we get
\[\tilde{\Theta}^* \sim \tilde{\mu} = \tfrac{1}{2}(\mu_{\Theta^*} + \mu_{-\Theta^*})\]
and
\begin{equation*}
f_{V, V^{r}, \sigma}(v, v^r, \sigma) = p(v) \cdot f_{\sigma|  Z^*}(\sigma | v) \cdot \frac{\int \varphi(v - \theta) \cdot \tfrac{1}{\sigma}\varphi\left (\tfrac{v^r-\theta}{\sigma}\right ) d\tilde{\mu}(\theta)}{\int p(v') \cdot \varphi(v' - \theta) dv' d\tilde{\mu}(\theta)}.
\end{equation*} 
This has the exact same form as the density of  $(Z,Z^r,\sigma)$ under the symmetric measure $\tilde{\mu}$. The claim follows, since identification of $\tilde\mu$ implies identification of the distribution of $|\Theta^*|$.
$\Box$

\paragraph{Proof of Theorem \ref{theo:identificationDependence}: }
Under the setup considered, using the implied conditional independence assumptions we get
\begin{align*}
f_{Z^r | Z, \sigma}(z^r, z, \sigma) &= 
\int f_{Z^{*r} | \sigma, Z^*, D, \Theta^*}(z^r | \sigma, z, 1, \theta) f_{\Theta^* |\sigma, Z^*, D}(\theta | \sigma, z, 1) d\theta\\
&=\int \varphi_\sigma(z^r - \theta) f_{\Theta^* | Z^*, D}(\theta | z, 1) d\theta\\
&= (f_{\Theta | Z} \ast \varphi_\sigma)(z^r | z).
\end{align*}
By deconvolution, this immediately implies that we can identify $f_{\Theta | Z} $.
Since $f_Z$ is directly identified, Bayes' rule yields the desired result via
\[f_{Z|\Theta}(z|\theta) = \frac{f_{\Theta | Z}(\theta | z) \cdot f_Z(z)}{\int f_{\Theta | Z}(\theta | z') \cdot f_Z(z') dz'}. \]
$\Box$

\paragraph{Proof of Theorem \ref{theo:identificationsigmanonpar}: }
Assume w.l.o.g. that $\sigma=1$ lies in the interior of the support of $\sigma,$ and let
\[h(z) = f_{Z^* |\sigma^*}(z|1).\]

\paragraph{If $h(\cdot)$ is identified, then so are $p(\cdot)$ and $\mu$.}
We will show that $h(\cdot)$ is identified, which immedaitely identifies $\mu$ by deconvolution, 
since $h = \mu \ast \varphi$.
We can then identify $p(z)$ as before, since the truncated conditional density of $Z$ is given by
\begin{equation}
f_{Z| \sigma}(z|\sigma)=  \frac{p\left(z\right)}{E\left[p\left(Z^*\right)|\sigma\right]}f_{Z^*| \sigma^*}(z|\sigma),
\label{eq:truncdensityvarsigma}
\end{equation}
and thus
\[p(z) = const. \cdot \frac{f_{Z| \sigma}(z| 1)}{h(z)}.\]

\paragraph{A second order ODE for $h(\cdot)$.}
Let $\pi=1/\sigma$ be the precision of an estimate.
Differentiating the $\log$ of expression \eqref{eq:truncdensityvarsigma} for the truncated density at $\pi=1$ yields
\begin{equation}
g(z) = \partial_\pi \log f_{Z| \sigma}(z|1) = C_1 + \partial_\pi \log f_{Z^*| \sigma^*}(z|1)
\label{eq:defg}
\end{equation}
for a constant $C_1$. 
Note how, as we differentiate $\log f_{Z| \sigma}(z|1)$ with respect to $\pi$ at a given value $z$, the term $p(z)$ drops out of the resulting equation.
The function $g$ is identified under our assumptions.

Recall now that the definition of the standard normal density implies $\varphi'(z) = - z  \varphi(z)$.
The density $f_{X^*| \sigma^*}$ is given by $\mu \ast \varphi_\sigma$, and thus
$f_{Z^*| \sigma^*}(z|1/\pi)	=\int  \varphi\left ( z- \theta \pi \right ) d\mu(\theta)$, which implies
\begin{align*}
\partial_z f_{Z^*| \sigma^*}(z|1)&= -\int \left ( z- \theta  \right )  \varphi\left ( z- \theta \right ) d\mu(\theta)\\
\partial_z^2 f_{Z^*| \sigma^*}(z|1)&=-f_{Z^*| \sigma^*}(z|1)+\int \left ( z- \theta  \right )^2  \varphi\left ( z- \theta \right ) d\mu(\theta)\\
\partial_\pi f_{Z^*| \sigma^*}(z|1)&=\int \theta \left ( z- \theta  \right )  \varphi\left ( z- \theta \right ) d\mu(\theta)\\
&=-\left [f_{Z^*| \sigma^*}(z|1) + z \cdot \partial_z f_{Z^*| \sigma^*}(z|1) + \partial_z^2 f_{Z^*| \sigma^*}(z|1)\right ],
\end{align*}
from which we conclude
\begin{equation}
h''(z) = (C_1- 1 - g(z)) \cdot h(z) - z \cdot h'(z).
\label{eq:diffeqh}
\end{equation}
Equation \eqref{eq:diffeqh} is a second order linear homogeneous ordinary differential equation.

\paragraph{Two free parameters}
Given the initial conditions $h(0)=h_0$ and $h'(0)=h'_0$, and given $C_1$, the solution to this equation exists and is unique, because all coefficients are continuous in $z$; cf. \cite{murphy2011ordinary}.
Furthermore, the general solution to this differential equation can be written in the form
$
h(z, C_1, h_0, h'_0) = h_0 \cdot h_1(z, C_1) + h'_0 \cdot h_2(z, C_1)$,
where the functions $h_1(\cdot)$ and $h_2(\cdot)$ are determined by equation \eqref{eq:diffeqh}; cf. \cite{murphy2011ordinary}, chapter B.
This leaves three free parameters to be determined,  $C_1, h_0$ and $h'_0$.
The constraint $\int h(z) dz=1$ pins down $h_0$ or $h'_0$ given the other two parameters, so that there remain two free parameters.

\paragraph{A fourth order ODE for $h(\cdot)$.}
We next turn to the second derivative $k(\cdot)$ defined by
$$
k(z) =\partial_\pi^2 \log f_{Z| \sigma}(z|1) = C_2 + \partial_\pi^2 \log f_{Z^*| \sigma^*}(z|1),
$$
which is identified under our assumptions, just like $g(\cdot)$.
Calculations similar to those for the first derivative with respect to $\pi$ yield the fourth order differential equation
\begin{equation}
h^{(4)}(z) =  \left (k(z)-C_2 +\left (g(z) -C_1\right )^2  -2 \right ) h(z) - 4z h'(z) - (z^2+5) h''(z) - 2z h^{(3)}(z) .
\label{eq:diffeqh4th}
\end{equation}

To complete this proof, we now (i) derive the fourth order differential equation \eqref{eq:diffeqh4th} and (ii) show that it allows us to pin down the remaining free parameters.
We provide further discussion immediately following the proof.

\paragraph{Derivation of the fourth order ODE for $h(\cdot)$}
Differentiating $\log f_{Z^*| \sigma^*}$ twice yields 
$$
\partial_\pi^2 \log f_{Z^*| \sigma^*}(z|1) = 
\dfrac{\partial_\pi^2  f_{Z^*| \sigma^* }(z|1)}{h(z)}
-\left (g(z) -C_1\right )^2,
$$
so that
$$\partial_\pi^2 f_{Z^*| \sigma^*}(z|1)
= h(z) \cdot \left (k(z)-C_2 +\left (g(z) -C_1\right )^2  \right ).$$
From $f_{Z^*| \sigma^*}(z|1/\pi)	=\int  \varphi\left ( z- \theta \pi \right ) d\mu(\theta)$ we note that
\begin{align*}
\partial_\pi^2 f_{Z^*| \sigma^*}(z|1)&=
\int \left (-\theta^2 + \theta^2 \left ( z- \theta  \right )^2\right )  \varphi\left ( z- \theta \right ) d\mu(\theta).
\end{align*}
We furthermore have
\begin{align*}
h^{(3)} &= - 3 h'(z) - \int \left ( z- \theta  \right )^3  \varphi\left ( z- \theta \right ) d\mu(\theta)\\
h^{(4)} &=-3 h''(z) - 3\int \left ( z- \theta  \right )^2  \varphi\left ( z- \theta \right ) d\mu(\theta)  + \int \left ( z- \theta  \right )^4  \varphi\left ( z- \theta \right ) d\mu(\theta)\\
&=-6 h''(z) - 3 h(z)  + \int \left ( z- \theta  \right )^4  \varphi\left ( z- \theta \right ) d\mu(\theta).
\end{align*}
Comparing coefficients on $\theta$ between $\partial_\pi^2 f_{Z^*| \sigma^*}$ and the derivatives of $h(\cdot)$, we get the fourth order differential equation \eqref{eq:diffeqh4th}.\\

\paragraph{The fourth order ODE pins down the remaining free parameters}
Our proof is complete once we have shown that there is at most one set of values $C_1, C_2, h_0$ and $h_0'$ such that the resulting $h$ satisfies the two differential equations \eqref{eq:diffeqh} and \eqref{eq:diffeqh4th}.
Differentiating equation \eqref{eq:diffeqh} three times yields
\footnotesize
$$
\begin{array}{rrrrr}
h''(z)=&(-1+C_1-g(z)) h(z)&-z h'(z)&\\
h^{(3)}(z)=&-g'(z) h(z)& +(-2+C_1-g(z)) h'(z)&-z h''(z)&\\
h^{(4)}(z)=&- g''(z) h(z)&-2 g'(z) h'(z)&+(-3+C_1-g(z)) h''(z)&-z h^{(3)}(z)\\
h^{(5)}(z)=&-  g^{(3)}(z) h(z)& -3  g''(z) h'(z)&-3 g'(z) h''(z)&\\
&&&+(-4+C_1-g(z)) h^{(3)}(z)&-z h^{(4)}(z),
\end{array}
$$
\normalsize
and differentiating equation \eqref{eq:diffeqh4th} yields
\footnotesize
$$
\begin{array}{rrrr}
h^{(4)}(z)=& \left(-2-C_2+(-C_1+g(z))^2+k(z)\right) h(z)&-4 z h'(z)&\\
&&-\left(5+z^2\right) h''(z)&-2 z h^{(3)}(z),\\
h^{(5)}(z)=& \left(2 (-C_1+g(z))
g'(z)+k'(z)\right) h(z)& + \left(-6-C_2+(C_1-g(z))^2+k(z)\right) h'(z)&\\
&-6 z h''(z)&+\left(-7-z^2\right) h^{(3)}(z)&-2 z 
h^{(4)}(z).
\end{array}
$$
\normalsize
We can iteratively eliminate the derivatives of $h(\cdot)$ from these equations by substitution. 
After doing so, we divide by $h(z)$, which is possible since $h(z)>0$ for all $z$ by construction.
This  yields the following equation involving the constants $C_1$ and $C_2$, but not involving the function $h(\cdot)$ or any of its derivatives:
\footnotesize
\begin{multline*}
C_1^2+C_2^2+g(z)^2+k(z)^2-z^2 g'(z)^2+4 k(z) g''(z)+3 
g''(z)^2\\
-2 C_2 \left(g(z)+k(z)+2 g''(z)\right)+2 g(z) 
\left(k(z)+2 \left(g'(z)^2+g''(z)\right)\right)\\
+C_1\left(2 C_2-2 \left(g(z)+k(z)+2 \left(g'(z)^2+g''(z)\right)\right)\right)-2 g'(z) g^{(3)}(z)
=2 g'(z) k'(z)
\end{multline*}
\normalsize
This equation again has to hold for all $z$.
Differentiating twice with respect to $z$ yields new equations where the constants $C_1$ and $C_2$ enter only linearly, and we can explicitly solve for them.\footnote{The resulting expressions are unwieldy and so are omitted here, but are available on request.}

Substituting the solutions $C_1$ and $C_2$ back into one of the
first order differential equations we obtained by substitution and elimination of higher order derivatives above, we obtain a solution for $h_0'$ given $h_0$.
Given $h_0$, $h_0'$ and the constants $C_1$ and $C_2$, equation \eqref{eq:diffeqh} yields a unique solution $h(z)$ for all $z$.
Rescaling any solution $h(\cdot)$ by a constant again yields a solution by linearity of the differential equations.
$h_0$ is finally pinned down by the constraint $\int h(z) dz =1$.
$\Box$\\[12pt]

\paragraph{Remarks:}
\begin{itemize}
\item The proof of Theorem \ref{theo:identificationsigmanonpar} shows that our model is overidentified. If we consider higher order derivatives of equations \eqref{eq:diffeqh} and \eqref{eq:diffeqh4th}, or alternatively evaluate them at different values $z$, we obtain infinitely many restrictions on a finite number of free parameters.

\item The proof of identification is considerably simplified if we restrict the model to a normal distribution for $\Theta^*$, $\Theta^*\sim N(\bar \mu, \tau^2)$, which implies 
$Z^*|\sigma^*=1 \sim N(\bar \mu, \tau^2 +1),$
and thus
$h(z) =const. \cdot \exp \left (-\tfrac{1}{2 (\tau^2+1)}(z-\bar{\mu})^2 \right )$.
Denoting $e(z) = \partial_z \log h(z)$, we can rewrite equation \eqref{eq:diffeqh} as
$$e'(z) =  C_1 - g(z) - 1 - z e(z)  -e^2(z),$$
while the normality assumption yields
$e(z) = -(z-\bar{\mu}) /(\tau^2+1) $ and $e'(z) = -\tfrac{1}{(\tau^2+1)}$.
Plugging in yields
$$-\tfrac{1}{(\tau^2+1)} =  C_1 - g(z) - 1 + z \tfrac{z-\bar{\mu}}{(\tau^2+1)}  -\left (\tfrac{z-\bar{\mu}}{(\tau^2+1)}\right )^2.$$
Evaluating this equation at different values $z$ pins down $\tau^2$ and $\bar{\mu}$.

\item The proof of Theorem \ref{theo:identificationsigmanonpar} could be equivalently stated in terms of linear operators rather than differential equations. In particular, the ordinary differential equations \eqref{eq:diffeqh} and \eqref{eq:diffeqh4th} are equivalent to the following two linear operator equations, indexed by $z$ and linear in $\mu$,
\begin{align*}
\int \left [\theta \left ( z- \theta  \right ) -(g(z) - C_1)  \right ]  \varphi\left ( z- \theta \right ) d\mu(\theta)
&=0\\
\int \left [\left (-\theta^2 + \theta^2 \left ( z- \theta  \right )^2\right )   -\left (k(z)-C_2 +\left (g(z) -C_1\right )^2  \right )\right ]  \varphi\left ( z- \theta \right ) d\mu(\theta)
&=0
\end{align*}
Identification is then equivalent to the ``completeness condition'' that there is at most one probability measure $\mu$ in the orthocomplement of the span of the functions of $\theta$
\begin{align*}
\left [\theta \left ( z- \theta  \right ) -(g(z) - C_1)  \right ] \varphi\left ( z- \theta \right ) & \textrm{ and}\\
\left [\left (-\theta^2 + \theta^2 \left ( z- \theta  \right )^2\right )   -\left (k(z)-C_2 +\left (g(z) -C_1\right )^2  \right )\right ] \varphi\left ( z- \theta \right )&.
\end{align*}
\end{itemize}
$•$

\paragraph{Proof of Corollary \ref{corr:metastudyNormalizedSign}: }
The proof proceeds like the proof of Corollary \ref{corr:replicationexperimentsNormalizedSign}.
Let $S_i^*=\pm 1$ with probability $0.5$, independently of  $(X^*_i,\sigma^*_i, \Theta^*_i)$, and $S_j= S^*_{I_j}$.
Define $V= S\cdot |X|$.
We show that $(V,\sigma)$ satisfies the assumptions of Theorem \ref{theo:identificationsigmanonpar}, from which the claim then follows.

Define $\tilde{S}^* =  S^* \cdot \sign(X^*)$, so that $V= \tilde{S}\cdot X$, and define $\tilde{\Theta}^* =\tilde{S}^* \cdot \Theta^*$.
Since $\tilde{S}$ is independent of $(Z,\sigma,\Theta)$, we get
$\tilde{\Theta}^* \sim \tilde{\mu} = \tfrac{1}{2}(\mu_{\Theta^*} + \mu_{-\Theta^*})$
and
\begin{equation*}
f_{V/\sigma | \sigma}(z| \sigma) = 
\frac{p(z) \cdot \int \varphi(z-\theta / \sigma) d\tilde{\mu}(\theta)}{\int p(z') \varphi(z'-\theta/\sigma) dz' d\tilde{\mu}(\theta)}.
\end{equation*} 
This has the exact same form as the density of  $Z$ given $\sigma$ under the symmetric measure $\tilde{\mu}$. The claim follows, where we again use the fact that identification of $\tilde\mu$ implies identification of the distribution of $|\Theta^*|$.
$\Box$

\paragraph{Proof of Theorem \ref{thm: quantile unbiased}}

For the first claim, note that since $F_{X|\Theta}(x|\theta)$ tends to zero as $\theta\to-\infty$ and tends to one as $\theta\to\infty,$ for any $x$ and any $\alpha\in(0,1)$ there exist $\theta_l(x)$ and $\theta_u(x)$ such that $$F_{X|\Theta}(x|\theta_u(x))<\alpha<F_{X|\Theta}(x|\theta_l(x)),$$ where since $F_{X|\Theta}(x|\theta)$ is decreasing in $\theta$ we know that $\theta_l(x)<\theta_u(x).$ 
  Thus, since $F_{X|\Theta}(x|\theta)$ is continuous in $\theta$,  the intermediate value theorem implies that there exists $\hat{\theta}_{\alpha}(x)\in(\theta_l(x),\theta_u(x))$ such that $F_{X|\Theta}\left (x|\hat{\theta}_{\alpha}\left(x\right)\right)=\alpha.$  Since $F_{X|\Theta}(x|\theta)$ is strictly decreasing we know this $\hat{\theta}_{\alpha}(x)$ is unique, while its strict monotonicity and continuity likewise follow from strict monotonicity and continuity of $F_{X|\Theta}$ in both arguments.

For the second claim, note that since $F_{X|\Theta}(x|\theta)$ is strictly decreasing in $\theta,$ we have $\hat\theta_\alpha(x)\le \theta$ if and only if $F_{X|\Theta}(x|\theta) \le \alpha.$  Continuity of $F_{X|\Theta}(x|\theta)$ in $x,$ however, means that $X$ is continuously distributed conditional on $\Theta=\theta$ for all $\theta,$ and thus that $F_{X|\Theta}(X|\theta)$ is uniformly distributed conditional on $\Theta=\theta.$  Thus, 
$$P\left(F_{X|\Theta}(x|\theta)\le \alpha | \Theta=\theta \right)=\alpha,$$ so \[
P \left( \hat{\theta}_{\alpha}\left(X\right)\le\theta | \Theta=\theta \right) =\alpha~\text{for all}~\theta,
\]
as we aimed to show. $\Box$

\paragraph{Proof of Lemma \ref{lem: quantile unbiased sufficient condition}}
Under the stated assumptions, Lemma \ref{lem:likelihood} implies that $X$ is continuously distributed under all $\theta\in\mathbb{R},$ with density given by (\ref{eq: truncated density}). 
To prove the strict monotonicity of $F_{X|\Theta}(x|\theta)$ in $\theta,$ we adapt the proof of Lemma A.1 in \cite{lee2016exact}.

In particular, note that for $x_1>x_0$ and $\theta_1>\theta_0$, 
$$\frac{f_{X|\Theta}(x_1|\theta_1)}{f_{X|\Theta}(x_0|\theta_1)}>\frac{f_{X|\Theta}(x_1|\theta_0)}{f_{X|\Theta}(x_0|\theta_0)},$$
as can be verified from multiplying out these expressions.  This means, however, that 
$$f_{X|\Theta}(x_1|\theta_1)f_{X|\Theta}(x_0|\theta_0)>f_{X|\Theta}(x_1|\theta_0)f_{X|\Theta}(x_0|\theta_1).$$
Integrating both sides with respect to $x_0$ from $-\infty$ to $x<x_1$, and with respect to $x_1$ from $x$ to $\infty$, we obtain that 
$$(1-F_{X|\Theta}(x|\theta_1))F_{X|\Theta}(x|\theta_0)>(1-F_{X|\Theta}(x|\theta_0))F_{X|\Theta}(x|\theta_1),$$
and thus that $F_{X|\Theta}(x|\theta_0)>F_{X|\Theta}(x|\theta_1).$ Since this argument applies for all $x$ and all $\theta_0,$ $\theta_1,$ we have shown that $F_{X|\Theta}(x|\theta)$ is strictly decreasing in $\theta$ for all $x.$

To prove that $F_{X|\Theta}(x|\theta)\to 0$ as $\theta \to \infty$, note that by our assumption that $p(x)$ is almost everywhere continuous, for any $x_0$ there exists a point $x_1>x_0$, and an open neighborhood $(x_1-\varepsilon,x_1+\varepsilon)$ of $x_1$ such that $p(\cdot)$ is continuous on the closure of this neighborhood, and $x_0<x_1-2\varepsilon$.  Note, however, that for $\theta>x_1+\varepsilon$, $f_{X|\Theta}(x|\theta)$ for $x\le x_0$ is bounded above by $\varphi((x-\theta)/\sigma)/(\sigma\cdot E[p(X)|\Theta^*=\theta])$.  On the other hand, the infimum of $f_{X|\Theta}(x|\theta)$ over $(x_1-\varepsilon,x_1+\varepsilon)$ is bounded below by $p_l \cdot \varphi((x_1-\varepsilon-\theta)/\sigma)/(\sigma\cdot E[p(X)|\Theta^*=\theta])$ for 
$$p_l =\inf_{x\in[x_1-\varepsilon,x_1+\varepsilon]}p(x)>0.$$
Integrating and taking the ratio, we see that 
$$\frac{P(x\le x_0|\Theta=\theta)}{P(x\in(x_1-\varepsilon,x_1+\varepsilon)|\Theta=\theta)}\le
\frac{\Phi((x_0-\theta)/\sigma)}{2\varepsilon p_l\cdot \varphi((x_1-\varepsilon-\theta)/\sigma)/\sigma}.$$
This expression can in turn be bounded above by 
$$\frac{\Phi((x_0-\theta)/\sigma)}{2\varepsilon p_l\cdot \varphi((x_0-\theta)/\sigma)/\sigma},$$ which is proportional to Mill's ratio and tends to zero and $\theta \to \infty$ (see, for example, \cite{baricz_mills_2008}). This immediately implies that $F_{X|\Theta}(x_0|\theta)\to 0$, as we aimed to show. The claim that $F_{X|\Theta}(x|\theta)\to 1$ as $\theta \to -\infty$ can be proved analogously.
$\Box$

\section{Interpretation of meta-regression coefficients}
\label{suppsec: meta-regression coefficients}

In Section \ref{ssec:identificationliterature} of the main text we discussed meta-regressions.
We noted that under our assumptions meta-regressions deliver a valid test of the null of no selectivity.
We also noted, however, that in the presence of selectivity the function $E[Z|1/\sigma=\pi]$ is in general non-linear, and the slope of the best linear predictor cannot be interpreted as a selection-corrected estimate of $E[\Theta^*]$.

To see this, consider the following simple example.
Suppose that $\Theta^*  \equiv \bar{\theta}>0$, so there is no parameter heterogeneity across latent studies, and that $p(Z) =\mathbf{1}(Z> z^c )$, so there is strict selection on significant, positive effects.
Let $\varepsilon\sim N(0,1)$, and let $m$ be the inverse Mill's ratio,
$m(x) = \frac{\varphi\left( x \right )}{1-\Phi\left (x \right )}$.
Then
\[
E[Z|1/\sigma=\pi] =
E[\pi \bar{\theta}  + \varepsilon | \pi \bar{\theta}  + \varepsilon > z^c     ] = \pi \bar{\theta} + m\left (z^c -\pi \bar{\theta} \right ).
\]
This is a nonlinear function of $\pi$, and the slope and intercept of the best linear predictor approximating this function both depend on the distribution of $\pi$ (that is, of $\sigma$).
If $\sigma$ takes on only small values, and thus $\pi$ only takes on large values, the Mill's ratio term is negligible, and
$E^*[Z|1/\sigma=\pi] \approx \pi \bar{\theta}$.
If $\sigma$ takes on only large values, a first order approximation around $\pi=0$ yields
$$E^*[Z|1/\sigma=\pi] \approx m(z^c) + \bar\theta ( 1- m'(z^c)) \cdot \pi.$$
This shows in particular that the slope, which in this example equals $\bar\theta ( 1- m'(z^c))$, is in general different from the average effect $\bar\theta$, so that meta-regressions cannot be expected to deliver bias-corrected estimates of $E[\Theta^*]$.

\section{Likelihood and parametric specifications}
\label{suppsec: Likelihood and parametric specifications}

%We begin by describing the parametric models we fit in our applications.  %We first discuss our specifications for settings with replication data and then turn to meta-studies.

\subsection{Systematic replications}
\label{sssec: replication specifications}

Under the replication setup of display (\ref{eq:replicationexperimentsSigma}), the marginal density of $Z, Z^{r}, \sigma$ is
\begin{equation}
f_{Z, Z^{r}, \sigma}(z, z^r, \sigma) = \frac{p(z)\int \varphi(z - \theta) \cdot \tfrac{1}{\sigma}\varphi\left (\tfrac{z^r-\theta}{\sigma}\right ) d\mu(\theta)}{\iint p(z') \cdot \varphi(z' - \theta) dz' d\mu(\theta)}f_{\sigma^*|Z^*}(\sigma|z).
\label{eq:replicationdensitynormal}  
\end{equation} 
Denoting the total number of observations by $J$, the joint likelihood of the observed sample $((z_1,z_1^r,\sigma_1),...,(z_J,z_J^r,\sigma_J))$ is $\mathcal{L}(p,\mu)=\prod_{j=1}^J f_{Z, Z^{r}, \sigma}(z_j, z_j^r, \sigma_j).$
To fit a given model, we maximize this likelihood with respect to $p(\cdot)$ and $\mu$.  Since $f_{\sigma^*|  Z^*}$ enters multiplicatively, it plays no role in maximum likelihood estimation of $p(\cdot)$ and $\mu.$  Hence, we drop this term from the likelihood used in estimation.

To model  $p(\cdot),$ similar to
 \cite{hedges1992modeling} we consider step functions
\[p(z) \propto \sum_{k=1}^K \beta_{p,k} \cdot \mathbf{1}\left (\zeta_{k-1} \leq z < \zeta_{k}\right ),\] 
where $-\infty=\zeta_0 < \zeta_1 < \ldots < \zeta_K=\infty$ are fixed cutoffs. Since $p(\cdot)$ is only identified up to scale, we normalize $\beta_{p,K}=1$ and estimate $\beta_{p,1},...,\beta_{p,K-1}$.  Thus $\beta_{p,k}$ can be interpreted as the publication probability for a latent study with $Z^*$ between $\zeta_{k-1}$ and $\zeta_{k},$ relative to a latent study with $Z^*\ge \zeta_{K-1}.$
%Finally, to model $\mu$ we assume that $\Theta^*$ is normally distributed with mean $\bar \theta$ and variance $\tau^2$, and hence that
%\[(Z^*,Z^{*r}, \sigma^*) \sim N\left (
%\begin{pmatrix}
%\bar \theta\\
%\bar \theta
%\end{pmatrix},
%\begin{pmatrix}
%\tau^2 +1 & \tau^2\\
%\tau^2 & \tau^2 + \sigma^{*2}
%\end{pmatrix}
%\right ) \cdot f_{\sigma^*|  Z^*} \]
%Under these assumptions the likelihood is available in closed form, simplifying estimation.

\paragraph{Sign normalization}

As noted in the discussion preceding Corollary \ref{corr:replicationexperimentsNormalizedSign}, the sign of the initial estimate is normalized to be positive in both of our replication datasets.  In these applications, we thus follow the approach of Corollary  \ref{corr:replicationexperimentsNormalizedSign} and assume that $p(\cdot)$ is symmetric around zero. We conduct estimation based on the normalized z-statistics  $(W,W^r)=\sign(Z) \cdot (Z, Z^r)$ using the marginal likelihood
\[f_{W, W^{r}, \sigma}(w, w^r, \sigma) =  f_{Z, Z^{r}, \sigma}(w, w^r, \sigma) + f_{Z, Z^{r}, \sigma}(-w, -w^r, \sigma).\]
In this setting, Corollary \ref{corr:replicationexperimentsNormalizedSign} implies that $\beta_1,...\beta_{k-1}$ and the distributuion of $|\Theta^*|$ are identified.

\paragraph{Specification test}

As noted in Section \ref{sssec: selection on theta}, replication data allows us to identify models where conditional publication probabilities may depend on both $Z^*$ and $\Theta^*.$  We use these models to check our baseline specifications.  
Note that in principle any model that nests the null of no dependence of $p(\cdot)$ on $\Theta^*$ given $Z^*$ can be used to construct a valid test of this null. The specific model we consider determines where power is directed.
In Section \ref{suppsec: latent selection model} we introduce a model where publication decisions depend both on $Z^*$ and on whether a 5\% z-test based on an unobserved independent normal estimate rejects $\Theta^*=0.$  
This yields a conditional publication probability of the form
\begin{equation}p\left(z,\theta\right)=\sum_{k=1}^K\left(\beta_{p,k}+\gamma_{p,k}\cdot\Psi(\theta)\right)\cdot1\left\{ \zeta_{k-1}\le z<\zeta_{k}\right\},\label{eq:p z theta spec}\end{equation}
for
$$\Psi(\theta)=\frac{\Phi(1.96-\theta)-\Phi(-1.96-\theta)-\Phi(1.96)+\Phi(-1.96)}{\Phi(1.96)+\Phi(-1.96)},$$
where $\Phi$ is the standard normal distribution function.
This model implies that the publication probability is $\beta_{p,k}$ when $Z^*$ is in bracket $k$ and $\Theta^*$ is zero, while the publication probability is approximately $\beta_{p,k}+\gamma_{p,k}$ when $Z^*$ is in bracket $k$ and $|\Theta^*|$ is large.
Setting  $\gamma_p=0$ recovers our baseline model, so testing $H_0:\gamma_p=0$ allows us to test our baseline specifications.

\subsection{Meta-studies}
 \label{suppsubsec: Meta-studies}
In the meta-study context, the marginal likelihood of $(X,\sigma)$ is 
\begin{equation}
f_{X, \sigma}(x,\sigma) =\frac{p(\tfrac{x}{\sigma}) \cdot \int   \varphi\left (\tfrac{x-\theta}{\sigma}\right ) d\mu(\theta)}{\int p(\tfrac{x'}{\sigma}) \cdot \varphi\left (\tfrac{x'-\theta}{\sigma}\right ) dx' d\mu(\theta)}f_\sigma^*(\sigma).
\label{eq:metastudydensitynormal}  
\end{equation} 
Again denoting the total number of observations by $J,$ this yields joint likelihood 
$\mathcal{L}(p,\mu)=\prod_{j=1}^J f_{X, \sigma}(x_j, \sigma_j),$
which we again use to estimate $p(\cdot)$ and $\mu.$  As before, $f_\sigma$ enters multiplicatively and need not be specified.
Also as before, we consider step function specifications for $p(\cdot).$
% and assume that $\Theta^*$ is $N(\bar\theta,\tau^2)$ distributed, so 
%$$(X^*,\sigma^*)\sim N(\bar\theta,\tau^2+\sigma^2)\cdot f_\sigma(\sigma^*).$$
%Under these assumptions, the marginal likelihood (\ref{eq:metastudydensitynormal}) is again available in closed form.

\paragraph{Sign normalization}

In contexts where the sign of the initial estimate has been normalized to be positive, we follow the analog of the approach described above, restricting $p(\cdot)$ to be symmetric and conducting estimation based on $|X|=W\cdot\sigma$ and $\sigma.$ 

Note that meta-regressions, as discussed in section \ref{ssec:identificationliterature}, do not yield a valid test of the null of no selectivity when using sign-normalized data. Regressions of $|X|$ on $\sigma$ can have a non-zero slope even when $p(\cdot)$ is constant, and regressions of $|Z|$ on $1/\sigma$ can have a non-zero intercept.
For this reason, we do not discuss meta-regression results in our sign-normalized applications.

\section{Latent selection model}
\label{suppsec: latent selection model}

The baseline model we consider assumes that $E[D=1| X^*, \Theta^*] = p(X^*)$, so there is no dependence of publication probabilities on the latent parameter given $X^*$.
In the context of systematic replication studies with normally distributed estimates, however, we showed that a more general class of models which allows for dependence of $p(\cdot)$ on $\Theta^*$ is identified.
In Section \ref{sssec: replication specifications} we introduced a parametric specification for such a more general model, which we then estimate to provide a specification check for our baseline model.

The parametric specification introduced in Section \ref{sssec: replication specifications} can be derived as follows. Assume that publication decisions are based on 
\[
\left(\begin{array}{c}
Z^{*}\\
V^{*}
\end{array}\right)|\Theta^{*}\sim N\left(\left(\begin{array}{c}
\Theta^{*}\\
\Theta^{*}
\end{array}\right),\left(\begin{array}{cc}
1 & 0\\
0 & 1
\end{array}\right)\right),
\]
where $V^{*}$ is a second, independent estimate of the true effect $\Theta^{*}$,
with the same variance as $Z^{*}$. Assume further that
\[
D_{i}|Z_{i}^{*},V_{i}^{*},\Theta_i^*\sim Ber\left(p\left(Z_{i}^{*},V_{i}^{*}\right)\right),
\]
so publication decisions are based on $Z_i^*$ and $V_i^*.$  Since $V_i^*$ is unobserved, integrating over its distribution gives publication probabilities of the form $p(Z^*,\Theta^*).$

We want our specification for $p\left(z,v\right)$ to nest our baseline
specifications, 
\[
p\left(z\right)=\sum_{k=1}^{K}\beta_{p,k}1\left\{ \zeta_{k-1}\le z<\zeta_{k}\right\}.
\]
To ensure this, we consider the generalized specification 
\[
p\left(z,v\right)=\begin{array}{c}
\sum_{k=1}^{K}\tilde{\beta}_{p,k}^{1}1\left\{ \zeta_{k-1}\le z<\zeta_{k},|v|\ge\zeta_{V}\right\} \\
+\sum_{k=1}^{K}\tilde{\beta}_{p,k}^{0}1\left\{ \zeta_{k-1}\le z<\zeta_{k},|v|<\zeta_{V}\right\} ,
\end{array}
\]
which allows publication probabilities to depend on whether two-sided
z-tests based on the latent variable $v$ reject $\Theta^{*}=0$.
Integrating over the distribution of $V^{*}$ yields the following
specification for $p\left(z,\theta\right)$: 
\[
p\left(z,\theta\right)=\begin{array}{c}
\sum_{k=1}^{K}\tilde{\beta}_{p,k}^{1}1\left\{ \zeta_{k-1}\le z<\zeta_{k}\right\} \left(1-\tilde\Psi\left(\zeta_{V},\theta\right)\right)\\
+\sum_{k=1}^{K}\tilde{\beta}_{p,k}^{0}1\left\{ \zeta_{k-1}\le z<\zeta_{k}\right\} \tilde\Psi\left(\zeta_{V},\theta\right),
\end{array}
\]
where 
\[
\tilde\Psi\left(\zeta_{V},\theta\right)=Pr\left\{ \left|V\right|<\zeta_{V}|\Theta^{*}=\theta\right\} =\Phi\left(\zeta_{V}-\theta\right)-\Phi\left(-\zeta_{V}-\theta\right).
\]

As noted in the main text, $p\left(z,\theta\right)$ is only nonparametrically
identified up to a normalization for each value $\theta$. Analogous to our baseline specifications, here we impose the normalization
$\tilde{\beta}_{p,K}^{1}=\tilde{\beta}_{p,K}^{0}=1.$ To obtain the specification discussed in Section \ref{sssec: replication specifications}, we then define 
$$\beta_{p,k}=\tilde\beta^1_{p,k}+\tilde\Psi(\zeta_V,0)\cdot(\tilde\beta^0_{p,k}-\tilde\beta^1_{p,k}),$$
 $$\gamma_{p,k}=\left(\tilde{\beta}_{p,k}^{1}-\tilde{\beta}_{p,k}^{0}\right)\cdot\tilde\Psi\left(\zeta_{V},0\right),$$ and 
$$\Psi(\zeta_V,\theta)=\frac{\tilde\Psi\left(\zeta_{V},\theta\right)-\tilde\Psi\left(\zeta_{V},0\right)}{-\tilde\Psi\left(\zeta_{V},0\right)},$$ which yields the specification
$$p\left(z,\theta\right)=\sum_{k=1}^K\left(\beta_{p,k}+\gamma_{p,k}\cdot\Psi(\zeta_V,\theta)\right)\cdot1\left\{ \zeta_{k-1}\le z<\zeta_{k}\right\} .$$
Note that our normalization now implies that $\beta_{p,K}=1$ and $\gamma_{p,K}=0.$
For our specification tests we set $\zeta_V=1.96,$ corresponding to a 5\% test based on $V^*.$   

\section{Details on data and variable construction}
\label{suppsec: Application details}

In this section, we give additional details on our applications in Section \ref{sec:applications} of the main text and discuss how we cast the data of \cite{camerer2016evaluating} and \cite{open2015estimating} into our framework.

\subsection{Details for economics laboratory experiments}

To apply our approach, we need z-statistics and standard errors for both the original and replication studies.  For the application to data from  \cite{camerer2016evaluating}, we first collect p-values and standardized effect sizes from table S1 in the supplement.  Some of the p-values are censored below at .001, so for these studies we also collect the original estimates and standard errors from the replication reports posted online by Camerer et al.\footnote{Available at \url{https://experimentaleconreplications.com/replicationreports.html}, accessed September 3, 2016.} and recompute the censored p-values.  We then construct z-statistics by inverting the p-value transformation, where $z=\Phi^{-1}(1-p/2)$.
To obtain effect size estimates, we apply the Fisher transformation to standardized effect sizes reported by  Camerer et al.  Dividing these estimates by the z-statistics finally recovers the standard error.

We can infer the sign of the z-statistics from the sign of the standardized effect. Since signs are arbitrary and not comparable across studies, however, we normalize all signs to be positive.

\subsection{Details for psychology laboratory experiments}

To apply our approach to the data from \cite{open2015estimating}, we again need z-statistics and standard errors for both the original and replication studies.  We draw the inputs for all of these calculations from the RPPdataConverted spreadsheet posted online by the Open Science Collaboration.\footnote{Available at \url{https://osf.io/ytpuq/files/}, accessed January 19, 2017.}  Since \cite{open2015estimating} report p-values for both the original and replication studies, we invert the p-value transform to obtain z statistics. We use the p-values reported in their columns T.pval.USE.O and T.pval.USE.R for the original and replication studies, respectively.  Since some of the p-values in this application  are based on one-sided tests, we account for this in the inversion step.  
To compute effect size estimates, we again apply the Fisher transformation to the standardized effect sizes (columns T.r.O and T.r.R of RPPdataConverted for the original and replication studies, respectively), and then divide these estimates by the z-statistics to construct standard errors.

\section{Additional maximum likelihood results}
\label{suppsec: Additional empirical results}

This section discusses results from additional specifications estimated by maximum likeihood, intended to complement the results discussed in the main text.

\subsection{Additional results for economics laboratory experiments}

Here we report results based on an alternative specification for the economics replication data from \cite{camerer2016evaluating}.  We consider specifications which allow the probability of publication to vary depending on whether a latent study is sent to the American Economic Review (AER) or Quarterly Journal of Economics (QJE).
The publication probability is identified up to scale separately for each journal.   We index the journal by $w,$ and set set $p(z,w)$ proportional to one for both journals when the result is significantly different from zero at the 5\% level.   This ensures that the $\beta$ parameters can be interpreted as publication probabilities for insignificant results relative to significant results at the same journal.  Our ultimate specification is
\begin{align*}
p(Z,S) \propto \begin{cases}
\beta_{p,1} & |Z|< 1.96, W=AER \\
\beta_{p,1}+\beta_{p,2} & |Z|< 1.96, W=QJE\\
1 & |Z|\geq 1.96.
\end{cases}
\end{align*}
 Results are reported in Table \ref{tab:EconControlsEstim}.  In both the replication and metastudy specifications we estimate that the QJE is more likely to publish insignificant results.  This makes sense given that the sample contains one significant result and one insignificant result published in the QJE, while it contains fifteen significant results and one insignificant result published in the AER.  The estimated publication probabilities for the QJE are quite noisy, however, and we cannot reject the hypothesis that $\beta_{p,2}=0,$ so the same publication rule is used at both journals.

\begin{table}[!htb]
    \begin{minipage}{.5\linewidth}
      \centering
      \textsc{Replication}\\[5pt]
\begin{tabular}{cc |cc}
$ \kappa$ & $\lambda$ & $\beta_{p,1}$ & $\beta_{p,2}$ \\ 
 \hline 
0.373 & 2.153 & 0.015 & 0.216 \\  
(0.267) & (1.029) & (0.021) & (0.333) \\  
\end{tabular}

    \end{minipage}%
    \begin{minipage}{.5\linewidth}
      \centering
      \textsc{Meta-study}\\[5pt]
\begin{tabular}{cc |cc}
$\tilde \kappa$ & $\tilde \lambda$ & $\beta_{p,1}$ & $\beta_{p,2}$ \\ 
 \hline 
1.847 & 0.131 & 0.021 & 0.786 \\  
(1.582) & (0.065) & (0.030) & (1.496) \\  
\end{tabular}

    \end{minipage}
\caption{Selection estimates from lab experiments in economics, allowing publication probability to vary by journal. The left panel reports estimates from replication specifications, while the right panel reports results from meta-study specifications.  Publication probability $\beta_p$ is measured relative to omitted category of studies significant at 5\% level.\label{tab:EconControlsEstim}} 
\end{table}

\subsection{Additional results for psychology laboratory experiments}

We next report results based on three alternative specifications for the psychology replication data from \cite{open2015estimating}.  First, we limit attention to studies with a large number of denominator degrees of freedom.  Second, we limit attention to studies where the replication protocols were approved by the original authors.  Third, we allow the publication rule to vary by journal.

\paragraph{Denominator degrees of freedom}

As noted in the main text, our baseline analysis of the \cite{open2015estimating} data focuses on studies that use z- or t-statistics (or the squares of these statistics).  Our analysis then treats these statistics as approximately normal.  A potential problem here is that t-distributions with a small number of degrees of freedom behave differently from normal distributions, and in particular have heavier tails.  While the smallest degrees of freedom in the \cite{open2015estimating} data is seven, this concern may still lead us to worry about the validity of our approach in this setting.  To address this concern, in Table \ref{tab:PsychLargedofEstim} we report parameter estimates using the replication and meta-study specifications discussed in Section \ref{ssec: psych application}, where
\begin{align*}
p(Z) \propto \begin{cases}
\beta_{p,1} & |Z|< 1.64\\
\beta_{p,2} & 1.64 \le|Z|< 1.96\\
1 & |Z|\geq 1.96,
\end{cases}
\end{align*}
except that we now limit attention to the 52 observations with denominator degrees of freedom at least 30 in the original study.\footnote{We screen only on the degrees of freedom in the original study since sample sizes, and thus degrees of freedom, in the replication studies depend on the results in the initial study.  Hence, screening on replication degrees of freedom has the potential to introduce additional selection on the results of the original study.}  Our results are broadly similar for this restricted sample and for the full data.

\begin{table}[!htb]
    \begin{minipage}{.5\linewidth}
      \centering
      \textsc{Replication}\\[5pt]
\begin{tabular}{cc |cc}
$ \kappa$ & $\lambda$ & $\beta_{p,1}$ & $\beta_{p,2}$ \\ 
 \hline 
0.174 & 1.602 & 0.007 & 0.142 \\  
(0.121) & (0.677) & (0.005) & (0.079) \\  
\end{tabular}

    \end{minipage}%
    \begin{minipage}{.5\linewidth}
      \centering
      \textsc{Meta-study}\\[5pt]
\begin{tabular}{cc |cc}
$\tilde \kappa$ & $\tilde \lambda$ & $\beta_{p,1}$ & $\beta_{p,2}$ \\ 
 \hline 
0.869 & 0.138 & 0.018 & 0.247 \\  
(0.657) & (0.059) & (0.012) & (0.142) \\  
\end{tabular}

    \end{minipage}
\caption{Selection estimates from lab experiments in psychology, restricted to observations with denominator degrees of freedom at least 30, with standard errors in parentheses. The left panel reports estimates from replication specifications, while the right panel reports results from meta-study specifications.  Publication probability $\beta_p$ is measured relative to omitted category of studies significant at 5\% level.\label{tab:PsychLargedofEstim}} 
\end{table}

\paragraph{Approved replications}

As discussed in the main text, \cite{Gilbertetal2016} argue that some of the replications in \cite{open2015estimating} deviated substantially from the protocol of the original studies, which might lead to a violation of our assumption that the replication and original results are generated by the same underlying parameter $\Theta$.  Before conducting their replications, however, \cite{open2015estimating} asked the authors of each original study to review the proposed replication protocol, and recorded whether the original authors endorsed the replication protocol.  We can thus partly address this critique by limiting attention to the subset of studies where the replication was endorsed by the authors of the original study.  Re-estimating the specifications of Section \ref{ssec: psych application} on the 51 endorsed replications, we obtain the estimates reported in Table \ref{tab:PsychApprovedRep}.  These estimates suggest a somewhat smaller degree of selection than our baseline estimates, consistent with a higher rate of replication for approved replications, but are broadly similar to our other estimates.

\begin{table}[!htb]
    \begin{minipage}{.5\linewidth}
      \centering
      \textsc{Replication}\\[5pt]
\begin{tabular}{cc |cc}
$ \kappa$ & $\lambda$ & $\beta_{p,1}$ & $\beta_{p,2}$ \\ 
 \hline 
0.490 & 1.159 & 0.017 & 0.365 \\  
(0.268) & (0.402) & (0.011) & (0.165) \\  
\end{tabular}

    \end{minipage}%
    \begin{minipage}{.5\linewidth}
      \centering
      \textsc{Meta-study}\\[5pt]
\begin{tabular}{cc |cc}
$\tilde \kappa$ & $\tilde \lambda$ & $\beta_{p,1}$ & $\beta_{p,2}$ \\ 
 \hline 
0.634 & 0.198 & 0.022 & 0.440 \\  
(0.502) & (0.078) & (0.014) & (0.217) \\  
\end{tabular}

    \end{minipage}
\caption{Selection estimates from lab experiments in psychology, approved replications, with standard errors in parentheses. The left panel reports estimates from replication specifications, while the right panel reports results from meta-study specifications.  Publication probability $\beta_p$ is measured relative to omitted category of studies significant at the 5\% level.\label{tab:PsychApprovedRep}} 
\end{table}

\paragraph{Publication rule varies by journal}

The published studies replicated in \cite{open2015estimating} are drawn from Psychological Science (PS), the Journal of Personality and Social Psychology (JPSP), and the Journal of Learning Memory and Cognition (JLMC).  In this section we estimate a model where we allow the publication rule to vary by journal, which we index by $W$.  In particular, we consider the publication rule:
\begin{align*}
p(Z,W) \propto \begin{cases}
\beta_{p,1} & |Z|< 1.64, W=JLMC \\
\beta_{p,1}+\beta_{p,2} & |Z|< 1.64, W=PS\\
\beta_{p,1}+\beta_{p,3} & |Z|< 1.64, W=JPSP\\
\beta_{p,4} & 1.64\le|Z|< 1.96, W=JLMC \\
\beta_{p,4}+\beta_{p,5} & 1.64\le |Z|< 1.96, W=PS\\
\beta_{p4}+\beta_{p,6} & 1.64\le|Z|< 1.96, W=JPSP\\
1 & |Z|\geq 1.96,
\end{cases}
\end{align*}
As discussed in the economics application above, we normalize the publication probability for studies significant at the 5\% level to be proportional to one, which allows us to interpret the $\beta$ coefficients in terms of the publication probability for insignificant studies relative to that for significant studies at the same journal.  Such a normalization is necessary since publication probabilities are only identified up to a journal-specific scaling factor.

Results from estimating this model are reported in Table \ref{tab:PsychControlsEstim}.  These are noisier than our baseline estimates, as is intuitive given the larger number of parameters, but the JLMC coefficients show roughly the same pattern as our baseline specifications.  None of the differences between journal publication probabilities are significant, and a joint test yields a p-value of .78 in the replication specification and .84 in the metastudy specification, so in neither case do we reject the null hypothesis that all the journals use the same publication rule.

\begin{table}[!htb]
 \centering
\textsc{Replication}\\[5pt]
\begin{tabular}{cc |cccccc}
$ \kappa$ & $\lambda$ & $\beta_{p,1}$ & $\beta_{p,2}$ & $\beta_{p,3}$ & $\beta_{p,4}$ & $\beta_{p,5}$ & $\beta_{p,6}$ \\ 
 \hline 
0.315 & 1.308 & 0.008 & 0.002 & -0.001 & 0.428 & -0.288 & -0.332 \\  
(0.140) & (0.330) & (0.008) & (0.011) & (0.011) & (0.245) & (0.264) & (0.260) \\  
\end{tabular}

 \centering
\textsc{Meta-study}\\[5pt]
\begin{tabular}{cc |cccccc}
$\tilde \kappa$ & $\tilde \lambda$ & $\beta_{p,1}$ & $\beta_{p,2}$ & $\beta_{p,3}$ & $\beta_{p,4}$ & $\beta_{p,5}$ & $\beta_{p,6}$ \\ 
 \hline 
0.966 & 0.154 & 0.013 & 0.005 & 0.008 & 0.555 & -0.360 & -0.368 \\  
(0.561) & (0.054) & (0.014) & (0.019) & (0.026) & (0.320) & (0.350) & (0.364) \\  
\end{tabular}

\caption{Selection estimates from lab experiments in psychology, allowing publication probability to vary by journal. The top panel reports estimates from replication specifications, while the bottom panel reports results from meta-study specifications.  Publication probability $\beta_p$ is measured relative to omitted category of studies significant at 5\% level.\label{tab:PsychControlsEstim}} 
\end{table}

\subsection{Additional results for minimum wage meta-study}

This section reports results based on two alternative specifications for the data from \cite{wolfson201515}.  Since \cite{wolfson201515} include estimates from both published and working papers, we first reanalyze the data limiting attention to published studies.  We then examine whether the publication rules appear to vary with time.

\paragraph{Published Studies}
Table \ref{tab:MinimumWagePublishedMeta} reports estimates based on the model
\begin{align*}
\Theta^* \sim  \bar\theta+t(\nu)\cdot\tilde\tau, ~~~ p(X/\sigma) \propto \begin{cases}
\beta_{p,1} & X/\sigma< -1.96\\
\beta_{p,2} & -1.96 \le X/\sigma < 0\\
\beta_{p,3} & 0 \le X/\sigma < 1.96\\
1 & X/\sigma \geq 1.96
\end{cases}
\end{align*}
based on the subset of published papers, consisting of 705 estimates drawn from 31 studies. As in the main text we cluster our standard errors at the study level.  The resulting estimates are broadly similar to those obtained on the full sample.

\begin{table}[h!]
\begin{center}
\begin{tabular}{ccc |ccc}
$\bar \theta$ & $\tilde \tau$ & $\tilde \nu$ & $\beta_{p,1}$ & $\beta_{p,2}$ & $\beta_{p,3}$ \\ 
 \hline 
0.022 & 0.044 & 1.697 & 0.838 & 0.365 & 0.387 \\  
(0.012) & (0.025) & (0.380) & (0.331) & (0.146) & (0.140) \\  
\end{tabular}

\caption{Meta-study selection estimates from minimum wage data, published studies, with standard errors in parentheses.  Publication probability $\beta_p$ is measured relative to omitted category of studies estimating a positive effect significant at the 5\% level.\label{tab:MinimumWagePublishedMeta}}
\end{center}
\end{table}

\paragraph{Time Trends}
We next examine whether publication rules appear to vary over time.  In particular, letting $T_i$ denote the year in which study $i$ was initially circulated, for $\varsigma(x)=\exp(x)/(1+\exp(x))$ the logistic function we consider the model
\begin{align*}
\Theta^* \sim  \bar\theta+t(\nu)\cdot\tilde\tau, ~~~ p(X/\sigma,T) \propto \begin{cases}
\varsigma\left(\beta_{p,1}+\beta_{p,2}(T-2013)\right) & X/\sigma< -1.96\\
\varsigma\left(\beta_{p,3}+ \beta_{p,4}(T-2013)\right) & -1.96 \le X/\sigma < 0\\
\varsigma\left(\beta_{p,5}+ \beta_{p,6}(T-2013)\right) & 0 \le X/\sigma < 1.96\\
\varsigma\left(1\right) & X/\sigma \geq 1.96
\end{cases}
\end{align*}
where we measure time in years relative to 2013, which is the median year observed in the data, and $T$ varies between 2000 and 2015.  We use the logistic function here to ensure that publication probabilities lie between zero and one, and without the time trend this would simply be a reparameterization of our baseline model. Publication probabilities are only identified up to a year-specific scaling, so by normalizing the publication coefficient for studies finding a negative and significant effect of the minimum wage on employment to be proportional to one, we again ensure that the $\beta_p$ coefficients can be interpreted as measuring publication probabilities relative to the publication  probability for studies finding a negative and significant effect within the same year.

\begin{table}[h!]
\begin{center}
\begin{tabular}{ccc |cccccc}
$\bar \theta$ & $\tilde \tau$ & $\tilde \nu$ & $\beta_{p,1}$ & $\beta_{p,2}$ & $\beta_{p,3}$ & $\beta_{p,4}$ & $\beta_{p,5}$ & $\beta_{p,6}$ \\ 
 \hline 
0.019 & 0.021 & 1.359 & 0.284 & 0.176 & -1.231 & 0.074 & -1.089 & 0.025 \\  
(0.009) & (0.013) & (0.300) & (0.845) & (0.178) & (0.602) & (0.117) & (0.478) & (0.113) \\  
\end{tabular}

\caption{Meta-study selection estimates from minimum wage data, published studies, with standard errors in parentheses.  Publication probability $\beta_p$ is measured relative to omitted category of studies estimating a positive effect significant at the 5\% level.\label{tab:MinimumWageControlsMeta}}
\end{center}
\end{table}

These estimates are consistent with our baseline model assuming that publication rules are constant over time, with a p-value of 0.7 for the test of the joint hypothesis that $\beta_{p,2}=\beta_{p,4}=\beta_{p,6}=0$.

\subsection{Additional results for deworming meta-study}

In the main text we report estimates for the deworming data of \cite{deworming2016} based on a specification that restricts $p(\cdot)$ to be symmetric around zero.  To complement those results, here we consider the more flexible specification
\begin{align*}
\Theta^* \sim N(\bar\theta, \tau^2), ~~~ p(X/\sigma) \propto \begin{cases}
\beta_{p,1} & X/\sigma< -1.96\\
\beta_{p,2} & -1.96 \le X/\sigma < 0\\
\beta_{p,3} & 0 \le X/\sigma < 1.96\\
1 & X/\sigma \geq 1.96.
\end{cases}
\end{align*}
Results based on this specification are reported in Table \ref{tab:DewormingAsymmetricMeta}.  These estimates differ substantially from those reported in the main text, and suggest strong selectivity against negative estimates, particularly negative and significant estimates.  However, as can be seen from Figure \ref{fig:DewormingScatter} in the main text there is only a single negative and statistically significant estimate in the sample, so the reliability of conventional large-sample approximations here is highly suspect.  
\begin{table}[h!]
\begin{center}
\begin{tabular}{cc |ccc}
$\bar \theta$ & $\tilde \tau$ & $\beta_{p,1}$ & $\beta_{p,2}$ & $\beta_{p,3}$ \\ 
 \hline 
-0.714 & 0.521 & 0.008 & 0.151 & 1.299 \\  
(0.626) & (0.206) & (0.025) & (0.207) & (1.113) \\  
\end{tabular}

\caption{Meta-study selection estimates from deworming wage data, flexible specification, with standard errors in parentheses.  Publication probability $\beta_p$ is measured relative to omitted category of studies estimating a positive effect significant at the 5\% level.\label{tab:DewormingAsymmetricMeta}}
\end{center}
\end{table}

To reduce the number of free parameters, we estimate a version of the model which does not allow discontinuities in $p(\cdot)$ based on statistical significance, but only based on the sign of the estimate,
\begin{align*}
\Theta^* \sim N(\bar\theta, \tau^2), ~~~ p(X/\sigma) \propto \begin{cases}
\beta_{p} & X/\sigma<0 \\
1 & X/\sigma \geq 0.
\end{cases}
\end{align*}
Fitting this model yields the estimates reported in Table \ref{tab:DewormingRestrictedAsymmetricMeta}.  These estimates suggest strong selectivity on the sign of the estimated effect, where positive effects are estimated to be ten times more likely to be published than negative effects.  While this is consistent with the distribution of observations in Figure \ref{fig:DewormingScatter}, our choice of this specification was driven by our results in Table \ref{tab:DewormingAsymmetricMeta}.  Given that this is a form of specification search, it suggests that conventional asymptotic approximations may be unreliable here, and thus that these results should be treated with caution.

\begin{table}[h!]
\begin{center}
\begin{tabular}{cc |c}
$\bar \theta$ & $\tilde \tau$ & $\beta_p$ \\ 
 \hline 
-0.217 & 0.365 & 0.094 \\  
(0.156) & (0.103) & (0.099) \\  
\end{tabular}

\caption{Meta-study selection estimates from deworming wage data, restricted asymmetric specification, with standard errors in parentheses.  Publication probability $\beta_p$ is measured relative to omitted category of studies estimating a positive effect significant at the 5\% level.\label{tab:DewormingRestrictedAsymmetricMeta}}
\end{center}
\end{table}

\section{Moment-based estimation results}
\label{suppsec:momentresults}

In the main text we report estimates based on parametric
specifications for the distribution $\mu$ of true effects $\Theta^{*}$
in latent studies. To confirm that our results are not sensitive to
the choice of parametric specification for $\mu$, in this section
we report results based on moment-based estimators that require only that
we specify a functional form for the publication probability $p$, and leave the
distribution of true effects fully nonparametric. 
The moments used to obtain these estimators are motivated by the identification
arguments in Section \ref{sec:identification} of the paper. 

We begin by introducing the
moments we consider in the replication and meta-study settings, respectively,
and then discuss results in our applications.  Overall, we find that while moment-based approaches often yield less precise conclusions, our main findings are robust to dropping our parametric specifications for $\mu.$ 

\subsection{Estimation Moments}

\subsubsection{Replication Moments }

In our discussion of identification for settings with replication data in Section \ref{ssec:identificationreplication} of the main text,
we noted that if
the original and replication estimates have the same distribution in the population of latent studies, then
 absent selective
publication the joint distribution of published and replication estimates will likewise be symmetric. %In particular, we should see an equal number of significant results in the original and replication studies. 
This observation implies
a moment restriction that can be used for estimation.

To derive our moments, we focus on the case where the original and
replication estimates $\left(Z_{i}^{*},Z_{i}^{*r}\right)$ are normally
distributed in the population of latent studies. First consider
the case where $\sigma_{i}^{*}\equiv1$ for all $i$, so the original
and replication studies have the same standard error. Note that for any
constants $c_1$, $c_2$ 
\[
E\left[1\left\{ \left|Z_{i}^{*}\right|> c_1, \left|Z_{i}^{r*}\right|\le c_2\right\}
-1\left\{\left|Z_{i}^{r*}\right|> c_1, \left|Z_{i}^{*}\right|\le c_2\right\} \right]=0,
\]
 in the population of latent studies  no matter the distribution $\mu$ of true effects.\footnote{Here we focus on the absolute value of the original and replication estimates to avoid complications from the sign normalization in our replication applications.}  In particular, this reflects our observation in the main text that, absent selection, we should observe an equal number of cases where the original results are significant and the replications are insignificant, and where the replication results are significant and the original results are insignificant, where we can consider results significant and insignificant at different levels.

We can recover the distribution of latent studies from the
distribution of published studies by weighting by the inverse of the
publication probability, $E\left[p\left(Z^{*}\right)\right]/p\left(Z\right).$
This implies the moment restriction
\[
E\left[\frac{E\left[p\left(Z_{i}^{*}\right)\right]}{p\left(Z_{j}\right)}\left(1\left\{ \left|Z_{j}\right|> c_1, \left|Z_{j}^{r}\right|\le c_2\right\}
-1\left\{\left|Z_{j}^{r}\right|> c_1, \left|Z_{j}\right|\le c_2\right\}  \right)\right]=0
\]
in the population of published studies. Since $E\left[p\left(Z_{i}^{*}\right)\right]$
does not vary across observations, the moment restriction continues
to hold if we drop this term yielding moments
\begin{equation}
E\left[\frac{1}{p\left(Z_{j}\right)}\left(\left\{ \left|Z_{j}\right|> c_1, \left|Z_{j}^{r}\right|\le c_2\right\}
-1\left\{\left|Z_{j}^{r}\right|> c_1, \left|Z_{j}\right|\le c_2\right\}\right)\right]=0\label{eq: symmetric replication moment}
\end{equation}
which depend only on observables and $p(\cdot)$  and so  can be used to estimate $p\left(\cdot\right).$

Thus far, in deriving moments we have assumed that $\sigma_{i}^{*}\equiv1$.
In our applications, however, we in fact have $\sigma_{i}^{*}|Z_{i}^{*}\sim f_{\sigma|X}\left(\sigma|Z_{i}^{*}\right)$.
If the distribution of $\sigma_{i}$ is bounded above
by some value $\sigma_{\max}$, we can adapt the moments (\ref{eq: symmetric replication moment})
to account for unequal variances by noising up both the original and
replication estimates to noise level $\sigma_{\max}.$ In particular,
for $\varepsilon_{j},\varepsilon_{j}^{r}$ iid $N\left(0,1\right)$
random variables, 
\[
E\left[\frac{1}{p\left(Z_{j}\right)}\left(\begin{array}{c}
1\left\{ \left|Z_{j}+\sqrt{\sigma_{\max}^{2}-1}\varepsilon_{j}\right|>c_{1},\left|Z_{j}^{r}+\sqrt{\sigma_{\max}^{2}-\sigma_{j}}\varepsilon_{j}^{r}\right|\le c_{2}\right\} -\\
1\left\{ \left|Z_{j}^{r}+\sqrt{\sigma_{\max}^{2}-\sigma_{j}}\varepsilon_{j}^{r}\right|>c_{1},\left|Z_{j}+\sqrt{\sigma_{\max}^{2}-1}\varepsilon_{j}\right|\le c_{2}\right\} 
\end{array}\right)\right]=0.
\]
To eliminate the added noise $(\varepsilon_{j},\varepsilon_j^r)$ in these moments,
we can take the conditional expectation of each component given the
data and define 
\[
h\left(z,1,z^{r},\sigma\right)=E\left[\left\{\left|Z_{j}^{*}+\sqrt{\sigma_{\max}^{2}-1}\varepsilon_{j}\right|>c_{1},\left|Z_{j}^{*r}+\sqrt{\sigma_{\max}^{2}-\sigma}\varepsilon_{j}^{r}\right|\le c_{2}\right\}|Z_{j}^{*}=z,Z_{j}^{r*}=z^{r}\right]
\]
\[
=\left(1-\Phi\left(\frac{c_{1}-z}{\sqrt{\sigma_{\max}^{2}-1}}\right)+\Phi\left(\frac{-c_{1}-z}{\sqrt{\sigma_{\max}^{2}-1}}\right)\right)\left(\Phi\left(\frac{c_{2}-z^{r}}{\sqrt{\sigma_{\max}^{2}-\sigma}}\right)-\Phi\left(\frac{-c_{1}-z^{r}}{\sqrt{\sigma_{\max}^{2}-\sigma}}\right)\right).
\]
By the law of iterated expectations, we obtain the moment restrictions
\begin{equation}
E\left[\frac{1}{p\left(Z_{j}\right)}\left(h\left(Z_{j},1,Z_{j}^{r},\sigma_{j}\right)-h\left(Z_{j}^{r},\sigma_{j},Z_{j},1\right)\right)\right]=0\label{eq: Replication moments}
\end{equation}
which depends only on observables and $p$ and so can be used for estimation. 

To use these moments in practice we need to choose a value of $\sigma_{\max}$ and values for $c.$
In our applications we below we take $\sigma_{\max}$ to equal sample maximum of $\sigma_{i}$,
which is about 2.5 for the economics replications and about 2 for
the psychology replications, and consider values $c$ in each specification corresponding to the critical values used in $p.$
Setting $\sigma_{\max}$ to the sample maximum is ad-hoc, so as a further check
we also report results based on the moments 
\begin{equation}
E\left[\frac{1}{p\left(Z_{j}\right)}\left(\left(Z_{j}^{2}-1\right)-\left(Z_{j}^{r2}-\sigma_{j}^{2}\right)\right)\right]=0\label{eq: Replication moments, simple}
\end{equation}
which can be shown to hold for any $\mu$ by arguments along the same
lines as above and do not require that we select a value $\sigma_{\max}$.

\subsubsection{Metastudy Moments}

The moments we consider in our metastudy applications are derived
using a similar approach. As noted in our discussion of metastudy identification
 in Section \ref{ssec:identificationmeta} of the text, absent selectivity
in the publication process our assumptions
imply that the distribution of effects for noisier studies is just
a noised-up version of the distribution for less noisy studies.
In particular, if we consider a pair of latent studies $\left(i,i'\right)$
with $\sigma_{i}^{*}>\sigma_{i'}^{*}$ then for any constant $c$
and $\varepsilon_{i}\sim N\left(0,1\right)$
\[
E\left[\left. 1\left\{ X_{i}^{*}<c\sigma_{i}^{*}\right\} -1\left\{ X_{i'}^{*}+\sqrt{\sigma_{i}^{*2}-\sigma_{i'}^{*2}}\varepsilon_{i}<c\sigma_{i}^{*}\right\} \right|\sigma_i^*,\sigma_{i'}^*\right]=0.
\]
As above we can eliminate the noise from the added error $\varepsilon_{i}$.
If we define 
\[
h\left(x,\sigma_{1},\sigma_{2}\right)=E\left[1\left\{ X_{i}^{*}+\sqrt{\sigma_{1}^{2}-\sigma_{2}^{2}}\varepsilon_{i}<c\sigma_{1}^{*}\right\} |X_{i}^{*}=x\right]=\Phi\left(\frac{c\sigma_{1}-x}{\sqrt{\sigma_{1}^{2}-\sigma_{2}^{2}}}\right)
\]
then the law of iterated expectations implies that 
\[
E\left[1\left\{ X_{i}^{*}<c\sigma_{i}^{*}\right\} -h\left(X_{i'}^{*},\sigma_{i'}^{*},\sigma_{i}^{*}\right)\right]=0.
\]

As in the replication setting, to obtain moments which hold in
the population of published studies, we can weight inversely by the
publication probability (now for the pair $X_{i}$, $X_{i'}$), again
dropping normalizing constants to obtain the moments
\begin{equation}
E\left[\frac{1}{p\left(X_{j}/\sigma_{j}\right)}\frac{1}{p\left(X_{j'}/\sigma_{j'}\right)}\left(1\left\{ X_{j}<c\sigma_{j}\right\} -h\left(X_{j'},\sigma_{j'},\sigma_{j}\right)\right)\right]=0\label{eq: Metastudy Moments}
\end{equation}
which depend only on $p(\cdot)$ and observables and so can be used for estimation.\footnote{In the sign-normalized case, as above we instead form moments based on the absolute value of $X_j$.}

For estimation, we again consider values of $c$ corresponding to
the thresholds used in $p\left(\cdot\right)$. Since our moments hold for each pair $\left(j,j'\right)$
with $\sigma_{j}>\sigma_{j'}$, we average over all pairs of observations
and obtain asymptotic distributions using results for estimators based on U-statistics from \cite{HonorePowell1994}.

\subsection{Empirical Applications}

\subsubsection{Economics laboratory experiments}

In our application to data on economics lab experiments from \cite{camerer2016evaluating}, we again model the publication probability as
\[
p\left(Z\right)\propto\begin{cases}
\beta_{p} & \text{if }\left|Z\right|\le1.96\\
1 & \text{otherwise.}
\end{cases}
\]
When we attempt to estimate $\beta_{p}$ 
based on moments (\ref{eq: Replication moments}), we find that while the system of moments is just-identified and can be solved exactly, the
zero of the sample moments corresponds to a negative value of $\beta_{p}$.  This occurs because, unlike in likelihood estimation, the GMM moments do not automatically rule out negative values of $\beta_p$, though such values are meaningless under our model.  Indeed, we see in simulation that even under correct specification negative point estimates arise with non-negligible probability for small sample sizes and small values of $\beta_p.$  To address this issue,  in Table \ref{tab:EconReplicationsGMM} we report 95\% confidence  sets based on \cite{StockWright2000}, which are robust both to weak-identification and to parameter-on-the-boundary issues.  

\begin{table}[!htb]
    \begin{minipage}{.5\linewidth}
      \centering
      \textsc{Robust CS, Baseline Moments}\\[5pt]
\begin{tabular}{cc}
$\beta_p$ Lower Bound & $\beta_p$ Upper Bound \\ 
 \hline 
0.000 & 0.049 \\  
\end{tabular}

    \end{minipage}%
    \begin{minipage}{.5\linewidth}
      \centering
      \textsc{Robust CS, Alternative Moments}\\[5pt]
\begin{tabular}{cc}
$\beta_p$ Lower Bound & $\beta_p$ Upper Bound \\ 
 \hline 
0.000 & $\infty$ \\  
\end{tabular}

    \end{minipage}
\caption{\cite{StockWright2000} 95\% confidence sets for $\beta_p$ for lab experiments in economics. The left panel reports results based on our baseline moments (\ref{eq: Replication moments}) for replication models, while the right panel reports results based on the alternative moments (\ref{eq: Replication moments, simple}). Publication probability $\beta_p$ is measured relative to omitted category of studies significant at the 5\% level.\label{tab:EconReplicationsGMM}} 
\end{table}

From these results, we see that when we consider our baseline moments (\ref{eq: Replication moments}) we obtain a robust confidence set roughly consistent with the estimate of $\beta_p$ reported in the main text, even though we are fully relaxing our assumption on the distribution $\mu$ of latent effects.  When we consider the alternative moments (\ref{eq: Replication moments, simple}), by contrast, the moments are less informative, and the robust confidence set covers the full parameter space.

As before, instead of using the replication data we can instead focus just on the initial estimates and standard errors and apply our meta-study approach based on the moments (\ref{eq: Metastudy Moments}).  The results from this approach are 
reported in Table \ref{tab:EconReplicationsGMMMetastudy}.  For comparability with the replication results above we include both a conventional point estimate and standard error and an identification-robust confidence based on the generalization of \cite{StockWright2000} to the present U-statistic setting.
These results are again broadly consistent with those obtained both from the replication moments above and from our likelihood estimates in the main text, showing strong selection in favor of statistically significant results.

\begin{table}[!htb]
    \begin{minipage}{.5\linewidth}
      \centering
      \textsc{Point Estimate}\\[5pt]
\begin{tabular}{c}
$\beta_p$ \\ 
 \hline 
0.040 \\  
(0.042) \\  
\end{tabular}

    \end{minipage}%
    \begin{minipage}{.5\linewidth}
      \centering
      \textsc{Robust CS}\\[5pt]
\begin{tabular}{cc}
$\beta_p$ Lower Bound & $\beta_p$ Upper Bound \\ 
 \hline 
0.000 & 0.177 \\  
\end{tabular}

    \end{minipage}
\caption{Moment-based results for lab experiments in economics. The left panel reports an estimate and standard error  based on our moments (\ref{eq: Metastudy Moments}) for metastudy models, while the right panel reports a 95\% identification-robust confidence set based on the same moments. Publication probability $\beta_p$ is measured relative to omitted category of studies significant at the 5\% level.\label{tab:EconReplicationsGMMMetastudy}} 
\end{table}

\subsubsection{Psychology laboratory experiments}

\begin{figure}[t!]
    \begin{minipage}{.5\linewidth}
      \centering
\includegraphics[scale=0.8]{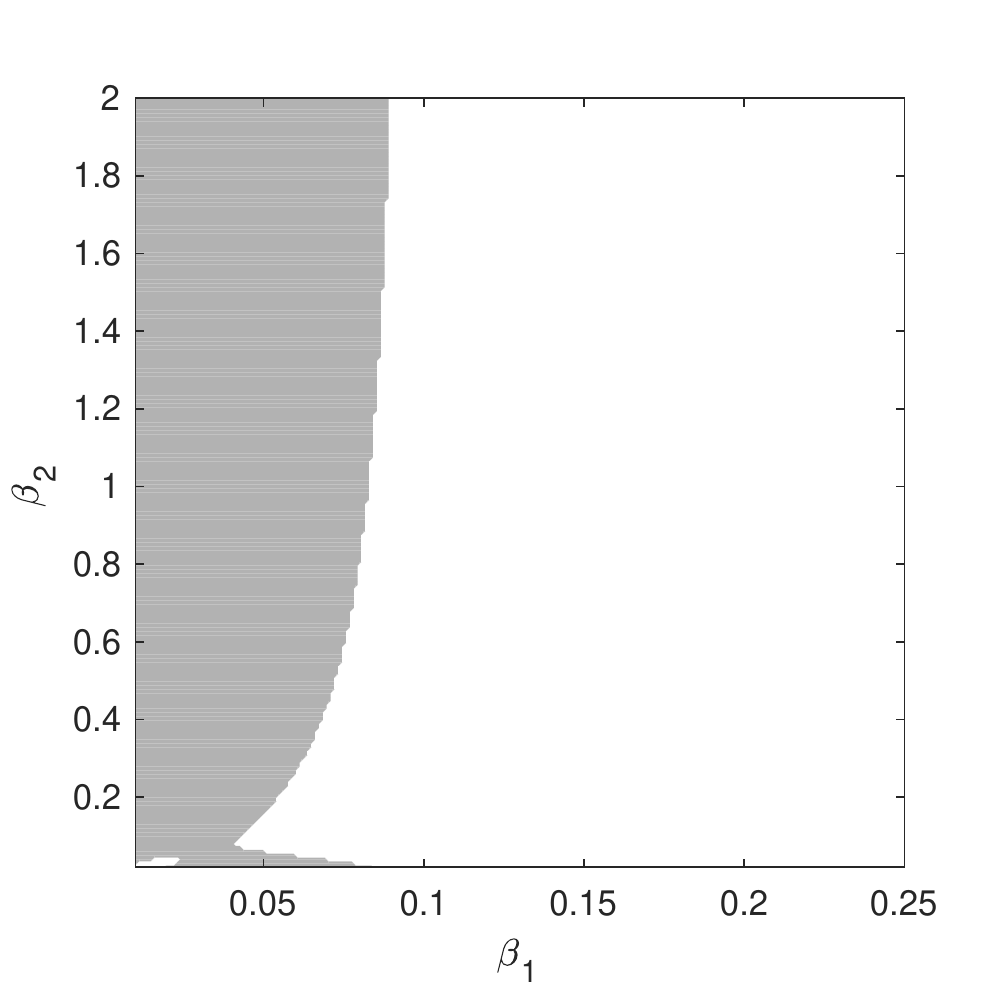}
    \end{minipage}
    \begin{minipage}{.5\linewidth}
      \centering
\includegraphics[scale=0.8]{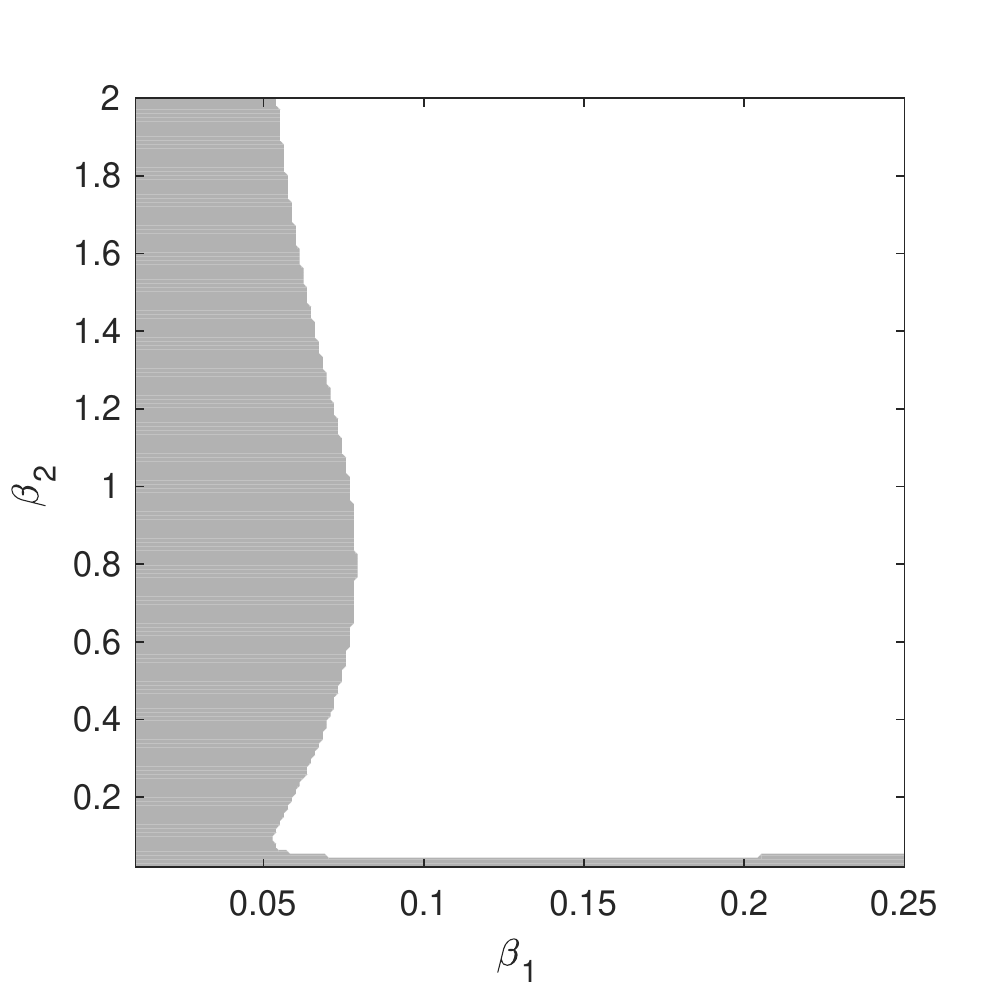}
    \end{minipage}
\caption{This figure plots 95\% \cite{StockWright2000} joint confidence sets for $\beta_{p,1}$ and $\beta_{p,2}$ using data from lab experiments in psychology.  The left panel plots results based on the baseline replication moments (\ref{eq: Replication moments}), while the right panel plots results based on the metastudy moments (\ref{eq: Metastudy Moments}).\label{fig: Psych Robust CS}}
\end{figure}

Turning next to the data on lab experiments in psychology from \cite{open2015estimating},  as in the main text we model the
publication probability as 
\[
p\left(Z\right)\propto\begin{cases}
\beta_{p1} & \text{if }\left|Z\right|\le1.64\\
\beta_{p2} & \text{if }1.64<\left|Z\right|\le1.96\\
1 & \text{otherwise}.
\end{cases}
\]
We find that identification
of $\beta_{p2}$ based on both our replication and metastudy moments appears weak in this setting. 
We report identification-robust joint confidence
sets for $\left(\beta_{p1},\beta_{p2}\right)$ based on \cite{StockWright2000}
in Figure \ref{fig: Psych Robust CS}.
While both confidence sets allow a wide range of possible values $\beta_{p2}$, only small values of $\beta_{p,1}$ are consistent with the confidence set based on replication data.  On the other hand, results based on our meta-study approach allow a wide range of values for either parameter, though they rule out cases where both are large simultaneously.  Both sets of results are consistent with our estimates in the main text, and in the case of the replications specification again provide evidence of selection against insignificant results.

To avoid specifying a value $\sigma_{\max}$ to use in the moments (\ref{eq: Replication moments}), we can instead consider the moments (\ref{eq: Replication moments, simple}).  Since yeilds only a single moment restriction, we consider  selection only on significance at the 5\% level, as in our application to economics
lab experiments above.  Robust confidence sets from this specification are reported in Table \ref{tab:PsychReplicationsGMMAlt}.
\begin{table}[!htb]
    \begin{minipage}{.5\linewidth}
      \centering
      \textsc{Robust CS, Alternative Replication Moments}\\[5pt]
\begin{tabular}{cc}
$\beta_p$ Lower Bound & $\beta_p$ Upper Bound \\ 
 \hline 
0.000 & 0.045 \\  
\end{tabular}

    \end{minipage}%
    \begin{minipage}{.5\linewidth}
      \centering
      \textsc{Robust CS, Metastudy Moments}\\[5pt]
\begin{tabular}{cc}
$\beta_p$ Lower Bound & $\beta_p$ Upper Bound \\ 
 \hline 
0.000 & 0.115 \\  
\end{tabular}
    \end{minipage}
\caption{\cite{StockWright2000} 95\% confidence sets for $\beta_p$ for lab experiments in psychology, assuming only selection on significance at the 5\% level. The left panel reports results based on our alternative moments (\ref{eq: Replication moments, simple}) for replication data, while the right panel reports results based on our metastudy moments. Publication probability $\beta_p$ is measured relative to omitted category of studies significant at the 5\% level.\label{tab:PsychReplicationsGMMAlt}} 
\end{table}
These results highlight that we still obtain informative results in this setting if we restrict attention to selection on significance at the 5\% level.

\subsubsection{Effect of minimum wage on employment}

For the data from \cite{wolfson201515} we consider the specification

\[
p\left(X/\sigma\right)\propto\begin{cases}
\beta_{p1} & \text{if }X/\sigma<-1.96\\
\beta_{p2} & \text{if }-1.96\le X/\sigma<0\\
\beta_{p3} & \text{if }0\le X/\sigma<1.96\\
1 & \text{if }X/\sigma\ge1/96.
\end{cases}
\]
Table \ref{tab:MinimumWageGMM} reports estimates and standard errors.
\begin{table}[h!]
\begin{center}
\begin{tabular}{ccc}
$\beta_{p,1}$ & $\beta_{p,2}$ & $\beta_{p,3}$ \\ 
 \hline 
1.174 & 0.231 & 0.235 \\  
(0.417) & (0.100) & (0.080) \\  
\end{tabular}

\caption{Meta-study selection estimates from GMM specifications for minimum wage data,  with standard errors in parentheses.  Publication probability $\beta_p$ is measured relative to omitted category of studies estimating a positive effect significant at the 5\% level.\label{tab:MinimumWageGMM}}
\end{center}
\end{table}
We see that the main message of our likelihood results in this setting, that results finding a significant and negative effect of the minimum wage on employment are favored over insignificant results, again comes through clearly.  In contrast to our likelihood results the point estimate for $\beta_{p1}$ also suggests selection in favor of significant results finding a positive effect of the minimum wage on employment, but given the large standard error associated with this coefficient the results are also consistent with selection on statistical significance alone ($\beta_{p1}=1$, $\beta_{p2}=\beta_{p3}$), with a p-value of .86 for the joint test.

\subsubsection{Deworming meta-study}

For the deworming data of \cite{deworming2016} we again consider the specification
\[
p\left(Z\right)\propto\begin{cases}
\beta_{p} & \text{if }\left|Z\right|\le1.96\\
1 & \text{otherwise}
\end{cases}.
\]
Estimating this model using our meta-study moments yields the point estimate and standard error reported in the left panel of Table \ref{tab:DewormingGMMMetastudy}.

\begin{table}[!htb]
    \begin{minipage}{.5\linewidth}
      \centering
      \textsc{Point Estimate}\\[5pt]
\begin{tabular}{c}
$\beta_p$ \\ 
 \hline 
0.251 \\  
(0.236) \\  
\end{tabular}

    \end{minipage}%
    \begin{minipage}{.5\linewidth}
      \centering
      \textsc{Robust CS}\\[5pt]
\begin{tabular}{cc}
$\beta_p$ Lower Bound & $\beta_p$ Upper Bound \\ 
 \hline 
0.048 & $\infty$ \\  
\end{tabular}

    \end{minipage}
\caption{Moment-based results for deworming data. Left panel reports an estimate and standard error based on our moments (\ref{eq: Metastudy Moments}) for metastudy models, while right panel reports a 95\% identification-robust confidence set based on the same moments. Publication probability $\beta_p$ is measured relative to omitted category of studies significant at the 5\% level.\label{tab:DewormingGMMMetastudy}} 
\end{table}
While the point estimate for $\beta_p$ obtained in this setting is quite different from that in our baseline specification, the robust confidence set is unbounded above, suggesting that identification is quite weak and the point estimate is likely unreliable.

\section{Bias corrections based on applications}
\label{suppsec: Inference corrections based on estimates}

\begin{figure}[p]
\minipage{0.45\textwidth}
\begin{center}
\includegraphics[width=\textwidth]{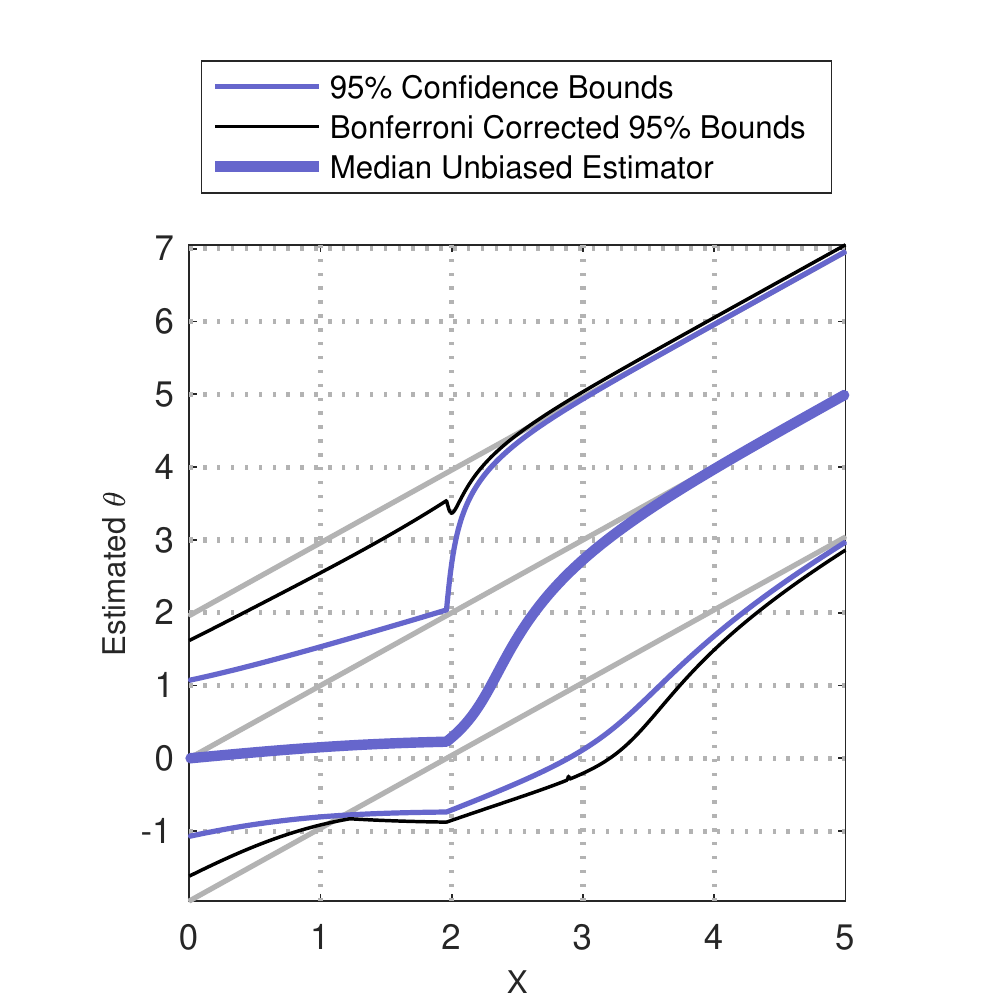}
\end{center}
\caption{This figure plots 95\% confidence bounds and the median unbiased estimator for the selection estimates based on replication data from \cite{camerer2016evaluating}. The usual (uncorrected) estimator and confidence bounds are plotted in grey for comparison.\label{fig: Econ confidence bounds}}
\endminipage\hfill
%\end{figure}
%
%\begin{figure}[t!]
\minipage{0.45\textwidth}
\begin{center}
\includegraphics[width=\textwidth]{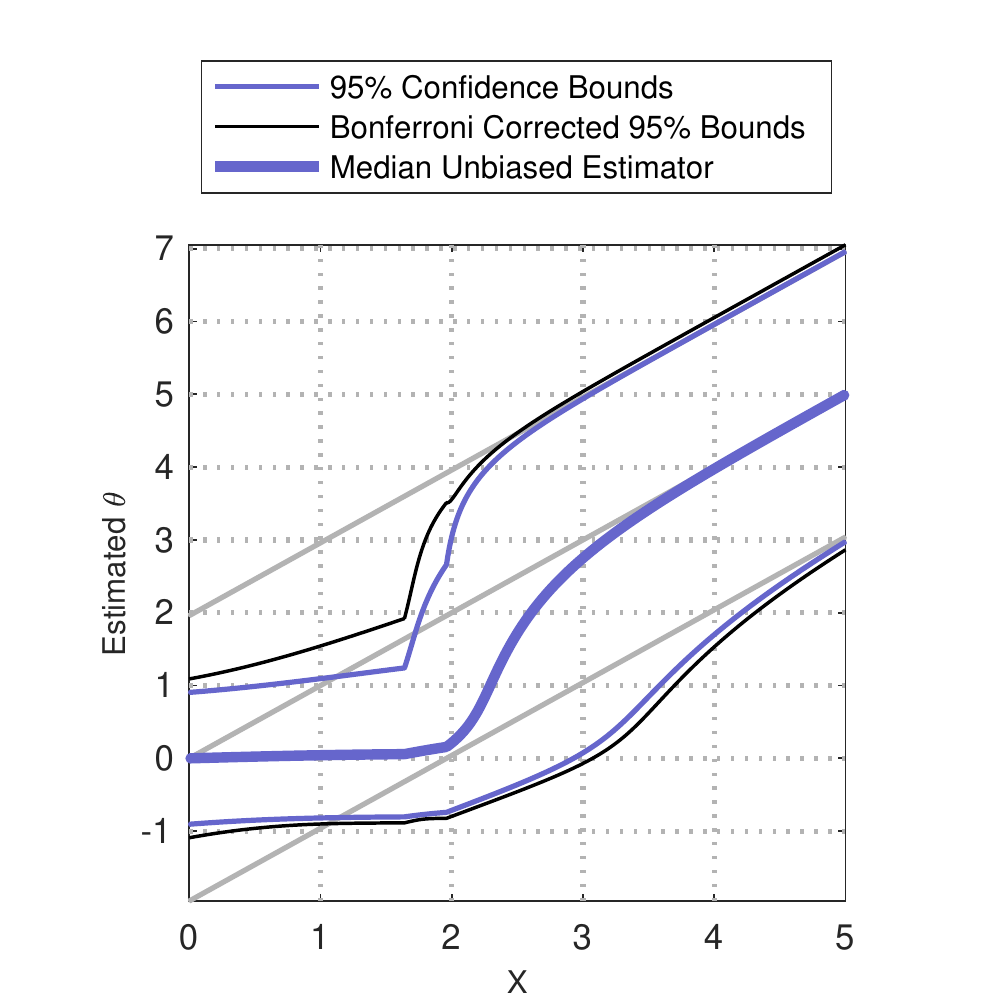}
\end{center}
\caption{This figure plots 95\% confidence bounds and the median unbiased estimator for the selection estimates based on replication data from \cite{open2015estimating}. The usual (uncorrected) estimator and confidence bounds are plotted in grey for comparison.\label{fig: Psych confidence bounds}}
%\end{figure}
%
%\begin{figure}[t!]
\endminipage\hfill

\minipage{.45\textwidth}
\begin{center}
\includegraphics[width=\textwidth]{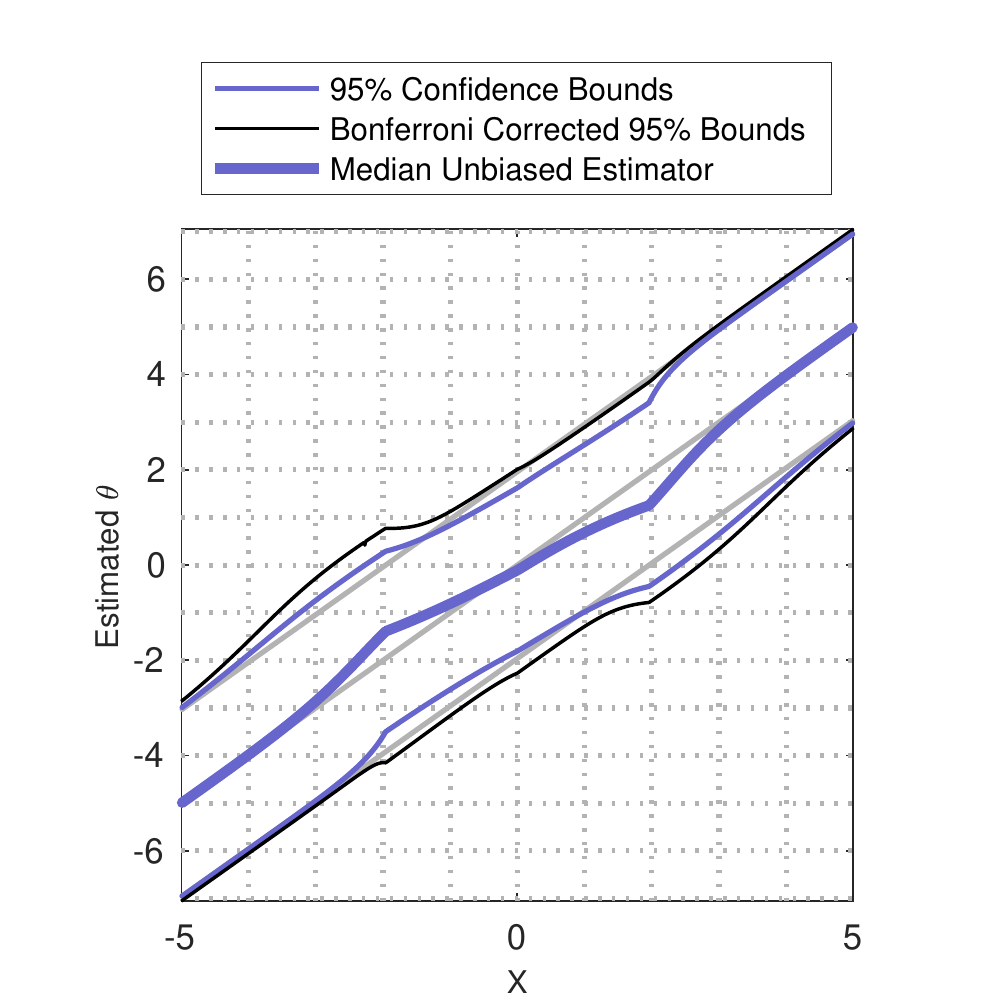}
\end{center}
\caption{The figure to the left plots 95\% confidence bounds and the median unbiased estimator for the selection estimates based on replication data from \cite{wolfson201515}. The usual (uncorrected) estimator and confidence bounds are plotted in grey for comparison. \label{fig: Min wage confidence bounds}}
\endminipage\hfill
\minipage{.45\textwidth}
\begin{center}
\includegraphics[width=\textwidth]{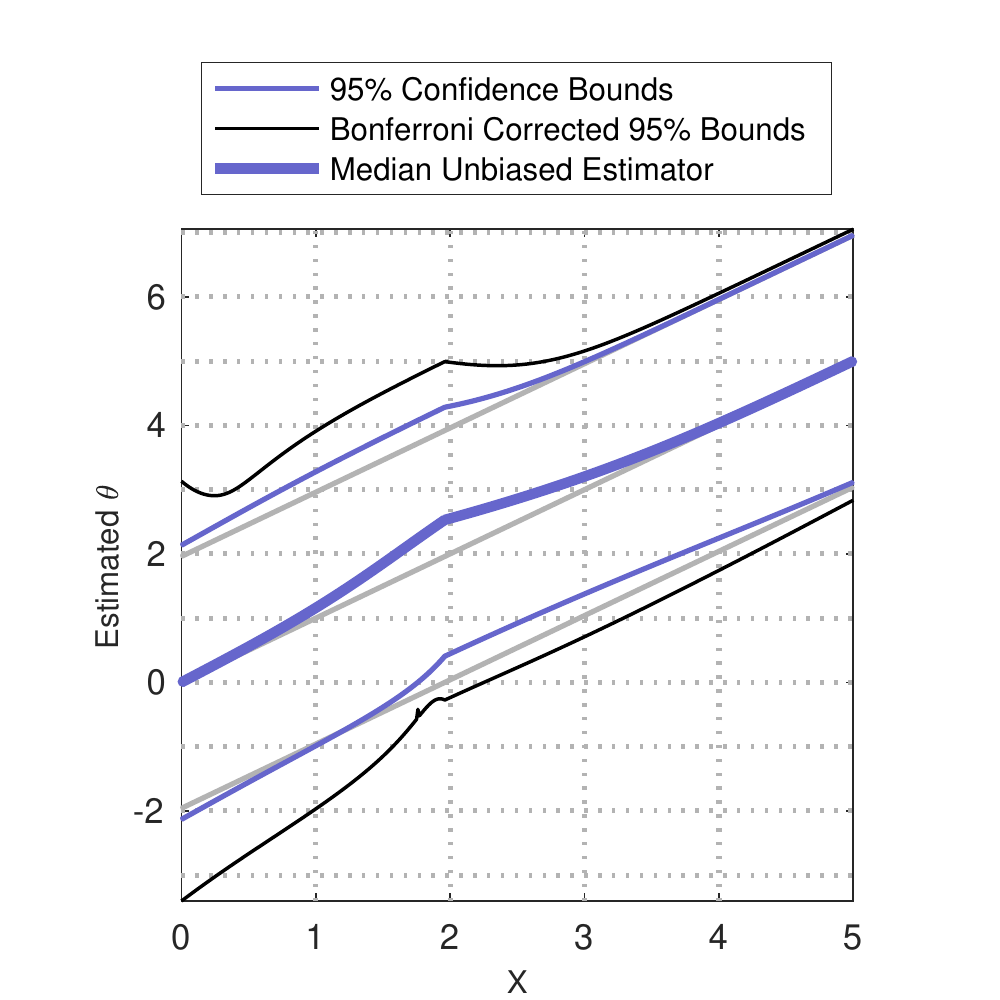}
\end{center}
\caption{This figure plots 95\% confidence bounds and the median unbiased estimator for the selection estimates based on replication data from \cite{deworming2016}. The usual (uncorrected) estimator and confidence bounds are plotted in grey for comparison.\label{fig: Deworming confidence bounds}}
\endminipage\hfill

\end{figure}

In this section, we plot our median unbiased estimators and corrected confidence sets, analogous to Figure \ref{fig: Confidence bounds} of the paper, based on the selection estimates from our applications.
Corrections based on replication estimates from the \cite{camerer2016evaluating} data are plotted in Figure \ref{fig: Econ confidence bounds}.  Corrections based on replication estimates from the \cite{open2015estimating} data are plotted in Figure \ref{fig: Psych confidence bounds}.  Corrections based on estimates using data from  \cite{wolfson201515} are reported in Figure \ref{fig: Min wage confidence bounds}. Finally, corrections based on estimates from the \cite{deworming2016} data are plotted in Figure \ref{fig: Deworming confidence bounds}.

\section{Inference when selection depends on multiple variables}
\label{suppsec: Inference in multivariate normal models}

In this section, we extend the frequentist inference results developed in the main text 
to cases where publication decisions are based not just on a scalar, but instead on a normally distributed vector of estimates.
Let
$X_{i}^{*}$ represent the estimates from study $i$, 
and assume that
\[
X_{i}^{*}|\Theta_{i}^{*}\sim N\left(\Theta_{i}^{*},\Sigma\right)
\]
for $\Sigma$ known.  Assume that $\Sigma$ is constant across latent studies $i$; the generalization to the case where latent study $i$ has variance $\Sigma^*_i$ is immediate.  
Since $X_{i}^{*}$ is a vector, $\Sigma$ is a matrix.  We thus obtain the following density for $X^*$ given $\Theta^*$:

\begin{Ass} \label{Ass: Normal Base Distribution} The distribution
$f_{X^{*}|\Theta^{*}}\left(x|\theta\right)$ is multivariate normal
with mean $\theta$ and variance $\Sigma$: 
\[
f_{X^{*}|\Theta^{*}}\left(x|\theta\right)=\left(2\pi\right)^{-\frac{k}{2}}\left|\Sigma\right|^{-\frac{1}{2}}\exp\left(-\frac{1}{2}\left(x-\theta\right)'\Sigma^{-1}\left(x-\theta\right)\right).
\]
\end{Ass}

We consider inference on $\Gamma=v'\Theta$ for a known non-zero
vector $v$, treating the other elements of $\Theta$, denoted $\Omega,$ as nuisance
parameters. %This corresponds to the common empirical practice of reporting separate standard errors and confidence intervals for each component of a parameter vector. 
To conduct inference on the
$i$th element of $\Theta$ we can simply take $v$ to be the $i$th
standard basis vector. 
To illustrate our results, we consider the example of difference in differences estimation, with selection on both statistical significance and a test for parallel trends.

\subsection{Illustrative example: difference in differences}

Suppose we observe
data from two states, $s\in\left\{ 1,2\right\} $ over three time
periods $t\in\left\{ 1,2,3\right\} $. Denote the average outcome for residents of state
$s$ at time $t$ by $Y_{st}$, and note that under regularity conditions, $Y_{st}$
will be approximately normally distributed
\[
Y_{st}\sim N\left(\mu_{st},\sigma_{st}^{2}\right).
\]
For simplicity we assume that $Y_{st}$ is independent of $Y_{s't'}$
if $s\neq s'$ or $t\neq t'$. 
%This corresponds to a simple model
%where we consider state-level average outcomes with no controls.

Suppose we are interested in estimating the effect of a particular
state-level policy, and let $D_{st}$ be a dummy for the presence
of the policy in state $s$ at time $t$. The difference in differences
model (with no control variables) assumes that 
\[
\mu_{st}=\alpha_{s}+\beta_{t}+D_{st}\gamma.
\]
If we are interested in the effect of a policy enacted in state $1$
in period $3$ and nowhere else in the sample, for example, we would
take 
\[
D_{st}=\left\{ s=1,t=3\right\} .
\]
A key identifying assumption in the difference-in-differences model
is that the only source of variation in $\mu_{st}$ at the state-by-time
level is the policy change of interest. In particular, while
we allow state fixed effects $\alpha_{s}$ and time fixed effects
$\beta_{t}$, we rule out state-time-specific effects other than
those acting through $D_{st}.$ This is known as the parallel trends
assumption. 

With only two periods of data this assumption is untestable, since
we have four free parameters $\left(\alpha_{1},\alpha_{2},\beta_{2},\gamma\right)$
and only four means $(\mu_{11},\mu_{12},\mu_{21},\mu_{22})$. With
data from an additional time period, however, we have five free parameters
and six means and so can instead consider the model 
\[
\mu_{st}=\alpha_{s}+\beta_{t}+\tilde{D}_{st}\lambda+D_{st}\gamma
\]
where 
\[
\tilde{D}_{st}=\left\{ s=1,t=2\right\} 
\]
and the parallel trends assumption implies that $\lambda=0$. Thus,
given data from two states in three time periods the parallel trends
assumption is testable.

Formal and informal tests of parallel trends are common in applications
of difference in differences strategies. To describe a formal test
in our setting, note that the natural 
estimator $\left(G,L\right)$ for $\left(\gamma,\lambda\right)$ has
a simple form,
\[
\left(G,L\right)=\left(\left(X_{13}-X_{12}\right)-\left(X_{23}-X_{22}\right),\left(X_{12}-X_{11}\right)-\left(X_{22}-X_{21}\right)\right).
\]
To test the parallel trends assumption in this setting we again want to test that  $\lambda$, the mean of $L$, is equal to zero.

Consider a population of latent studies with the structure just described,
and let us further simplify the model by setting $\sigma_{st}=1$
for all $t$. For latent estimates $X^{*}=\left(G^{*},L^{*}\right)$
and latent true effects $\Theta^{*}=\left(\Gamma^{*},\Lambda^{*}\right)$,
\[
\left.\left(\begin{array}{c}
G^{*}\\
L^{*}
\end{array}\right)\right|\left(\begin{array}{c}
\Gamma^{*}\\
\Lambda^{*}
\end{array}\right)\sim N\left(\left(\begin{array}{c}
\text{\ensuremath{\Gamma^{*}}}\\
\Lambda^{*}
\end{array}\right),\left(\begin{array}{cc}
4 & 2\\
2 & 4
\end{array}\right)\right)
\]
where the covariance matrix is known. 
%Note that while we have derived this joint
%normal distribution in a stylized setting, difference-in-differences
%and pre-trend estimates are jointly asymptotically normal with consistently
%estimable variance for a broad range of empirically relevant models
%that allow for more states, more periods, control variables, clustering at the state
%level, and so on. Consequently, analogs of the finite-sample results
%we derive here apply asymptotically in wide variety of settings.

As in our illustrative example in the main text, assume studies
that reject $\gamma=0$ at the 5\% level are
ten times more likely to be published than studies that do not. In
addition, assume studies that reject $\lambda=0$ at the 5\%
level are ten times \emph{less}
likely to be published than studies that do not.
This leads to publication probability 
\[
p\left(X\right)\propto 1\left\{ \frac{\left|G^{*}\right|}{\sigma_{G}}>1.96,\frac{\left|L^{*}\right|}{\sigma_{L}}\le1.96\right\} 1+1\left\{ \frac{\left|G^{*}\right|}{\sigma_{G}}>1.96,\frac{\left|L^{*}\right|}{\sigma_{L}}\ge1.96\right\} 0.1
\]
\[
+1\left\{ \frac{\left|G^{*}\right|}{\sigma_{G}}\le1.96,\frac{\left|L^{*}\right|}{\sigma_{L}}\le1.96\right\} 0.1+1\left\{ \frac{\left|G^{*}\right|}{\sigma_{G}}\le1.96,\frac{\left|L^{*}\right|}{\sigma_{L}}>1.96\right\} 0.01.
\]
This publication rule favors studies that find significant difference in difference estimates, and disfavors studies that reject the parallel trends assumption.

\begin{figure}
\minipage{0.45\textwidth}
\begin{center}
\includegraphics[width=\textwidth]{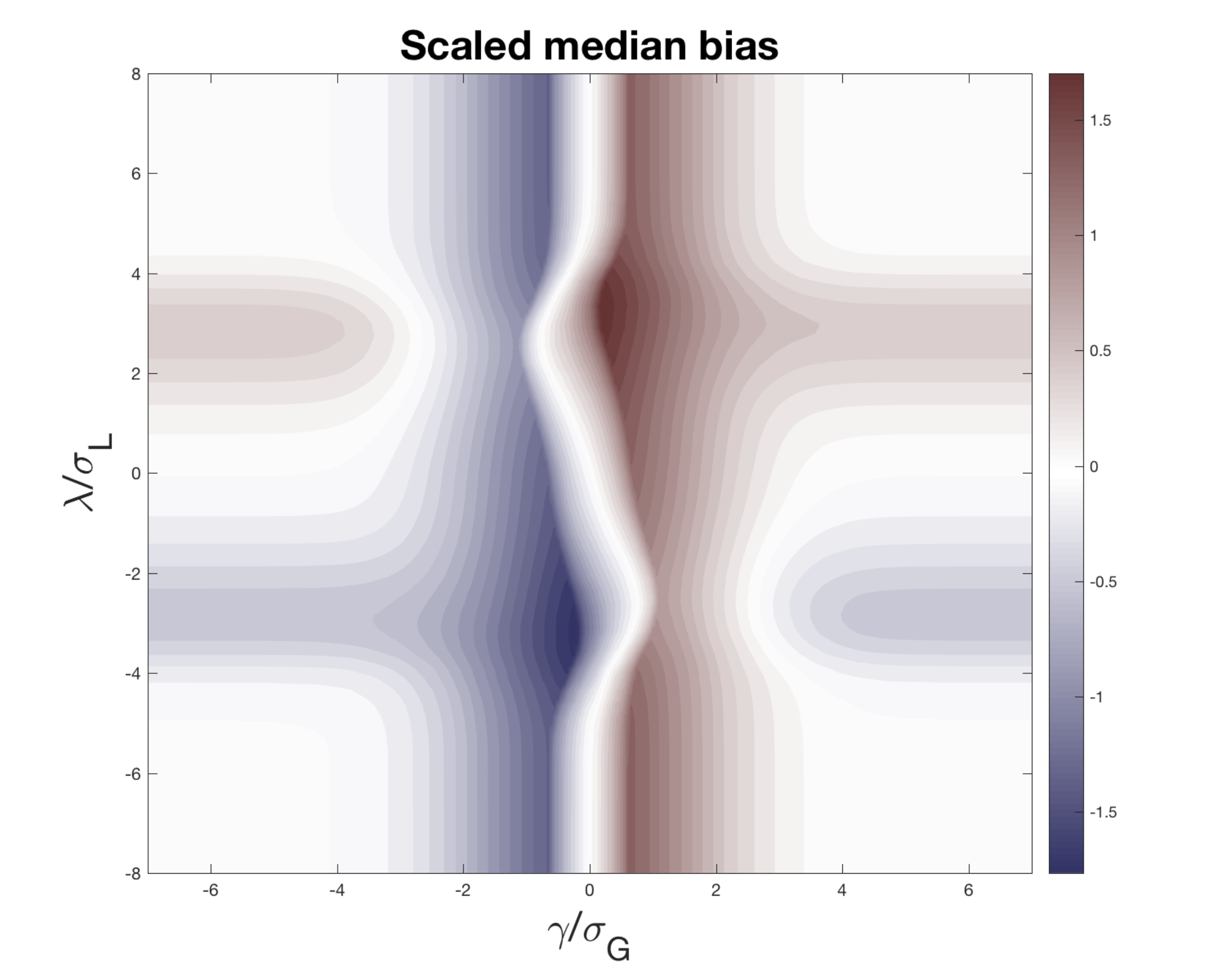}
\end{center}
\caption{This figure plots the median bias of $\left(G\right)/\sigma_{G}$ in the difference
in differences example.\label{fig:DiffinDiff Median Bias}}
\endminipage\hfill
\minipage{0.45\textwidth}
\begin{center}
\includegraphics[width=\textwidth]{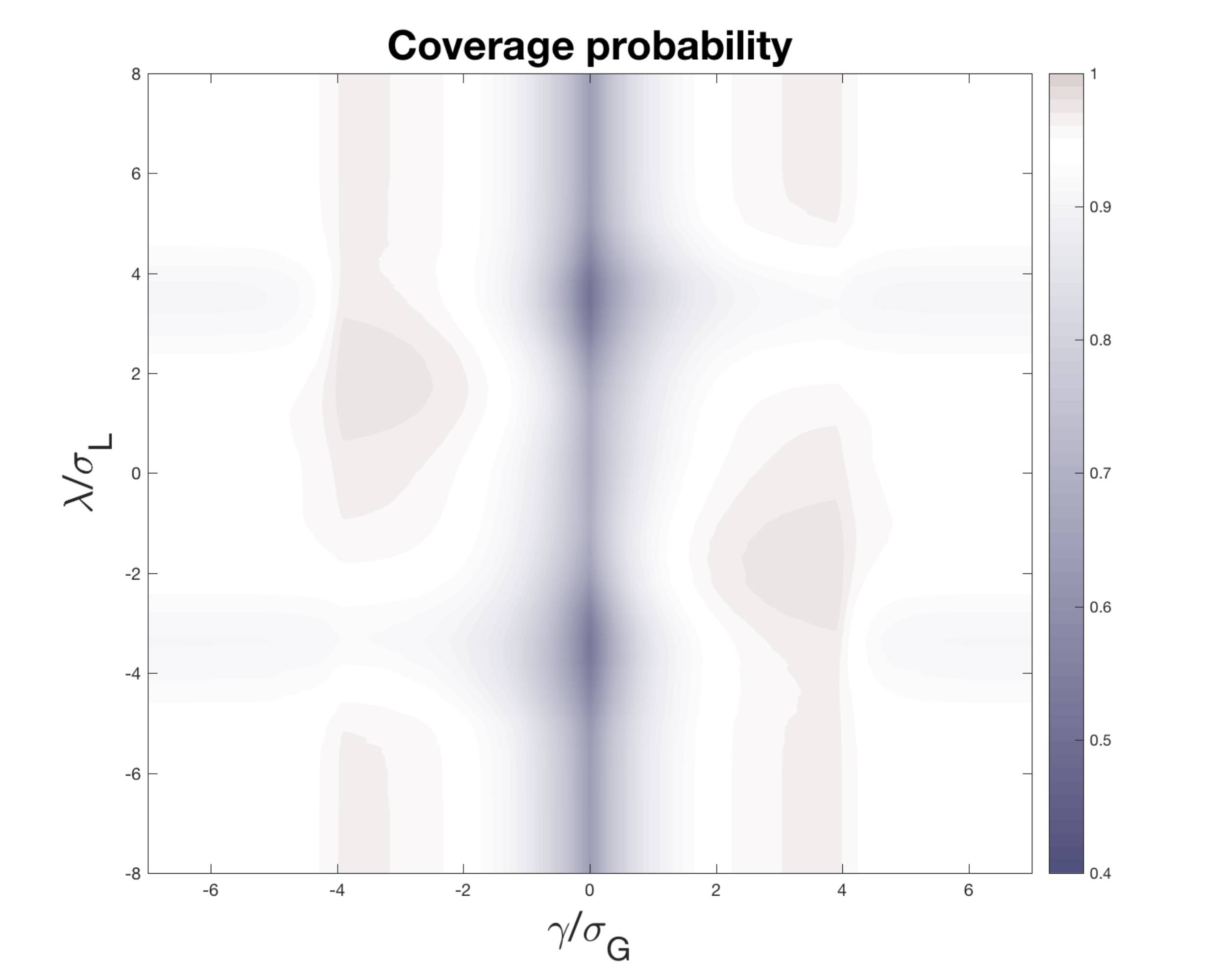}
\end{center}
\caption{This figure plots the coverage of conventional 95\% confidence sets in the difference
in differences example.\label{fig:DiffinDiff Coverage}}
\endminipage\hfill
\end{figure}

To illustrate the effect of selective publication in this setting,
Figure \ref{fig:DiffinDiff Median Bias} plots the median bias of $G$ as an
estimator for $\gamma$ (scaled by the standard deviation $\sigma_{G}$
of $G^*$).  Selective publication results in large bias for the conventional estimator $G$, which depends on both the parameter of interest $\gamma$ and the nuisance parameter $\lambda.$  Analogously, Figure \ref{fig:DiffinDiff Coverage} plots the coverage of the usual two-sided confidence
set $G^*\pm1.96\sigma_{G}$, and shows that selective publication results in substantial coverage distortions.

\subsection{Sufficient statistic for nuisance parameter}

To conduct inference on $\gamma,$ treating $\omega$ as a nuisance parameter, it will be
helpful to derive a sufficient statistic for $\omega.$  Note that for $M\left(v\right)$
a $\left(\dim\left(X\right)-1\right)\times\dim\left(X\right)$ matrix
such that $M\left(v\right)\left(I-\frac{\Sigma vv'}{v'\Sigma v}\right)$
has full row-rank, 
\[
\left(G\left(x\right),W\left(x\right)\right)=\left(v'x,M\left(v\right)\left(I-\frac{\Sigma vv'}{v'\Sigma v}\right)x\right)
\]
is a one-to one transformation of $x$. Thus $\left(G,W\right)=\left(G\left(X\right),W\left(X\right)\right)$
are jointly sufficient for $\theta$, and rather than basing inference
on $X$ we can equally well base inference on $\left(G,W\right)$.
Note moreover that for $G^{*}=G\left(X^{*}\right)$ and $W^{*}=W\left(X^{*}\right),$
$X^{*}\sim N\left(\theta,\Sigma\right)$ implies that

\begin{equation}
\left(\begin{array}{c}
G^{*}\\
W^{*}
\end{array}\right)\sim N\left(\left(\begin{array}{c}
\gamma\\
\omega
\end{array}\right),\left(\begin{array}{cc}
\sigma_{G}^{2} & 0\\
0 & \Sigma_{W}
\end{array}\right)\right)\label{eq: Normal G W Distribution}
\end{equation}
for $\omega=M\left(v\right)\left(I-\frac{\Sigma vv'}{v'\Sigma v}\right)\theta,$
$\sigma_{G}^{2}=v'\Sigma v$, and $\Sigma_{W}=M\left(v\right)\left(I-\frac{\Sigma vv'}{v'\Sigma v}\right)\Sigma\left(I-\frac{vv'\Sigma}{v'\Sigma v}\right)M\left(v\right)'$.
Thus the conditional distribution of $G^{*}$ given $W^{*}$ depends
only on $\gamma,$
\[
G^{*}|W^{*}\sim N\left(\gamma,\sigma_{G}^{*}\right),
\]
and by conditioning on $W^*$ we can eliminate dependence on the
nuisance parameter $\omega$. This property continues to hold for the
conditional distribution of published $G$ given $W$, as the following lemma
shows.

\begin{lem} \label{lem: conditional exp family structure} Under
Assumption \ref{Ass: Normal Base Distribution}, the conditional density
$G|W,\Gamma$ is given by 
\begin{equation}
f_{G|W,\Gamma}\left(g|w,\gamma\right)=\frac{p\left(g,w\right)}{E\left[p\left(G^{*},W^{*}\right)|W^{*}=w,\Gamma^{*}=\gamma\right]}\frac{1}{\sigma_{G}}\phi\left(\frac{g-\gamma}{\sigma_{G}}\right)\label{eq: conditional density of G given W}
\end{equation}
for $\phi$ the standard normal density, where we use the
fact that $\left(g,w\right)$ is a one-to-one transformation of $x$
to write $p\left(g,w\right)=p\left(x\left(g,w\right)\right).$

\end{lem}

\paragraph{Proof of Lemma \ref{lem: conditional exp family structure}}

Note that we can draw from the conditional distribution $G|W=w,\Gamma=\gamma$
by drawing from the conditional distribution $G^{*}|W^{*}=w,\Gamma^{*}=\gamma$
and discarding the draw $G^{*}$ with probability $1-p\left(G^{*},w\right)$.
The result then follows by the same argument as Lemma \ref{lem:likelihood}. $\Box$

Thus, we see that the conditional density of $G$ given $W$ depends
only on the parameter of interest $\gamma$ and not on the nuisance
parameter $\omega$. Hence, by conditioning on $W$ we can eliminate
the nuisance parameter and conduct inference
on $\gamma$ alone.

\subsection{Optimal quantile-unbiased estimates}

To conduct frequentist inference, we generalize the median-unbiased
estimator and equal-tailed confidence set proposed in Section \ref{sec:inference}
to the present setting.  Using a result from \cite{Pfanzagl1994}
we show that the resulting quantile-unbiased estimators are optimal
in a strong sense.

Formally, define $\hat{\gamma}_{\alpha}\left(X\right)$ by 
\[
F_{G\left(X\right)|W\left(X\right),\Gamma}\left(G|W,\hat{\gamma}_{\alpha}\left(X\right)\right)=\alpha.
\]
This estimator is simply the value $\gamma$ such that the observed
 $G$ lies at the $\alpha$ quantile of the corresponding conditional
distribution given $W$. The following theorem, based on the results
of \cite{Pfanzagl1994}, shows that this estimator is both quantile-unbiased
and, in a strong sense, optimal in the class of quantile-unbiased
estimators.

\begin{theorem} \label{thm: optimal quant unbiased} Let Assumption
\ref{Ass: Normal Base Distribution} hold, and assume further that
the conditional distribution of $G$ given $W$ is absolutely continuous
for all $\gamma$ and almost every $W$, and that the parameter space
for $\omega$ given $\gamma$ contains an open set for all $\gamma$.
Then
\begin{enumerate}
\item The estimator $\hat{\gamma}_{\alpha}\left(X\right)$ is level-$\alpha$
quantile unbiased: 
\[
Pr\left\{ \hat{\gamma}_{\alpha}\left(X\right)\le\gamma|\Theta=\left(\gamma,\omega\right)\right\} =\alpha\,\mbox{for all }\gamma,\omega,
\]

\item This estimator is uniformly most concentrated in the class of level-$\alpha$
quantile-unbiased estimators, in the sense that for any other level-$\alpha$
quantile unbiased estimator $\tilde{\gamma}\left(X\right)$ and any
loss function $L\left(d,\gamma\right)$ that attains its minimum at
$d=\gamma$ and is increasing as $d$ moves away from $\gamma,$ 
\[
E\left[L\left(\hat{\gamma}_{\alpha}\left(X\right),\gamma\right)|\Theta=\left(\gamma,\omega\right)\right]\le E\left[L\left(\tilde{\gamma}\left(X\right),\gamma\right)|\Theta=\left(\gamma,\omega\right)\right]\,\mbox{for all }\gamma,\omega.
\]

\end{enumerate}
\end{theorem}

\paragraph{Proof of Theorem \ref{thm: optimal quant unbiased}}

Since the multivariate normal distribution belongs to the exponential family,
we can write 
\[
f_{G^{*},W^{*}|\Theta^{*}}\left(g,w|\theta\right)=\tilde{h}\left(g,w\right)\tilde{r}\left(\gamma\left(\theta\right),\omega\left(\theta\right)\right)\exp\left(\gamma\left(\theta\right)g+\omega\left(\theta\right)'w\right).
\]
By the same argument as in the proof of Lemma \ref{lem:likelihood}, this implies that 
\begin{equation}
f_{G,W|\Theta}\left(g,w|\theta\right)=h\left(g,w\right)r\left(\gamma\left(\theta\right),\omega\left(\theta\right)\right)\exp\left(\gamma\left(\theta\right)g\right)\exp\left(\omega\left(\theta\right)'w\right)\label{eq: G W density}
\end{equation}
for $h\left(g,w\right)=p\left(g,w\right)\tilde{h}\left(g,w\right)$
and 
\[
r\left(\gamma,\omega\right)=\frac{\tilde{r}\left(\gamma,\omega\right)}{E\left[p\left(X_{i}^{*}\right)|\Theta_{i}^{*}=\theta\left(\gamma,\omega\right)\right]}.
\]

The density (\ref{eq: G W density}) has the same structure
as (5.5.14) of \cite{Pfanzagl1994}, and satisfies properties (5.5.1)-(5.5.3)
of \cite{Pfanzagl1994} as well. Part 1 of the theorem then follows
immediately Theorem 5.5.9 of \cite{Pfanzagl1994}.

Part 2 of the theorem follows by using Theorem 5.5.9 of \cite{Pfanzagl1994}
along with (\ref{eq: G W density}) to verify the conditions of Theorem
5.5.15 of \cite{Pfanzagl1994}. $\Box$

Using this result we see that $\hat{\gamma}_{\frac{1}{2}}\left(X\right)$
is the optimal median-unbiased estimator for the parameter of interest
$\gamma$. A natural level-$\alpha$ confidence interval to accompany
this estimator is then the equal-tailed confidence interval 
\[
CS=\left[\hat{\gamma}_{1-\frac{\alpha}{2}}\left(X\right),\hat{\gamma}_{\frac{\alpha}{2}}\left(X\right)\right].
\]

\paragraph{Difference in differences example (continued)}

\begin{figure}
\minipage{0.45\textwidth}
\begin{center}
\includegraphics[width=\textwidth]{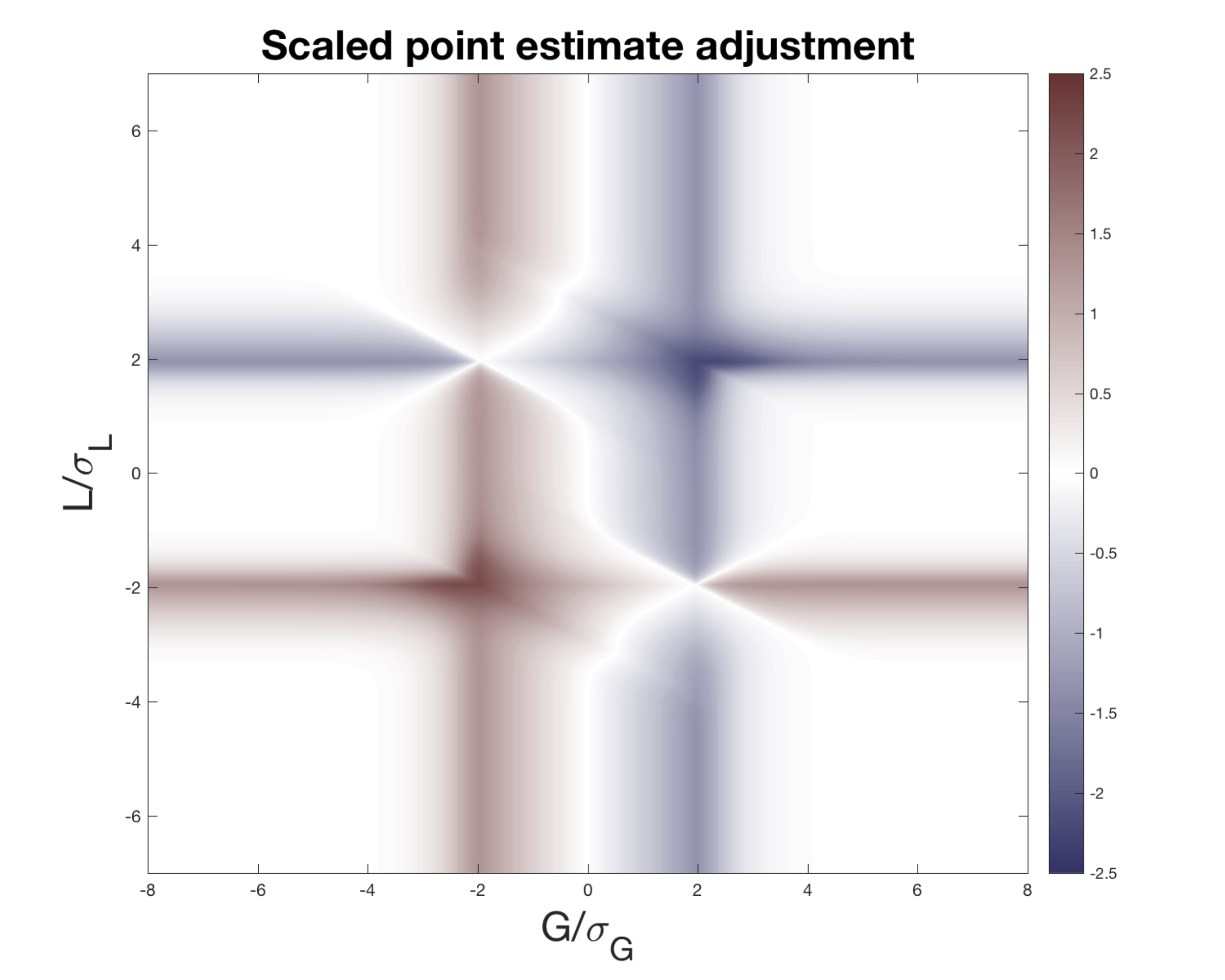}
\end{center}
\caption{This figure plots the difference between the median-unbiased estiamtor $\hat\gamma_{\frac{1}{2}}(X)$ and the conventional estimator $G$ in the difference-in-differences example.\label{fig:DiffinDiff Median Unbiased}}
\endminipage\hfill
\minipage{0.45\textwidth}
\begin{center}
\includegraphics[width=\textwidth]{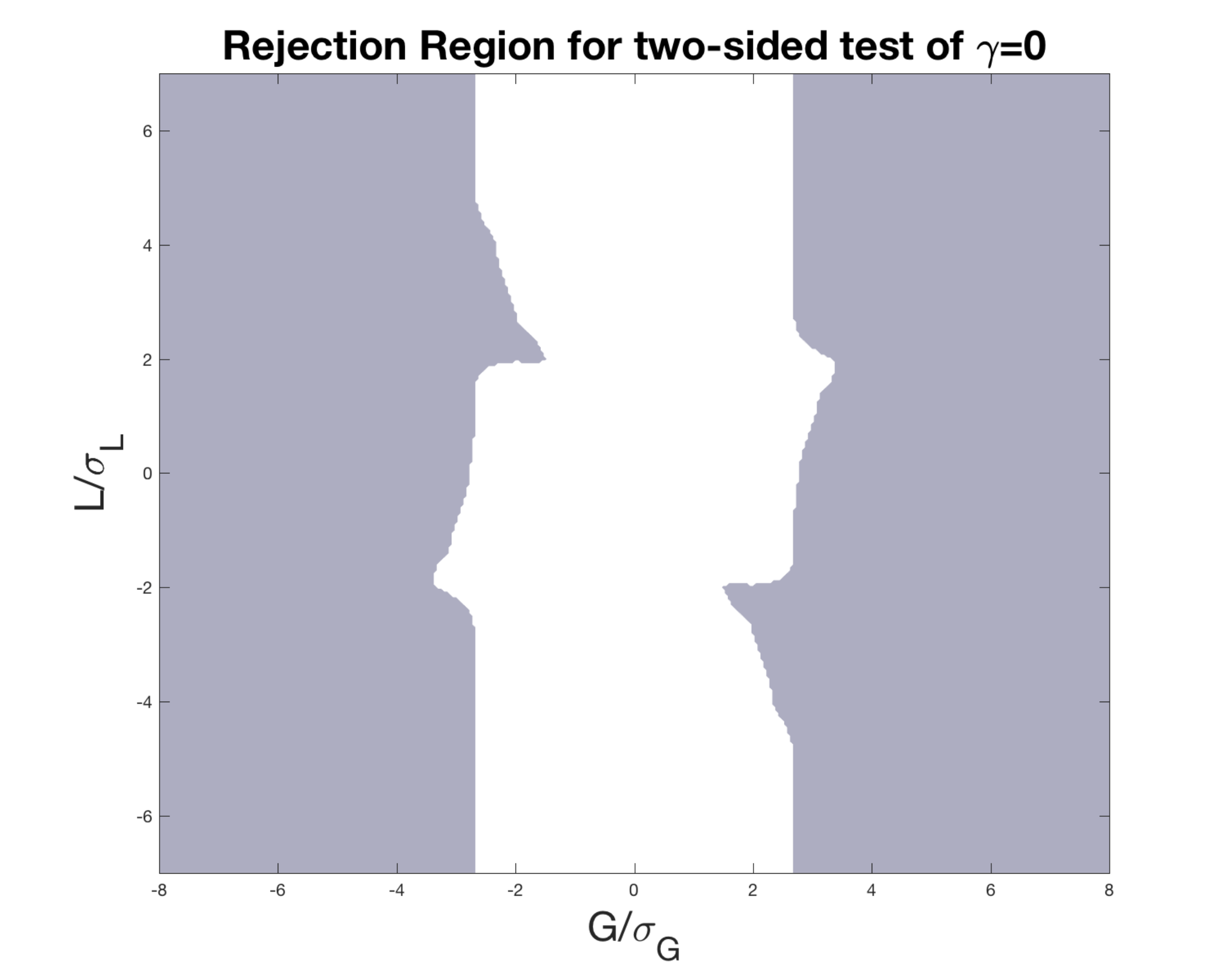}
\end{center}
\caption{This figure plots the (shaded) rejection region for a 5\% test of $H_0:\gamma=0$ based on equal-tailed confidence sets for $\gamma$ in the differences in differences example.\label{fig:DiffinDiff Rejection}}
\endminipage\hfill
\end{figure}

To illustrate our corrections in a multivariate setting, Figure \ref{fig:DiffinDiff Median Unbiased} plots the difference between our median-unbiased estimator $\hat\gamma_{\frac{1}{2}}(X)$ and the conventional estimator $\hat\gamma=G$ in the difference-in-differences example.  As this plot makes clear, $\hat\gamma_{\frac{1}{2}}(X)$ depends on both $G$ and $L.$  Thus, while we are interested only in the difference-in-differences parameter $\gamma,$ the result for the pretest of parallel trends also plays a role in our estimate.  Figure \ref{fig:DiffinDiff Rejection} plots the rejection region for a 5\% test of $H_0:\gamma=0$ based on our equal-tailed confidence interval for $\gamma$.  As this plot shows, the results of this test likewise depend on both $G$ and $L.$

\section{Bayesian inference}
\label{suppsec: Bayesian inference}

In the main text we discuss the effect of selective publication on frequentist inference on $\theta$ under known $p(\cdot)$.  The effect of selective publication on Bayesian inference is more subtle, and depends on the prior.  Here we briefly discuss Bayesian inference on $\theta$ under known $p(\cdot)$  for two natural classes of priors.
These priors can be thought of as two extreme points of the set of relevant priors.

\begin{defi}[Two classes of priors]
\label{def:specialpriors}
Consider the following two classes of prior distributions $\pi_\mu$ for $\mu$:
\begin{enumerate}
\item Unrelated Parameters: $\pi_\mu$ is a point mass at some $\mu$, so that $\mu$ is known and the prior distribution of $\Theta^*_i$ is i.i.d. across $i$.
\item Common Parameters: $\pi_\mu$ assigns positive probability only to point-measures $\mu$, so that $\Theta^*_i$ is constant across $i$ (equal to $\Theta^*_0$) with probability $1$.
\end{enumerate}
\end{defi}

The unrelated parameters prior corresponds to the case where each latent study considers a different parameter.  Thus, under priors in this class, learning the true parameter value $\Theta^*_i$ in latent study $i$ conveys no information about the true parameter value $\Theta^*_{i'}$ in latent study $i',$ and $\Theta^*_i$ is iid across $i$.
The common parameters prior, by contrast, assumes that all latent studies attempt to estimate the same parameter $\Theta^*_0.$ Thus, priors in this class imply that $\Theta_i^*$ is perfectly dependent across $i.$

For both the unrelated and common parameters classes, the marginal prior $\pi_{\Theta^*}$ for $\Theta^*$ is unrestricted.  For any  $\pi_{\Theta^*}$ there is a unique prior in each class consistent with this marginal distribution.  

If we observe a single draw $X^*,$ our posterior for $\Theta^*$ depends only on the marginal prior  $\pi_{\Theta^*},$ and so is the same whether we consider the unrelated or common parameters priors.  By contrast, when we observe a single draw $X$ from the distribution of published papers, which class of priors we use turns out to be important.  The following result is closely related to the findings of \cite{Yekutieli2012}.

\begin{lem}[Two posterior distributions]
Based on single observation of $X,$ we obtain the following posteriors:
\label{lem:posterior}
\begin{enumerate}
\item Under unrelated parameters priors:
\begin{equation*}
f_{\Theta|X}(\theta | x) = f_{X^*|\Theta^*}(x|\theta) \cdot \pi_{\Theta^*}(\theta) / \pi_{X^*}(x)
\end{equation*}
\item Under common parameters priors:
\begin{equation*}
f_{\Theta|X}(\theta | x) =  \frac{p\left(x\right)}{E\left[p\left(X^*_i\right)|\Theta_i^*=\theta\right]} f_{X^*|\Theta^*}(x|\theta) \cdot \pi_{\Theta^*}(\theta) / \pi_{X^*}(x)
\end{equation*}
\begin{equation*}
\propto f_{X|\Theta}(x|\theta)\cdot \pi_{\Theta^*}(\theta)
\end{equation*}
\end{enumerate}
\end{lem}

\paragraph{Proof of Lemma \ref{lem:posterior}: }
\begin{enumerate}
\item Unrelated parameters: By construction $D_i \perp \Theta_i | X^*_i,$
and thus
\begin{align*}
f_{\Theta|X} (\theta | x) &= f_{\Theta^*_i|X^*_i,D_i}(\theta | x, d=1)\\
&= f_{\Theta^*_i|X^*_i}(\theta | x)\\
&= f_{X^*|\Theta^*}(x|\theta) \cdot \pi_{\Theta^*}(\theta) / f_{X^*}(x).
\end{align*}
\item Common parameters: 
This follows immediately from the truncated likelihood derived in Lemma \ref{lem:likelihood} of the main text.
\end{enumerate}
$\Box$

Under the unrelated parameters prior, our posterior $f_{\Theta|X}(\theta | x)$ after observing $X=x$ is the same as our posterior had we observed $X^*=x.$  The form of $p(\cdot)$ has no effect on our posterior distribution, and inference proceeds exactly as in the case without selection.  Under the common parameters prior, by contrast, our posterior $f_{\Theta|X}(\theta | x)$ corresponds to updating our marginal prior $\pi_{\Theta^*}$ using the truncated likelihood $f_{X|\Theta}(x|\theta)$ derived in Lemma \ref{lem:likelihood}.

The fact that selection has no effect on our posterior under the common parameters prior may be surprising, but reflects the fact that under this prior, selection changes the marginal prior $\pi_\Theta$ for true effects in published studies.  In particular, under this prior we have
$$\pi_\Theta(\theta)=\frac{E\left[p\left(X^*_i\right)|\Theta_i^*=\theta\right]}{E\left[p\left(X^*_i\right)\right]}\pi_{\Theta^*}(\theta),$$ which reflects the fact that the distribution of true effects for published studies differs from that for latent studies under this prior.  When we update this prior based on observation of $X,$ however, the adjustment by $E\left[p\left(X^*_i\right)|\Theta_i^*=\theta\right]$ in the prior cancels that in the likelihood, and selection has no net effect on the posterior.   Under the common parameters prior, by contrast, $\pi_{\Theta^*}=\pi_{\Theta},$ so the adjustment term in the prior due to selective inference continues to play a role in the posterior.  For related discussion, see  \cite{Yekutieli2012}.

\section{Optimal selection for publication in a simple model}
\label{suppsec:Optimal selection in a simple model}

In the main text we discuss how to account for selective publication in inference and how to identify selectivity. It is natural to ask, however, whether selective publication is a good idea in the first place or just a misguided application of statistics leading to either publication bias or needlessly complicated inference.  The answer to this question depends on the journal's objective function.
One possibility is as follows.  Suppose that published estimates are inputs into policy decisions, for instance in development economics, education, public finance, or medicine. If there are constraints on how many studies are published and read, then selectivity of the sort we observe might be justified.

We discuss a stylized version of this idea in a development economics context, though our model might also be considered a stylized description of medical publishing and doctors' prescriptions of treatments for patients.
Suppose that each $i$ corresponds to a different policy intervention.
Suppose the distribution $\mu$ of true treatment effects $\Theta_i^*$ is known to journal editors and readers, and that the expected effect $E[\Theta_i^*]$ of a randomly chosen treatment on the likelihood of escaping poverty is non-positive.
Suppose further that the journal is read by  policy makers who aim to minimize poverty. Assume finally that each treatment is relevant for a population of equal size, normalized to $1$.
A policy maker wishes to implement a given treatment $j$ if the expected impact on the outcome considered is positive, conditional on the observed estimate $X_j=x.$  Thus, their optimal treatment assignment rule is
\begin{equation}
t(x) = \mathbf{1}(E[\Theta_j | X_j=x] > 0),
\end{equation}
which results in the expected outcome
\begin{equation}
v(x) = \max(0, E[\Theta_j | X_j=x])
\end{equation}
where $E[\Theta_j | X_j]$ is the policymakers' posterior expectation of $\Theta_j$ after observing $X_j$.\footnote{Perhaps surprisingly, truncation is irrelevant for this posterior expectation.  This stems from the fact that we  assume policy makers have unrelated parameters priors as in Definition \ref{def:specialpriors} above.}
Suppose the journal also aims to minimize poverty, but faces a marginal (opportunity) cost of $c$, in units comparable to treatment outcomes, when publishing a given study. 
Policymakers update their behavior only for published studies with $E[\Theta_j | X_j]>0$. This updated behavior results in an expected poverty reduction of $E[\Theta_j | X_j]$ relative to the status quo.
It follows that the optimal publication rule for the journal is
\begin{equation}
p(X_i^*)=\mathbf{1}(E[\Theta_i^* | X_i^*] > c).
\end{equation}
If the conditional expectation is monotonic in $X_i^*$, this rule is equivalent to
\[p(X_i^*)=\mathbf{1}(X_i^* > x_c),\]
so that results should get published if they are positive and ``significant''  relative to the critical value $x_c$, defined via $E[\Theta_i^* | X_i^* =x_c] = c$. 

This result rationalizes selectivity in the publication process: the optimal rule derived here corresponds to one-sided testing.
A more realistic version of this story allows for variation across $i$ in the variance of $X_i^*$, the cost of implementing treatment, the size of the populations to be treated, etc. All of these would affect the critical value $x_c$, which thus should vary across $i$ and need not be equal to conventional critical values of hypothesis tests. What remains true, however, is that publication decisions that are optimal according to the above model are selective in a way which leads to publication bias, and correct inference needs to account for this selectivity.

\end{document}